\documentclass[fleqn,usenatbib]{mnras}
\usepackage[T1]{fontenc}
\usepackage{graphicx}
\usepackage{amsmath}
\usepackage{amssymb}
\usepackage{newtxtext,newtxmath}
\usepackage{comment}

\title[Blazar X-ray variability]{X-ray timing and spectral variability properties of blazars \\S5 0716+714, OJ 287, Mrk 501, and RBS 2070}

\author[Maksym Mohorian et al.]{
Maksym Mohorian,$^{1,2}$\thanks{E-mail: mogormax@gmail.com, gopal.bhatta@ifj.edu.pl}
Gopal Bhatta,$^{3}$
Tek P. Adhikari,$^{4}$
Niraj Dhital,$^{5}$
Radim P{\'a}nis,$^{6}$
\newauthor Adithiya Dinesh,$^{7}$
Suvas C. Chaudhary,$^{8}$
Rajesh K. Bachchan$^{9}$
and Zden\v{e}k Stuchl{\'i}k$^{6}$
\\
$^{1}$Department of Physics \& Astronomy, Macquarie University, Sydney, NSW 2109, Australia\\
$^{2}$Astronomy, Astrophysics and Astrophotonics Research Centre, Macquarie University, Sydney, NSW 2109, Australia\\
$^{3}$Institute of Nuclear Physics Polish Academy of Sciences PL-31342 Krak\'ow, Poland\\
$^{4}$Inter-University Centre for Astronomy and Astrophysics (IUCAA), Pune 411007, India\\
$^{5}$Central Department of Physics, Tribhuvan University, Kirtipur 44613, Nepal\\
$^{6}$Research Centre for Theoretical Physics and Astrophysics, Institute of Physics, Silesian University in Opava,\\
Bezru{\v c}ovo n{\'a}m.13, CZ-74601 Opava, Czech Republic\\
$^{7}$SUPA, School of Physics \& Astronomy, University of St Andrews, North Haugh, St Andrews, KY16 9SS, United Kingdom\\
$^{8}$Indian Institute of Technology Bombay, Powai, Mumbai 400076, India\\
$^{9}$Department of Physics, Patan Multiple Campus, Nepal
}

\date{Accepted 2021 December 20. Received 2021 December 3; in original form 2021 July 15.}

\pubyear{2021}

\begin{document}
\label{firstpage}
\pagerange{\pageref{firstpage}--\pageref{lastpage}}
\maketitle
\begin{abstract}
The X-ray emission from blazars has been widely investigated using several space telescopes. In this work, we explored statistical properties of the X-ray variability in the blazars S5 0716+714, OJ 287, Mkn 501 and RBS 2070 using the archival observations from the XMM-Newton telescope between the period 2002--2020. Several methods of timing and spectral analyses including fractional variability, minimum variability timescale, power spectral density analyses and countrate distribution were performed. In addition, we fitted various spectral models to the observations as well as estimated hardness ratio. The results show that the sources are moderately variable within the intra-day timescale. Three of the four sources exhibited a clear bi-modal pattern in their countrate distribution revealing possible indication of two distinct countrate states, that is, hard and soft countrate states. The slope indices of the power spectral density were found to be centered around 0.5. Furthermore, the spectra of the sources were fitted with single power-law, broken power-law, log-parabolic and black-body+log-parabolic models (the latter only for OJ 287). We conclude that for most of the observations log-parabolic model was the best fit. The power spectral density analysis revealed the variable nature of PSD slopes in the source light curves. The results of this analysis could indicate the non-stationary nature of the blazar processes on intra-day timescales. The observed features can be explained within the context of current blazar models, in which the non-thermal emission mostly arises from kilo-parsec scale relativistic jets.
\end{abstract}


\section{Introduction}
Blazars are an extreme class of radio-loud active galactic nuclei, characterized by high luminosity, broad-band continuum emission, rapid variability in flux and polarization. Blazar emission mainly consists of the Doppler boosted non-thermal emission from the relativistic jets. Blazar spectral energy distribution that extends from radio to $\gamma$-ray energies can be identified as the two bumps in the $\log\nu f_\nu$ vs. $\log\nu$ plane: the lower energy bump can be found in a large spectral range from radio up to X-rays, this peak originates from synchrotron emission of relativistic electrons in the blazar jets; yet the higher energy bump's nature (UV--$\gamma$) is still argued about. The widely discussed scenario on the origin of the latter bump is: inverse Compton scattering of low-energy photons by relativistic electrons (IC, see \cite{1992ApJ...397L...5M, 1993ApJ...416..458D, 1994ApJ...421..153S}); relativistic proton synchrotron radiation \citep{2003APh....18..593M}; and synchrotron radiation of secondary particles produced in proton-photon interactions \citep{2015MNRAS.448..910C}.

Blazars are classified into two categories based on the presence of emission lines in their SEDs: a class of more luminous sources called the flat-spectrum radio quasars (FSRQ) which show emission lines over the continuum and a class of less luminous BL Lacertae (BL Lac) sources which show weak or no such lines. BL Lacs are further classified by the synchrotron bump's peak frequency $\nu_{\rm p}$. The subclasses are as follows (see \cite{Fan2016}): low synchrotron peaked blazars (LSP; $\nu_{\rm p}<10^{14}$~Hz), intermediate synchrotron peaked blazars (ISP; $10^{14}<\nu_{\rm p}<10^{15.3}$~Hz) and high synchrotron peaked blazars (HSP; $\nu_{\rm p}>10^{15.3}$~Hz). Moreover, FSRQs also have the synchrotron peak frequency $\nu_{\rm p}<10^{14}$~Hz, however, they are easily distinguished from BL Lacs due to the difference in spectral properties such as bolometric luminosity (higher for FSRQs; lower for BL Lacs) and peak frequency of the $\gamma$-ray emission (lower for FSRQs; higher for BL Lacs).

The variability of blazar fluxes is strong across the entire electromagnetic  spectrum with the variability timescales varying from minutes to several decades \citep[see e.g.][]{Bhatta2018a,Bhatta2020,2021ApJ...909..103Z,Bhatta2021}. In particular, the X-ray variability properties of blazars are broadly studied using the observations from the space-bound telescopes such as RXTE, Chandra, XMM-Newton, NuSTAR, and Swift/XRT \citep[][and references therein]{2010A&A...516A..56H, 2018ApJ...859...49P, 2020arXiv200902289W, 1989A&AS...80..103S}. The multi-wavelength (MWL) variability properties of blazars usually can be ascribed to both the relativistic jets e. g. by shock-in-jet model \citep{1985ApJ...298..114M} and accretion disc based models \citep{2007A&A...473..819R}. However, the details of the processes leading to such MWL and multi-timescale variability are still under discussion.

In this paper, we conduct a comprehensive analysis of all X-ray light curves and spectra of the blazar sources RBS 2070, OJ 287, Mrk 501, and S5 0716+714 from XMM-Newton archive. We give brief information about the studied sources in Section \ref{sec:src}. We describe the observations and the corresponding data processing in Section \ref{sec:obs}. Section \ref{sec:mth} contains methodology of our analysis. We present our results in Section \ref{sec:res}, whereas the interpretation is discussed in Section \ref{sec:dsc}. The conclusions are discussed in Section \ref{sec:con}.

\begin{table*}
    \caption{General information considering the blazar sources studied in this paper. The penultimate column provides the redshift values (a is \protect\cite{2011NewA...16..503M}, b is \protect\cite{2001A&A...375..739D}). The last column shows the values of galactic absorption column density in [$10^{20}$ cm$^{-2}$] \protect\citep{1990ARA&A..28..215D,2005A&A...440..775K,2016yCat..35940116H}.}
    \label{tab:gen}
    \begin{tabular}{cccccc}\hline
        Source name & Source class & R.A. (J2000) & Dec. (J2000) & z & $N_H$ \\\hline
        S5 0716+714 & ISP, TeV & $07^h21^m53.45^s$ & $+71^\circ20'36.4''$ & 0.300$^b$ & 2.88 \\
        OJ 287 & ISP, TeV & $08^h54^m48.88^s$ & $+20^\circ06'30.6''$ & 0.306$^a$ & 2.37 \\
        Mrk 501 & HSP, TeV & $16^h53^m52.22^s$ & $+39^\circ45'36.6''$ & 0.033$^a$ & 1.69 \\
        RBS 2070 & HSP, TeV & $23^h59^m07.90^s$ & $-30^\circ37'40.7''$ & 0.165$^b$ & 1.18 \\\hline
    \end{tabular}
\end{table*}

\section{Sample sources}\label{sec:src}
The X-ray observations of studied blazars namely, RBS 2070, OJ 287, Mrk 501 and S5 0716+714, span a large period of time from 2004 to 2020. This work is a part of a larger project consisting of the XMM-Newton archival data analysis of the several other sources, which will be extended in the works currently in progress. The main characteristics of sample sources are shown in Table \ref{tab:gen}. A brief introduction of the sample sources is presented below.

\subsection{RBS 2070 (H 2356-309)}
\cite{2010A&A...516A..56H} conducted a thorough analysis of RBS 2070 using the observations made by RXTE Satellite, HESS, ROTSE-IIIc, Nancay radio telescope, XMM-Newton and ATOM telescope. They suggested that the broadband spectral energy distribution (SED) of the source blazar can be simply fitted with a synchrotron self-Compton (SSC) model with a double-peak structure, in which the peaks in X-ray and very high energies (VHE) are produced via synchrotron radiation and inverse Compton scattering, respectively.

\cite{2018ApJ...859...49P} studied an observation made by NuSTAR (3--79 keV) and did not find any significant intra-day variability or spectral variation of the source. \cite{2020arXiv200902289W} presented the study of all nine XMM-Newton observations of this source taken during the period of June 2005–December 2013 with a goal to study the flux and spectral variability in the 0.3–10 keV energy range. Additionally, they fitted the spectra of RBS 2070 using various models to study the source's X-ray spectral curvature and to constrain its break energy. They concluded that spectra in three observations are fitted well using the power-law model, and the remaining six observations are well-fitted using log-parabolic model. Moreover, they found significant flux variability with small amplitude variations only in five light curves. In this paper, we expand and test the results obtained by \cite{2020arXiv200902289W} for this source.

\subsection{OJ 287}
This source is one of the most luminous and rapidly variable BL Lacs at radio to optical frequencies \citep{1989A&AS...80..103S}. It is also one of the most extensively studied extra-galactic AGNs over the entire electromagnetic spectrum. It is highly polarized in the optical band and exhibits a flat-spectrum radio core with a superluminal parsec-scale radio jet, both being characteristic of blazars \citep{2016AJ....152...12L}. Apart from the typical stochastic variability of blazars and favorable observational properties, the most prominent feature responsible for making the source famous is the presence of regular optical double-peaked quasi-periodic outbursts recurring every $\sim$12 years. As of now, these outbursts are well described by the disk-impact binary SMBH model \citep{2016ApJ...819L..37V}.

\cite{2018ApJ...863..175G} presented their results of power spectral density analysis for this source. They posited that the X-ray power spectrum of OJ 287 resembles the radio and optical power spectra on the analogous timescales ranging from tens of years down to months. \cite{2020Galax...8...58K} reported short-term variability in the optical to soft X-ray bands at a moderate level, while the 2–10 keV hard X-ray band was significantly less variable as compared to observed Swift/XRT data. Moreover, most of their X-ray spectra were well described by an absorbed power-law model in the 0.3–10 keV energy range.

\subsection{Mrk 501}
Mrk 501 is one of the most favored targets for multi-frequency observations. This source has minutes variability at TeV band as revealed by MAGIC Telescope \citep{2007ApJ...669..862A}. Such fast variability indicates substantial substructure in jets, which may be due to turbulence or as a result of magnetic reconnection \citep{2012A&A...545A.125R}. \cite{2004ApJ...600..127G} reported the presence of observational signatures of limb brightening in the very-long-baseline interferometry (VLBI) images. In the $\gamma$-ray observations from the Fermi/LAT, the source was reported to display quasi-periodic oscillations with a characteristic timescale of $\sim 300$ day \citep{Bhatta2019}.

\subsection{S5 0716+714}
S5 0716+714 is a blazar with extreme variability, prominent jet component, and negligible host contribution. \cite{2018A&A...619A..93B} reported rapid variability of this source with a significant flux change based on the \textit{NuSTAR} data (3 -- 79 keV). They also did not find any obvious flux-HR (flux-hardness ratio) relation. \cite{2006A&A...453..829F} approximated spectra of this source with a broken power law model using XMM-Newton observations and obtained the break energy at 2.3 keV.

\cite{2006A&A...453..829F} reported results from the analysis of quasi-simultaneous target-of-opportunity observations with the XMM-Newton and \textit{INTEGRAL OMC} data after the optical outburst of the source at the end of March 2004. The authors state that the significant variability and the short flares in S5 0716+714 are due to the most energetic electrons of the synchrotron spectrum, cooling much faster than the electrons producing the inverse Compton emission, which are likely to be at the low end of the distribution.

The gradual decay afterburst is instead likely to be attributed to the escape of electrons from the processing region or to a decrease in the soft seed photons or both. \cite{2006A&A...457..133F} have found that both the synchrotron and inverse Compton components vary on time scales of hours and have estimated the size of the emitting region. The synchrotron emission was discovered to become dominant during episodes of flaring activity, following a harder-when-brighter trend. A similar trend was also observed in the optical observations that were taken during the 5-day photo-polarimetric observation campaign by \citet{Bhatta2016}. \cite{2016MNRAS.458.2350W} reported the highest break energy for S5 0716+714. According to the authors, a spectral break between low- and high-energy components appears at about 8 keV, and this feature is present in our results as well.

\begin{table*}
  \caption{Observation logs of the XMM-Newton EPIC PN observations of the studied sources.}
  \label{tab:obs}
    \begin{tabular}{cccccccc}
      \hline
      Object & Obs. ID & Obs. date & Mode & Time (ks) & \shortstack{Exposure\\ID} & Pile-up & Region\\ 
      \hline
      RBS 2070 & 0304080501 & 2005-06-12 & T & 16.739 & S003 & no & $25\leq$RAWX$\leq50$ \\
      (H 2356-309) & 0304080601 & 2005-06-14 & I & 17.036 & S003 & no & $r\leq600$ \\
      & 0504370701 & 2007-06-02 & T & 129.590 & S003 & yes & 30 (38) $\leq $RAWX$\leq$ 36 (44) \\
      & 0693500101 & 2012-11-18 & I & 118.070 & S003 & yes & $100\leq r\leq700$ \\
      & 0722860101 & 2013-12-02 & I & 22.194 & S003 & no & $r\leq560$ \\
      & 0722860701 & 2013-12-03 & I & 63.470 & S003 & yes & $125\leq r\leq750$ \\
      & 0722860201 & 2013-12-10 & I & 105.470 & S003 & yes & $250\leq r\leq750$ \\
      & 0722860301 & 2013-12-12 & I & 107.739 & S003 & yes & $300\leq r\leq750$ \\
      & 0722860401 & 2013-12-24 & I & 99.029 & U002 & yes & $250\leq r\leq750$ \\
      \hline
      OJ 287 & 0300480201 & 2005-04-12 & I & 18.482 & S003 & no & $r\leq570$ \\ 
      & 0300480301 & 2005-11-03 & I & 38.990 & U002 & no & $r\leq650$ \\
      & 0401060201 & 2006-11-17 & I & 44.970 & S001 & no & $r\leq500$ \\
      & 0502630201 & 2008-04-22 & I & 53.564 & S001 & no & $r\leq650$ \\
      & 0679380701 & 2011-10-15 & I & 21.667 & S003 & no & $r\leq660$ \\
      & 0761500201 & 2015-05-07 & I & 121.438 & S003 & no & $r\leq550$ \\
      & 0830190501 & 2018-04-18 & I & 22.333 & S003 & no & $r\leq500$ \\
      & 0854591201 & 2020-04-24 & I & 15.000 & S003 & no & $r\leq700$ \\
      \hline
      Mrk 501 & 0113060201 & 2002-07-12 & I & 15.773 & S006 & yes & $100\leq r\leq850$ \\
      & 0113060401 & 2002-07-14 & I & 11.111 & S006 & yes & $160\leq r\leq960$ \\
      & 0652570101 & 2010-09-08 & I & 44.464 & S009 & no & $r\leq750$ \\
      & 0652570201 & 2010-09-10 & I & 44.464 & S003 & yes & $75\leq r\leq950$ \\
      & 0652570301 & 2011-02-11 & I & 40.464 & S003 & yes & $100\leq r\leq900$ \\
      & 0652570401 & 2011-02-15 & I & 40.264 & S003 & yes & $100\leq r\leq700$ \\
      \hline
      S5 0716+714 & 0150495601 & 2004-04-04 & T & 55.601 & S003 & yes & 28 (40) $\leq$RAWX$\leq$ 37 (49) \\
      & 0502271401 & 2007-09-24 & I & 71.624 & U002 & yes & $150\leq r\leq450$ \\\hline
    \end{tabular}
\end{table*}

\section{XMM Observations and Data Reduction}\label{sec:obs}
The sample sources were observed with XMM-Newton EPIC-PN CCD camera \citep{2001A&A...365L..18S}. We considered only EPIC-PN data due to its high signal-to-noise ratio, high quantum efficiency and high effective area compared with the EPIC-MOS\footnote{https://www.cosmos.esa.int/web/xmm-newton/technical-details-epic}. Extractions of light curves, source, and background spectra were done with the XMM-Newton Science Analysis System (SAS) v18.0.0. The Calibration Index File (CIF) and the summary file of the Observation Data Files (ODFs) were generated using Updated Calibration Files (CCF) following the ``User's Guide to the XMM-Newton Science Analysis System''\footnote{Issue 15.0, de la Calle et al. 2019}. Event files were produced by the \textit{EPPROC} pipeline. Pile-up effects were corrected iteratively following standard procedure using \textit{EPATPLOT} task of SAS to identify the good regions for source extraction. For image mode, the region is an annular area; for timing mode, each region comprises two columns. The selected source regions are listed in Table \ref{tab:obs} (column 7). Before the light curves and spectra extraction, we set the extracted event pattern and flag to be ``FLAG==0 \&\& PATTERN$<=$4" for all PN data. The background regions are selected by identifying the source free part of each region from the same event maps. Then we selected good time intervals (GTI) with the background countrate threshold ``RATE$<=$0.4". A standard energy range of 0.3–7 keV is used in our study. To ensure the validity of Gaussian statistics, data have been grouped by combining instrumental channels so that each bin contains 100 entries. We created response matrix files and ancillary response files for the extracted spectra using the \textit{RMFGEN} and \textit{ARFGEN} tasks. The spectra prepared for analysis are grouped in order to make sure that there are at least 25 counts for each spectral channel.

The observations for our research were selected using these criteria:

\begin{itemize}
    \item the observation must have been obtained for scientific purpose (not for calibration, etc.);
    \item the exposition time provided in \textit{XMM-MASTER} (\textit{XMM-Newton Master Catalog and Public Archive} table which has been created from information supplied to the \textit{HEASARC} by the \textit{XMM-Newton Project}) must have exceeded 10 ks;
    \item the source should not lie at the edge of the detector (for example, OBSID 0012850101 of S5 0716+714 was excluded from the study).
\end{itemize}

\begin{figure*}
	\centering
	\begin{minipage}{.45\textwidth}
		\centering 
		\includegraphics[width=.99\linewidth]{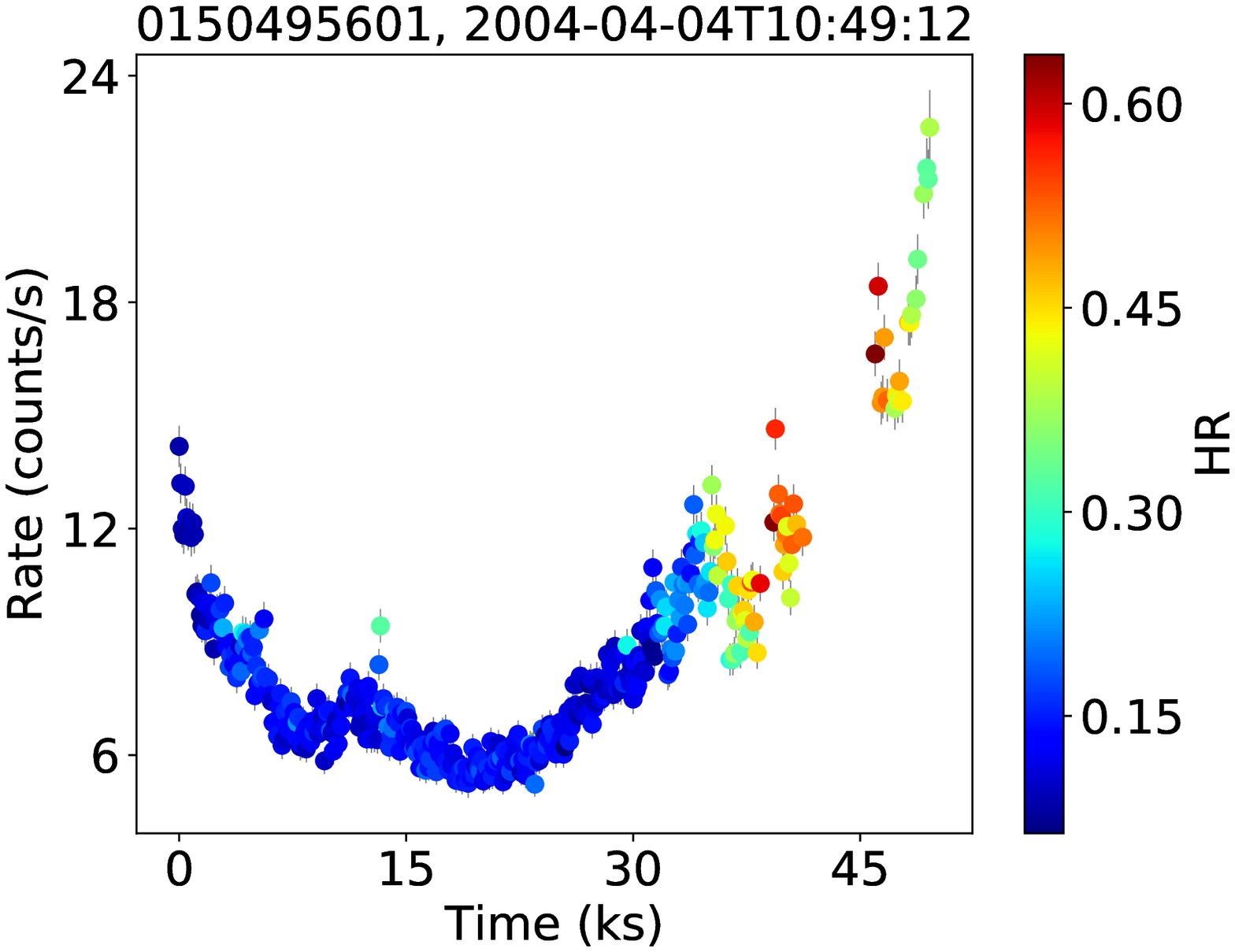}
	\end{minipage}
	\begin{minipage}{.45\textwidth}
		\centering 
		\includegraphics[width=.99\linewidth]{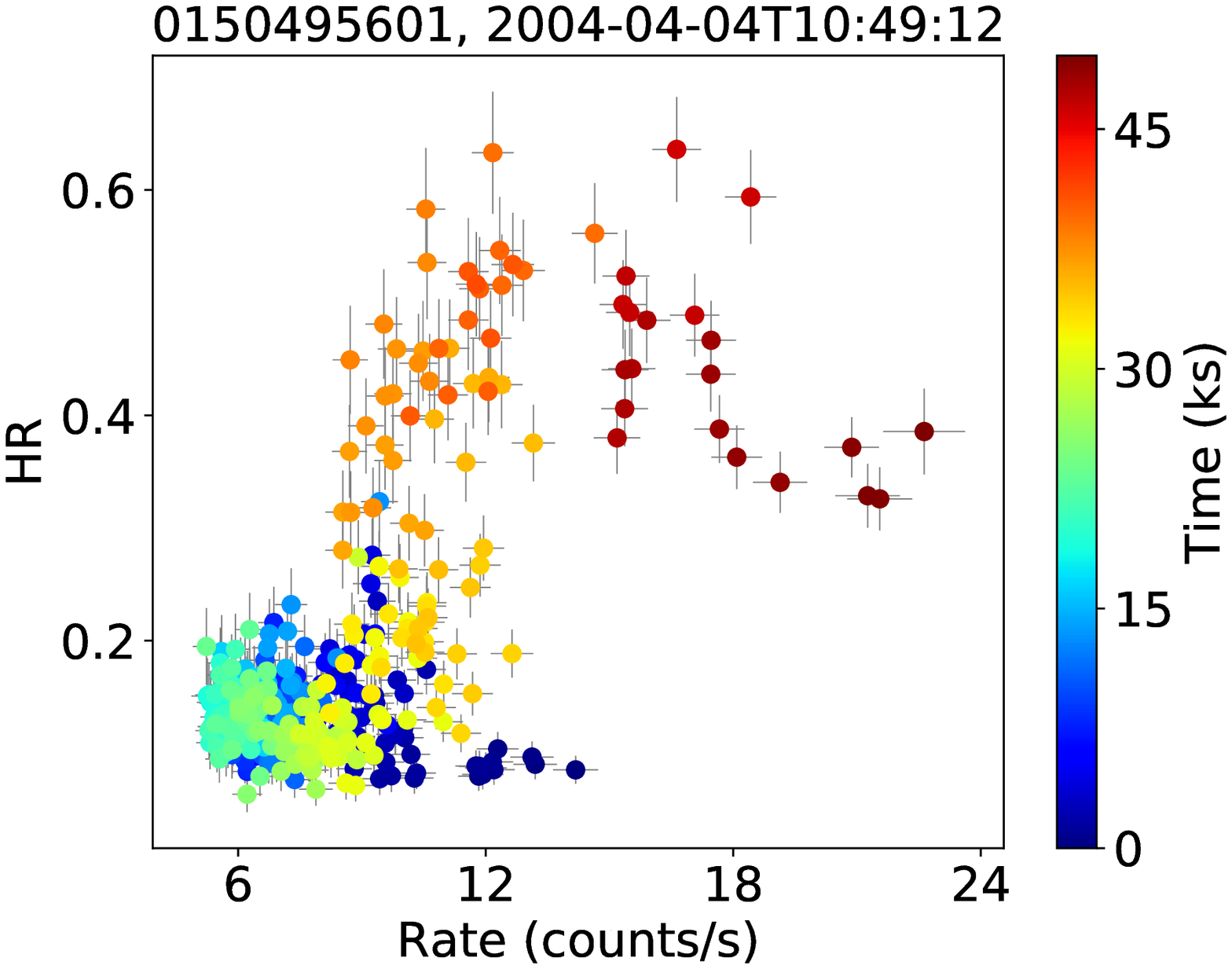}
	\end{minipage}
	\begin{minipage}{.45\textwidth}
		\centering 
		\includegraphics[height=.99\linewidth, angle=-90]{PlotsPersonal/0150495601_EPLP.eps}
	\end{minipage}
	\begin{minipage}{.45\textwidth}
		\centering
		\includegraphics[width=.99\linewidth]{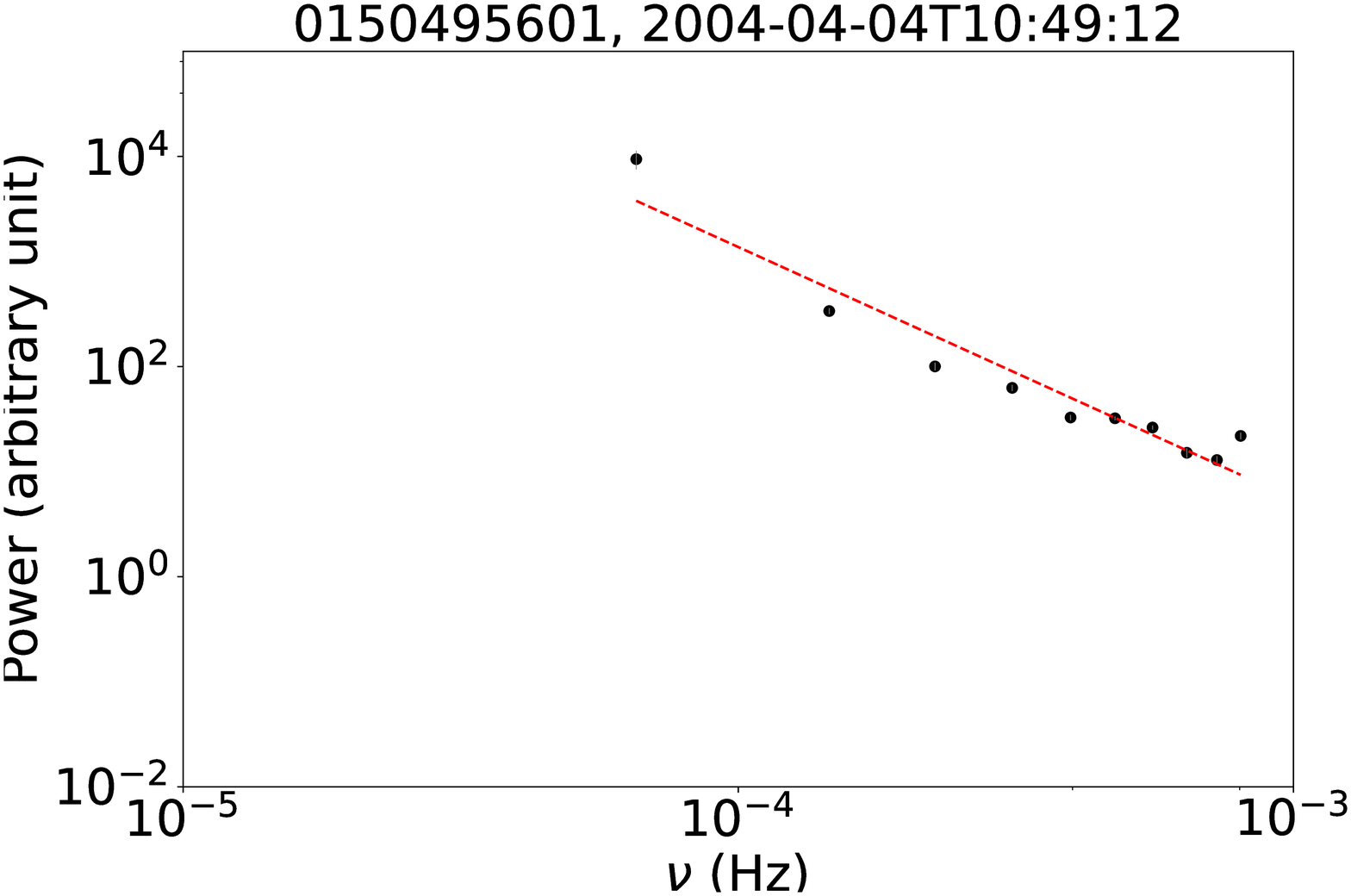}
	\end{minipage}
	\caption{Light curve (\textit{upper left panel}), countrate-HR relation (\textit{upper right panel}), spectral fit (\textit{lower left panel}) and discrete Fourier periodogram in log-log space (\textit{lower right panel}) for the source S5 0716+714 (OBS ID: 0150495601). The color bars provide the third axis in light curve and the countrate-HR plots (HR and time, respectively). Similar plots for all observations are presented in Appendix.}
	\label{fig:show}
\end{figure*}

\section{Analysis}\label{sec:mth}
We present the XMM-Newton observations of the blazars studied in this paper (including their observation IDs and observation dates) in Table \ref{tab:obs}. An example of light curve, countrate-HR plot and spectral fitting for one of the observations of the source S5 0716+714 is shown in Figure \ref{fig:show}. Similar figures for the rest of the observations are presented in Appendix. For a better visualisation of the spectral states at each countrate point, the circle symbols are color-coded according to the hardness ratio (defined below). To study the properties of X-ray variability of our selected sources, we implemented several methods of timing and spectral analyses, which are discussed below (see Table \ref{tab:var}). The methods of the conducted analysis can broadly be classified as timing and spectral analysis.

\subsection{Timing analysis}
\subsubsection{Countrate Distribution: Fractional Variability and RMS-Countrate Relation}
The observed variability can be simply estimated by variability amplitude (\textit{VA}) which describes peak-to-peak countrate variation. Such \textit{VA} can be written as
\begin{equation}
    VA = \dfrac{F_{\mathrm{max}}-F_{\mathrm{min}}}{F_{\mathrm{min}}},
\end{equation}
where $F_{\mathrm{max}}$ and $F_{\mathrm{min}}$ are the maximum and minimum countrate, respectively, in counts/s. The uncertainty can be calculated using the general error propagation rule as

\begin{equation}
    \sigma_{\mathrm{VA}} = (VA+1)\cdot\sqrt{\left(\dfrac{\sigma_{F_{\mathrm{max}}}}{F_{\mathrm{max}}}\right)^2+\left(\dfrac{\sigma_{F_{\mathrm{min}}}}{F_{\mathrm{min}}}\right)^2}.
\end{equation}

\begin{table*}
    \centering
    \caption{Variability properties of the \textit{XMM-Newton EPIC PN} observations of the studied sources. Col. 1: source name; Col. 2: observation ID; Col. 3: mean countrate in counts/s; Col.4: fractional variability; Col. 5: variability amplitude; Col. 6: variability timescale; Col. 7: negative spectral power index (see Fig. \ref{fig:DFT_RBS}--\ref{fig:DFT_S5}). The missing $F_{\mathrm{var}}$ values indicate that the sample variance is smaller than mean square error (see Eq. \ref{eq:rms}).}
    \label{tab:var}
    \begin{tabular}{|c|c|c|c|c|c|c|}\hline
        Object & Obs. ID & \shortstack{Mean countrate\\(counts/s)} & $F_{\mathrm{var}}$ (\%) & VA & $\tau_{\mathrm{var}}$ (ks) & $-\beta_\mathrm{P}$ \\\hline
        RBS 2070 & 0304080501 & $11.45\pm0.40$ & $1.52\pm0.38$ & $0.71\pm0.08$ & $1.40\pm0.87$ & 0.137\\
        & 0304080601 & $11.70\pm0.43$ & -- & $0.71\pm0.09$ & $1.12\pm0.68$ & 0.288\\
        & 0504370701 & $15.63\pm0.96$ & $5.25\pm0.10$ & $1.08\pm0.09$ & $1.98\pm1.70$ & 1.186\\
        & 0693500101 & $15.46\pm0.81$ & $3.37\pm0.15$ & $0.92\pm0.11$ & $1.34\pm0.96$ & 0.873\\
        & 0722860101 & $11.70\pm0.43$ & -- & $0.73\pm0.10$ & $1.06\pm0.61$ & 0.230\\
        & 0722860701 & $11.46\pm0.60$ & $1.48\pm0.44$ & $1.00\pm0.14$ & $1.33\pm1.21$ & 0.021\\
        & 0722860201 & $10.49\pm0.87$ & $2.97\pm0.46$ & $1.41\pm0.28$ & $0.61\pm0.49$ & 0.426\\
        & 0722860301 & $10.32\pm0.95$ & $2.55\pm0.64$ & $1.61\pm0.35$ & $0.43\pm0.29$ & 0.293\\
        & 0722860401 & $11.97\pm0.88$ & $1.90\pm0.56$ & $1.34\pm0.25$ & $0.86\pm0.79$ & 0.291\\\hline
        OJ 287 & 0300480201 & $1.25\pm0.14$ & $4.86\pm1.81$ & $1.28\pm0.34$ & $0.37\pm0.21$ & 1.380\\
        & 0300480301 & $1.21\pm0.13$ & $2.56\pm1.44$ & $1.48\pm0.37$ & $0.54\pm0.42$ & 0.323\\
        & 0401060201 & $0.99\pm0.12$ & $1.18\pm2.06$ & $2.01\pm0.55$ & $0.40\pm0.27$ & 0.157\\
        & 0502630201 & $0.99\pm0.12$ & $2.53\pm1.32$ & $1.77\pm0.47$ & $0.50\pm0.38$ & 0.509\\
        & 0679380701 & $3.29\pm0.20$ & -- & $0.92\pm0.17$ & $0.75\pm0.49$ & 0.438\\
        & 0761500201 & $2.49\pm0.18$ & $1.37\pm0.68$ & $1.29\pm0.24$ & $0.96\pm0.93$ & 0.681\\
        & 0830190501 & $2.28\pm0.18$ & $1.32\pm1.48$ & $1.11\pm0.24$ & $0.67\pm0.48$ & 0.002\\
        & 0854591201 & $13.97\pm0.43$ & -- & $0.59\pm0.08$ & $1.68\pm1.38$ & --0.267\\
        \hline
        Mrk 501 & 0113060201 & $43.16\pm1.11$ & $0.92\pm0.48$ & $0.51\pm0.05$ & $2.05\pm1.39$ & --0.100 \\
        & 0113060401 & $43.52\pm1.17$ & -- & $0.57\pm0.06$ & $2.53\pm2.49$ & --1.371\\
        & 0652570101 & $29.72\pm0.65$ & -- & $0.54\pm0.05$ & $2.21\pm1.62$ & 0.269\\
        & 0652570201 & $31.50\pm0.78$ & -- & $0.59\pm0.06$ & $1.35\pm0.66$ & --0.113\\
        & 0652570301 & $32.70\pm1.12$ & $2.12\pm0.17$ & $0.68\pm0.06$ & $1.79\pm1.22$ & 0.933\\
        & 0652570401 & $42.50\pm1.15$ & $1.21\pm0.19$ & $0.60\pm0.05$ & $1.09\pm0.40$ & 0.465\\\hline
        S5 0716+714 & 0150495601 & $8.45\pm2.87$ & $33.63\pm0.23$ & $4.21\pm0.40$ & $1.40\pm1.08$ & 2.396\\
        & 0502271401 & $4.97\pm1.24$ & $23.25\pm0.36$ & $4.42\pm1.12$ & $0.70\pm0.51$ & 1.849\\\hline
    \end{tabular}
\end{table*}

This measure is derived only from the extreme countrates. Consequently, \textit{VA} may not fully describe the variability in general. A measure for the normalized excess variance, called the fractional variability which considers all countrates in the light curves may be more suitable to represent the observed variability. Fractional variability (FV or $F_{\mathrm{var}}$, see \cite{2003MNRAS.345.1271V}) can be given as

\begin{equation}
    F_{\mathrm{var}} = \dfrac{\sqrt{S^2-\langle\sigma_{\mathrm{err}}^2 \rangle}}{\langle F\rangle},
\end{equation}
where $S^2$ represents the sample variance and $\langle\sigma_\mathrm{err}^2\rangle$ is the mean square of measurement error given by $\langle\sigma_\mathrm{err}^2\rangle = \frac{1}{N}\Sigma_{i=1}^N \sigma_\mathrm{err,i}^2$. The uncertainty of $F_{\mathrm{var}}$ can be estimated as
\begin{equation}
    \sigma_{F_{\mathrm{var}}} = \sqrt{F_{\mathrm{var}}^2+\sqrt{\dfrac{2}{N}\dfrac{\langle\sigma_{\mathrm{err}}^2\rangle^2}{\langle F\rangle^4}+\dfrac{4}{N}\dfrac{\langle\sigma_{\mathrm{err}}^2\rangle^2}{\langle F\rangle^2}F_{\mathrm{var}}^2}} - F_{\mathrm{var}},
\end{equation}

From the variable flux (or countrate) features of the source light curve, a measure for the minimum timescale of the variability the minimum variability timescale can be determined using the expression
\begin{equation}
    \tau_{\mathrm{var}} = \left|\dfrac{\Delta t}{\Delta\ln F}\right|,
\end{equation}
\cite[see][]{1974ApJ...193...43B}, where $\Delta t$ is the time interval between flux or countrate measurements. To compute the uncertainty in variability timescale, we followed the general error propagation rule again and estimated this uncertainty as
\begin{equation}
    \sigma_{\tau_{\mathrm{var}}} \approx \sqrt{\dfrac{F_1^2\sigma_{F_2}^2+F_2^2\sigma_{F_1}^2}{F_1^2 F_2^2 (\ln[F_1/F_2])^4}}\cdot\Delta t,
\end{equation}
where $F_1$ and $F_2$ are the countrates used to estimate $\tau_{\mathrm{var}}$, $\sigma_{F_1}$ and $\sigma_{F_2}$ are the corresponding errors \citep[see][]{2018A&A...619A..93B}. The parameters that characterize countrate variability in the sources are listed in Table \ref{tab:var}. These parameters include fractional variability, variability amplitude, and minimum variability timescales.

\begin{figure*}
	\centering
	\begin{minipage}{.45\textwidth} 
		\centering
		\includegraphics[width=.99\linewidth]{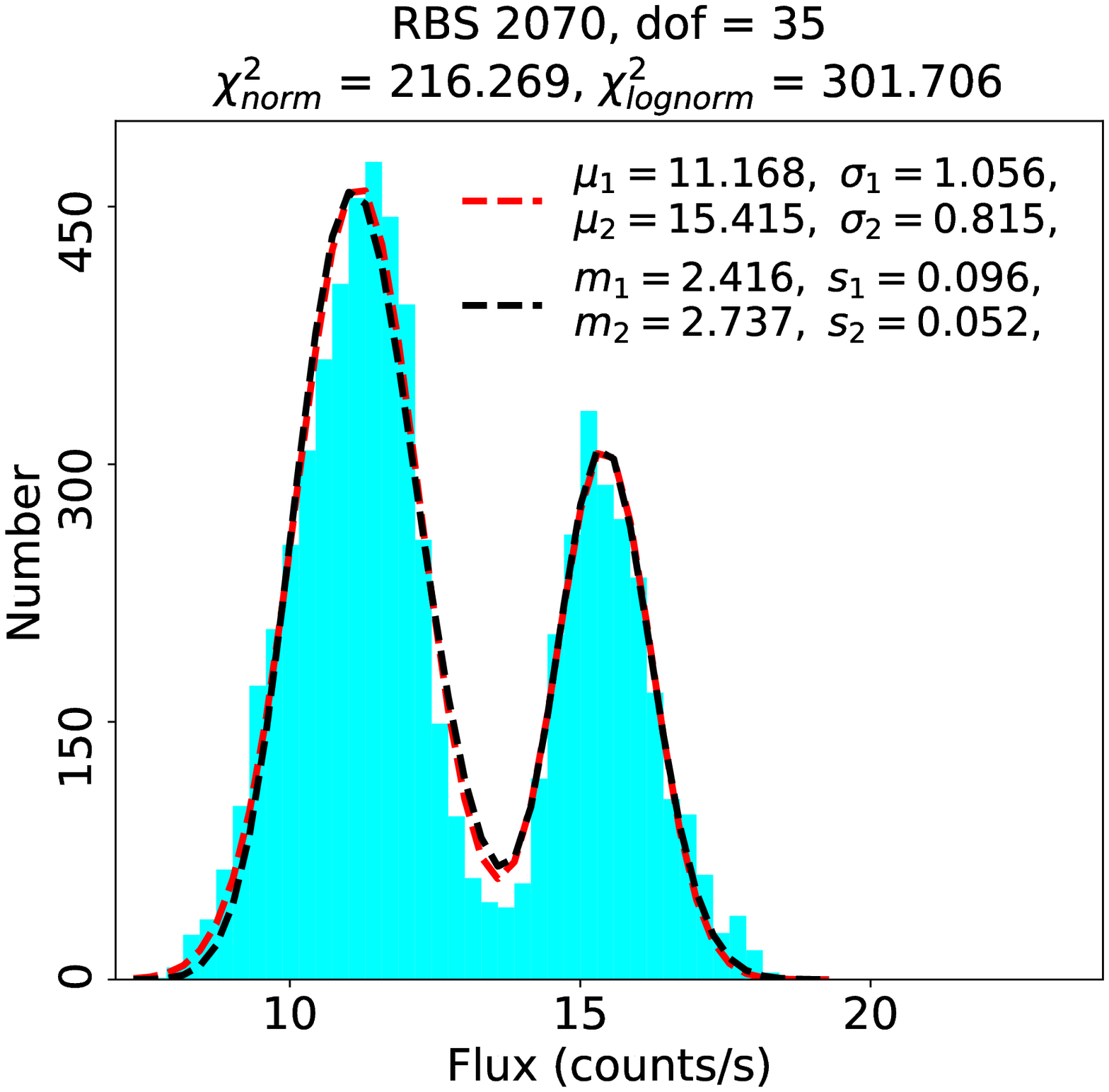}
	\end{minipage}
	\begin{minipage}{.45\textwidth} 
		\centering
		\includegraphics[width=.99\linewidth]{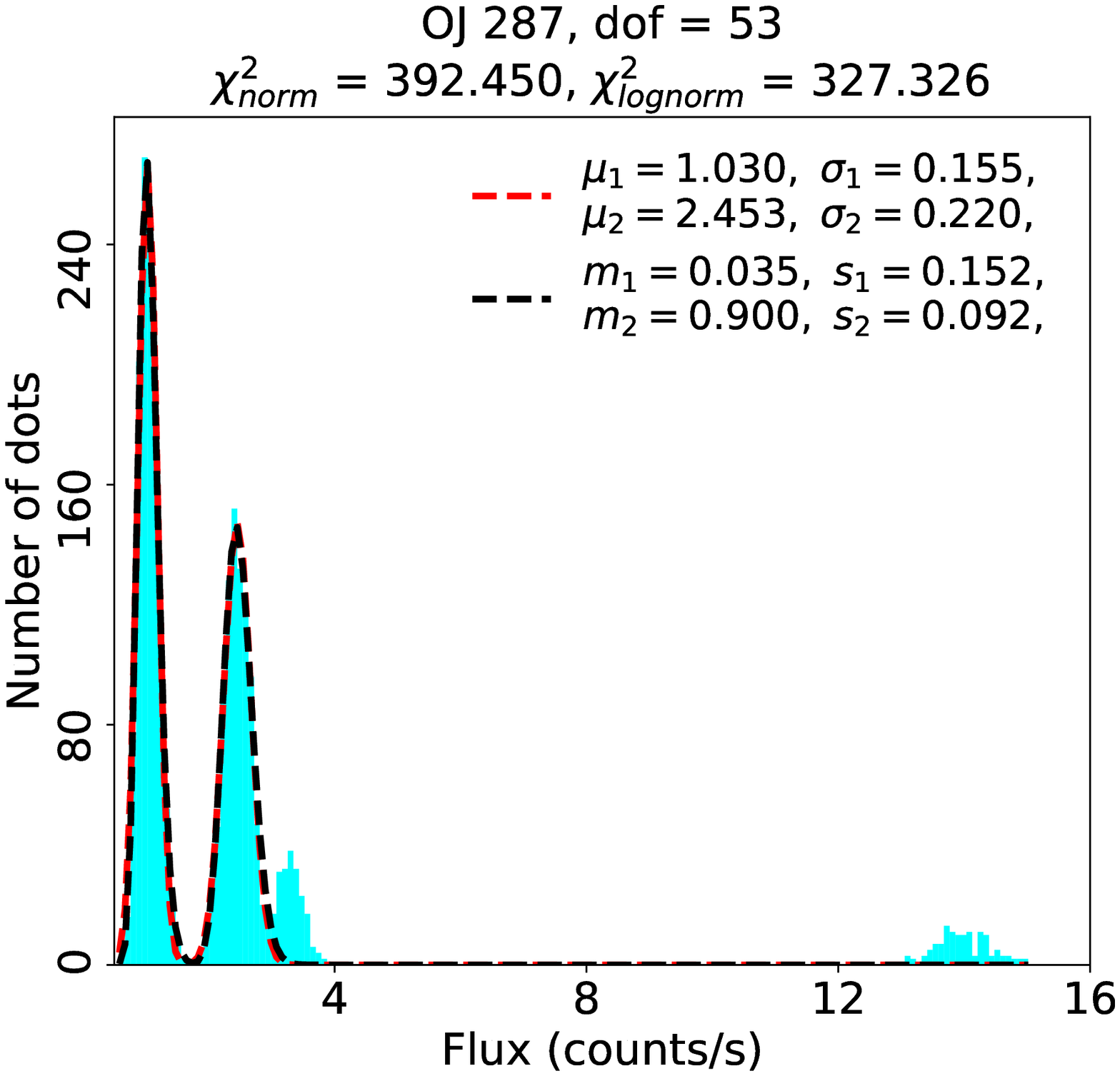}
	\end{minipage}
	\begin{minipage}{.45\textwidth} 
		\centering
		\includegraphics[width=.99\linewidth]{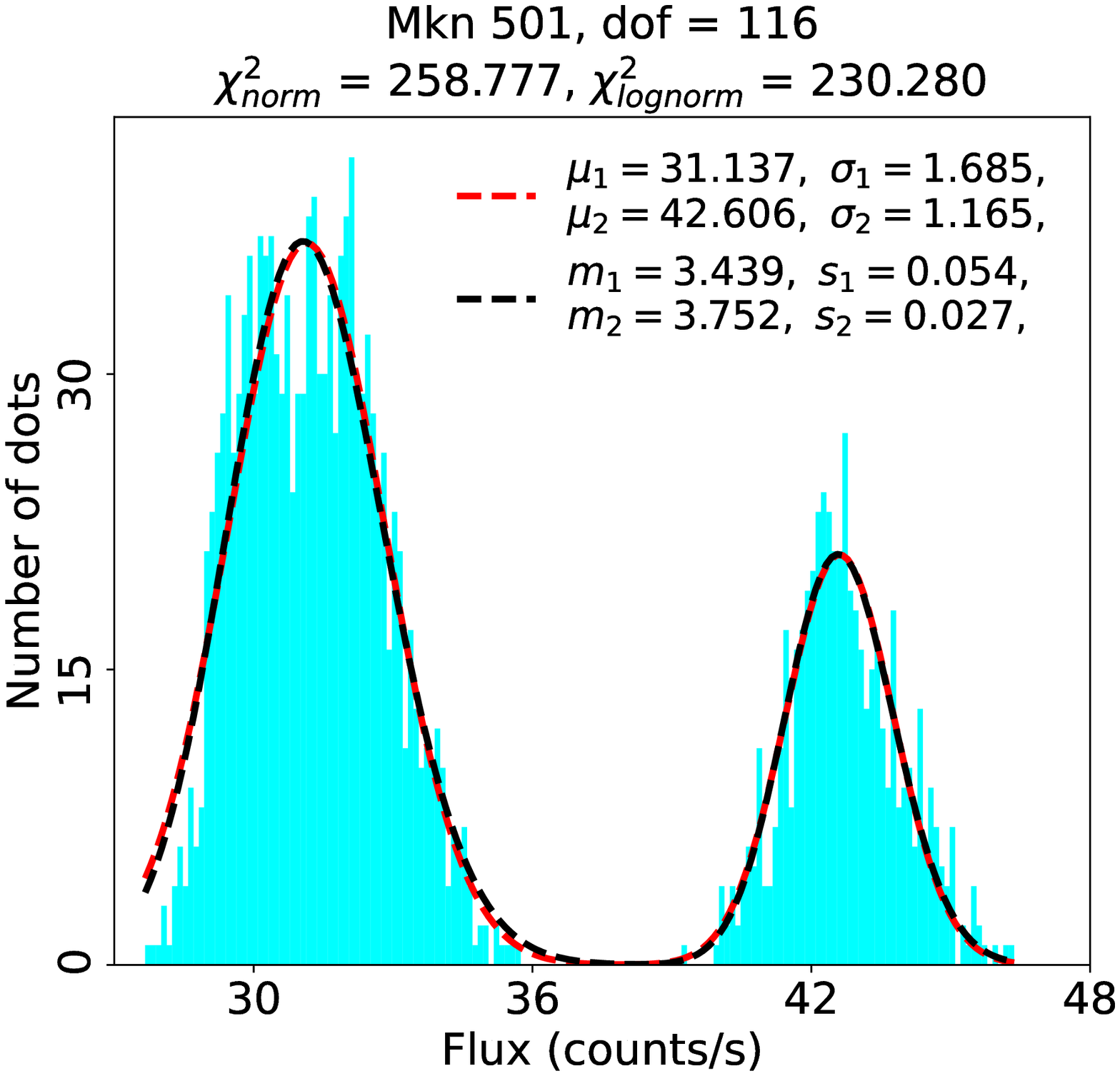}
	\end{minipage}
	\begin{minipage}{.45\textwidth} 
		\centering
		\includegraphics[width=.99\linewidth]{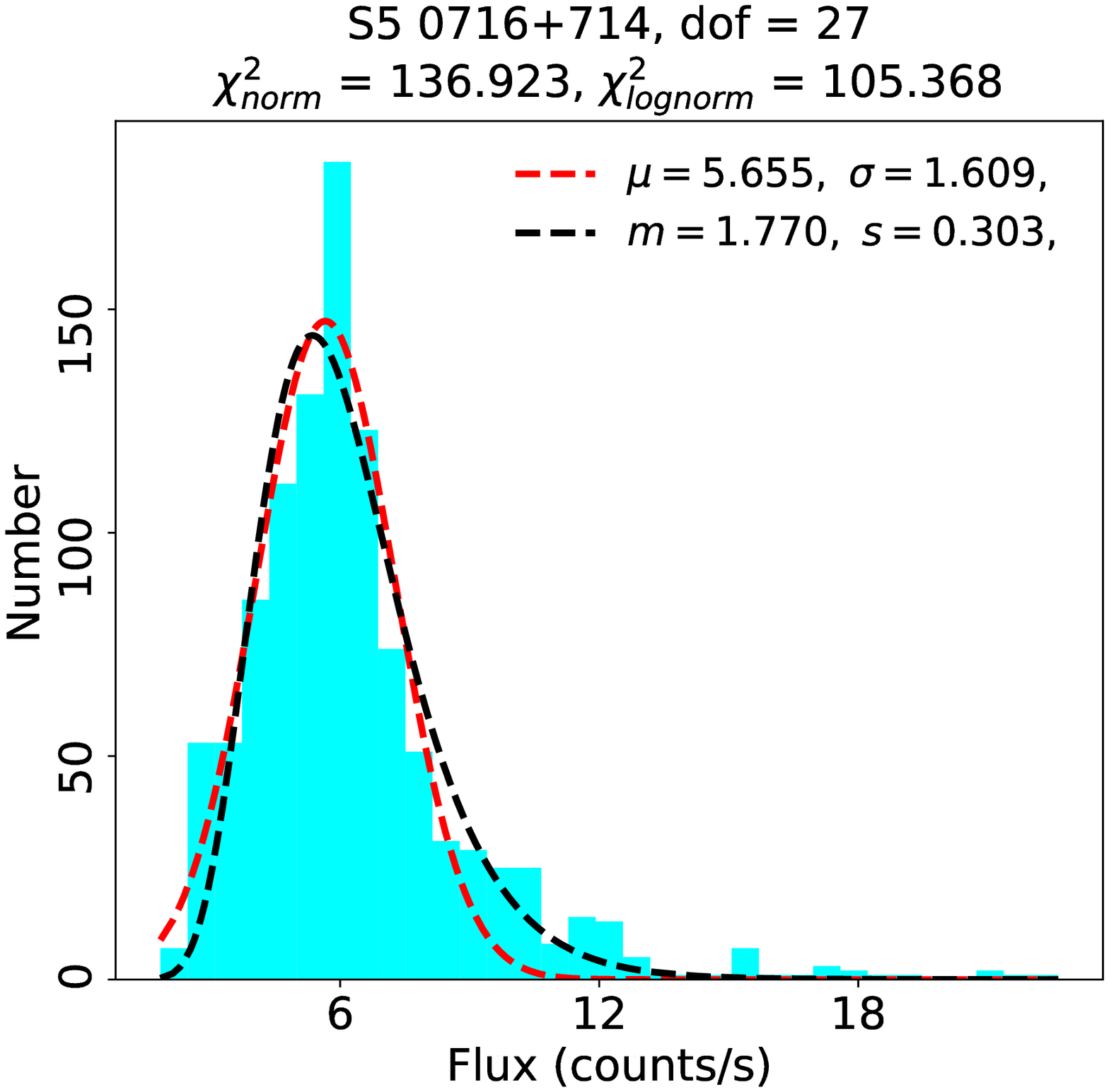}
	\end{minipage}
	\caption{The countrate histograms of the studied sources are fitted using four model PDFs: normal, log-normal, bimodal normal and bimodal log-normal.}
	\label{fig:hist}
\end{figure*}

Yet another way to study the variability properties of the flux can be the RMS-flux relation. That is a correlation between the flux (or countrate) states and source activity represented by root mean square (RMS). The RMS is defined as the square-root of the Poisson noise corrected excess variance given by
\begin{equation}\label{eq:rms}
    \sigma_\mathrm{XS}^2 = S^2-\langle\sigma_\mathrm{err}^2\rangle.
\end{equation}
If the RMS value is below zero, it means that the variability could not be detected by this measure.

\subsubsection{Countrate Distribution: Histograms}
Flux (or countrate) distribution of blazars can unveil some of the important information about the origin and the nature of their variability. A statistical study of probability density function (PDF) of the X-ray countrate was one of the goals of this study. Thus, we studied X-ray countrate distribution of the blazar samples by constructing the histograms to ascertain the PDF of the distribution, which can be approximated by the model fit to the distribution of the countrates from the long-term light curves. We mainly attempted fitting normal and log-normal PDFs (see Fig. \ref{fig:hist}).

PDFs represented by normal distribution could be produced in linear additive process, whereas log-normal distribution with a heavy tail represents non-linear multiplicative process \citep[see][more in Section 6.2]{Uttley2005}. A normal distribution is defined by
\begin{equation}
    f_{\mathrm{norm}}(x) = \dfrac{1}{\sqrt{2\pi}\sigma}\exp\left(-\dfrac{(x-\mu)^2}{2\sigma^2}\right),
\end{equation}
where $\mu$ and $\sigma$ are the mean and the standard deviation of the normal distribution, respectively, expressed in the unit of countrate, i. e., counts/sec. In three of the sources we studied (RBS 2070, OJ 287 and Mrk 501), the countrate distributions were clearly bimodal and thus we attempted fitting binormal PDFs given by,
\begin{multline}
    f_{\mathrm{binorm}}(x) = \dfrac{1}{\sqrt{2\pi}\sigma_1}\exp\left(-\dfrac{(x-\mu_1)^2}{2\sigma_1^2}\right)\\
    +\dfrac{1}{\sqrt{2\pi}\sigma_2}\exp\left(-\dfrac{(x-\mu_2)^2}{2\sigma_2^2}\right),
\end{multline}
where $\mu_1$, $\mu_2$, $\sigma_1$ and $\sigma_2$ are the same values as in the normal PDF.

Similarly, log-normal distribution is defined by
\begin{equation}
    f_{\mathrm{lognorm}}(x) = \dfrac{1}{\sqrt{2\pi}sx}\exp\left(-\dfrac{(\ln x-m)^2}{2s^2}\right),
\end{equation}
where $m$ and $s$ are the mean location and the scale parameter of the distribution, respectively, and $m$ is expressed in the unit of the natural logarithm of countrate. Finally, bi-modal log-normal distribution can be described as
\begin{multline}
    f_{\mathrm{logbinorm}}(x) = \dfrac{1}{\sqrt{2\pi}s_1x}\exp\left(-\dfrac{(\ln x-m_1)^2}{2s_1^2}\right)\\
    +\dfrac{1}{\sqrt{2\pi}s_2x}\exp\left(-\dfrac{(\ln x-m_2)^2}{2s_2^2}\right),
\end{multline}
where $m_1$, $m_2$, $s_1$ and $s_2$ are the same values as in the log-normal PDF.

We observed that normal and log-normal distributions both explain the countrate distribution of every source well, although with moderate reduced $\chi^2$ values.

\subsubsection{Power Spectral Density Analysis}
Discrete Fourier periodogram (DFP) of a light curve of a variable source gives an opportunity to measure the variability power
at a given temporal frequency (or at a given timescale). The mentioned power may be estimated as the square of absolute value of discrete Fourier transform:
\begin{equation}
    P(\nu) = \dfrac{T}{n^2\langle F\rangle^2}\left|\sum_{j=1}^n x(t_j)e^{-i2\pi\nu t_j}\right|^2,
\end{equation}
where $x(t_j)$ is a time series sampled at times $t_j$ with $j = 1, 2, ..., N$ and $\nu$ is a temporal frequency. The periodogram is normalized so it is expressed in the units of (rms/mean)$^2$ Hz$^{-1}$ \citep{2002MNRAS.332..231U}. The periodograms are computed for $n/2$ frequencies that are evenly spaced between the minimum $\nu_{\mathrm{min}} = \frac{1}{T}$ and $\nu_{\mathrm{max}} = \frac{1}{2\Delta t}$, where T is a total duration of the observation, $\Delta t$ is a mean sampling step in the light curve.

Many astrophysical sources show erratic, aperiodic brightness fluctuations with steep power spectra. In general, blazar periodograms have been found to be best approximated by power-law function of the form $P(\nu)\propto\nu^{-\beta_{\mathrm{P}}}$ with spectral power index $\beta_\mathrm{P}$ \citep[e. g.][]{Bhatta2020,2018A&A...620A.185N}. During this type of variability (also known as red noise), the intrinsic variations in the source brightness are random -- this has nothing to do with measurement errors. \cite{2005A&A...431..391V} proposed a method that can be used to test the significance of candidate periodicities superposed on a red noise spectrum that has an approximately power-law shape. In our case, there are no hidden periodicities (DFPs of every observation are shown in on-line material), yet the $\beta_{\mathrm{P}}$ values display correlation in both source types (see Fig. \ref{fig:flux_DFT}).

\subsection{Spectral analysis}
\subsubsection{Spectral Distribution: Hardness Ratio}
We produced the source light curves in two energy bands: soft energy band which includes photons with energies 0.3 -- 2 keV and hard energy band which includes photons with energies 2 -- 7 keV. This gives us an opportunity to study the spectral variability of the X-ray emission coming from sample sources. Then we calculated hardness ratio (HR) as HR = $F_{\mathrm{hard}}/F_{\mathrm{soft}}$, where $F_{\mathrm{hard}}$ and $F_{\mathrm{soft}}$ are the countrates in counts/s in the hard and soft bands, respectively. The hardness ratio is generally used as a model-independent method to examine spectral variations at different times and countrate states via, for example, HR-plots.

In this work, we inspect in detail the relation between countrate and HRs through the observation period to constrain the underlying emission processes. To seek any credible hysteresis loops in the countrate-HR plane, the symbols in the plots were color-coded by the observation period of time. Following the general error propagation rule, the uncertainty in HR is dependent on the corresponding countrate uncertainties $\sigma_{F_{\mathrm{hard}}}$ and $\sigma_{F_{\mathrm{soft}}}$, and can be evaluated as
\begin{equation}
    \sigma_{HR} = HR\cdot\sqrt{\left(\dfrac{\sigma_{F_{\mathrm{hard}}}}{F_{\mathrm{hard}}}\right)^2+\left(\dfrac{\sigma_{F_{\mathrm{soft}}}}{F_{\mathrm{soft}}}\right)^2}.
\end{equation}

\subsubsection{Spectral Fitting}
To carry out the spectral analysis of the XMM-Newton blazars, we fit the source spectra using XSPEC models \citep{1996ASPC..101...17A} by applying the $\chi^2$ minimization statistics. We fitted each spectrum with three different spectral models: power-law (PL), log-parabolic with peak energy (EPLP), and broken power-law (BPL).

When studying the spectra of OJ 287, we also used a fourth model -- an addition of black-body and log-parabolic with peak energy (BBEL). The model was fitted to investigate whether the source X-ray emission contains any signatures of thermal emission from the accretion disk.

The power-law model is described as
\begin{equation}
    \dfrac{dN}{dE} = N_0\cdot E^{-\Gamma},
\end{equation}
where $N_0$ is normalization, E is photon energy and $\Gamma$ is photon index. Likewise, the log-parabolic model with a continuous break is evaluated as
\begin{equation}
    \dfrac{dN}{dE} = N_0\cdot10^{-\beta\left(\log\ (E/E_{\rm p})\right)^2}/E^2,
\end{equation}
where $E_{\rm p}$ is peak energy and $\beta$ is curvature parameter. 
Thirdly, the broken power-law model is given as
\begin{equation}
    \dfrac{dN}{dE} = 
    \begin{cases}
      N_0\cdot E^{-\Gamma_1} & \text{if $E\geq E_{b}$},\\
      N_0\cdot E^{-\Gamma_2} & \text{otherwise},
    \end{cases}
\end{equation}
where $N_0$ is normalization, $E_{\rm b}$ is break energy, $\Gamma_1$ and $\Gamma_2$ represent the high- and low-energy photon indexes, respectively.

\begin{figure}
	\centering
	\begin{minipage}{.45\textwidth}
		\centering 
		\includegraphics[width=.99\linewidth]{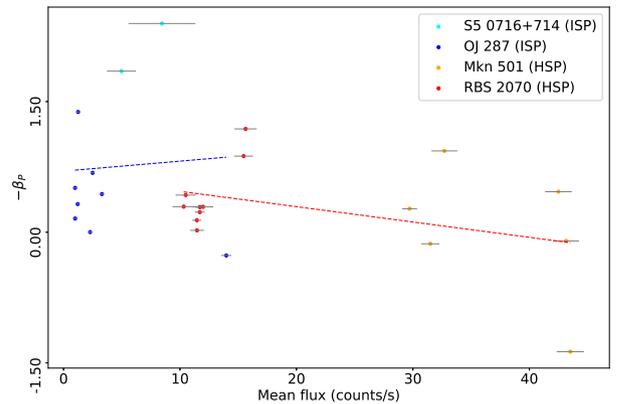}
	\end{minipage}
	\caption{The plot showing the relation of mean countrate and DFP slope index in the studied sources. The dashed lines display the linear fitting (blue for ISPs and red for HSPs).}
	\label{fig:flux_DFT}
\end{figure}

\begin{figure}
	\includegraphics[width=.99\linewidth]{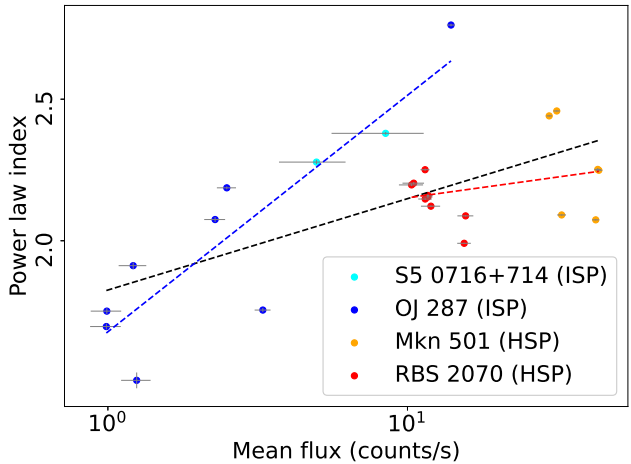}
	\caption{Distribution of the XMM-Newton power-law photon indexes over the mean countrates. The dashed lines display the linear fitting (black for all sources, blue for ISPs and red for HSPs).}
	\label{fig:flux_index}
\end{figure}

\begin{figure}
	\begin{minipage}{.45\textwidth}
		\centering 
		\includegraphics[width=.99\linewidth]{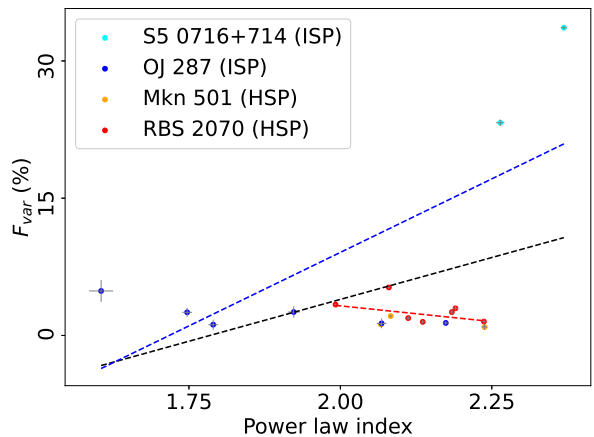}
	    \caption{Distribution of the XMM-Newton fractional variabilities over the power-law photon indexes. The dashed lines display the linear fitting (black for all sources, blue for ISPs and red for HSPs).}
	    \label{fig:index_Fvar}
	\end{minipage}
\end{figure}
Finally, the black-body model can be expressed as
\begin{equation}
    \dfrac{dN}{dE} = \dfrac{N_0\cdot E^2}{(kT)^4\cdot(\exp(E/kT)-1)},
\end{equation}
where $kT$ is temperature in keV and $N_0$ is the normalization parameter. To be more precise, the latter is $L_{39}/D_{10}^2$, where $L_{39}$ is the source luminosity in units of $10^{39}$ erg/s and $D_{10}$ is the distance from the observer to the source in units of 10 kpc.

To resolve the galactic absorption, Tuebingen-Boulder ISM absorption model (TBABS; \cite{2000ApJ...542..914W}) was multiplied with the models above. The hydrogen column density values used for this part are shown in Table \ref{tab:gen}. Of the three (four in case of OJ 287) models, we selected the best-fit one by executing the F-test available in XSPEC. In particular, we evaluated the significance of LP and BPL (and BBEL) against PL which is assumed to be null hypothesis (NH). We imposed the model was best-fit if the probability under the null hypothesis was equal or smaller than 0.1 (equivalently, significance was equal or greater than 90\%). Else, PL was assumed to characterize the spectrum better. Furthermore, if two (or three) models overcome the significance level of 90\%, the model with higher significance was chosen to be the best fit. Based on the mentioned criteria, 5, 11, 8, and 1 spectra were found to be best characterized by PL, EPLP, BPL, and BBEL models, respectively. It should be pointed out that several AGNs might show considerable residuals above 7 keV in their X-ray emission. These instrumental effects correspond to the Ni-K, Cu-K, and Zn-K lines in PN spectra \citep{2004A&A...414..767K}. Therefore, the higher energy limit of our spectra was lowered from 10 keV to 7 keV in order to discard the region affected by the instrumental lines.

We present the fitting parameters for each observation in Table \ref{tab:spec}. The corresponding spectral fittings are shown in Appendix. We provide the distribution of the PL photon indices over the mean countrate in Figure \ref{fig:flux_index}.

\begin{table*}
    \caption{Spectral properties of studied AGNs. Col. 1: source name; Col. 2: observation ID; Col. 3: spectral models, namely power-law (PL), log-parabolic (EPLP), broken power-law (BPL), black-body + log-parabolic (BBEL); Col. 4: photon index (PL), curvature parameter (EPLP and BBEL), high-energy photon index (BPL); Col. 5: low-energy photon index (BPL); Col. 6: peak energy in keV (EPLP and BBEL), break energy in keV (BPL); Col. 7: black-body temperature (keV); Col. 8: $\chi^2$/degrees of freedom; Col. 9: F-test result along with probability value. The best-fit models in Col. 3 are emphasized with bold font.}
    \label{tab:spec}
    \begin{tabular}{|c|c|c|c|c|c|c|c|c|}
      Object & Obs. ID & Model & $\Gamma$/$\beta$/$\Gamma_1$ & $\Gamma_2$ & $E_{p}/E_{b}$ (keV) & kT (keV) & $\chi^2$/dof & F-test (prob.)\\\hline\hline
      RBS 2070 & 0304080501 & \textbf{PL} & 2.249(4) & -- & -- & -- & 134.57/125 & NH\\
      (H 2356-309) && EPLP & --0.066(13) & -- & 99.82(88.26) & -- & 143.98/124 & --\\
      && BPL & 2.257(6) & 2.210(30) & 2.30(86) & -- & 129.86/123 & 2.23 (1.12$\times10^{-1}$)\\\hline
      & 0304080601 & PL & 2.144(4) & -- & -- & -- & 352.33/127 & NH\\
      && \textbf{EPLP} & 0.193(13) & -- & 0.51(3) & -- & 130.50/126 & 214.18 ($<0.001$)\\
      && BPL & 2.024(13) & 2.240(10) & 1.03(5) & -- & 156.94/125 & 77.81 ($<0.001$)\\\hline
      & 0504370701 & PL & 2.098(2) & -- & -- & -- & 1057.53/131 & NH\\
      && EPLP & --0.145(5) & -- & 2.90(8) & -- & 222.94/130 & 486.66 ($<0.001$)\\
      && \textbf{BPL} & 2.154(3) & 1.959(7) & 1.83(5) & -- & 146.40/129 & 401.42 ($<0.001$)\\\hline
      & 0693500101 & PL & 1.991(2) & -- & -- & -- & 375.41/134 & NH\\
      && EPLP & 0.064(5) & -- & 1.44(5) & -- & 231.52/133 & 82.66 ($<0.001$)\\
      && \textbf{BPL} & 1.776(32) & 2.011(2) & 0.55(2) & -- & 171.06/132 & 78.84 ($<0.001$)\\\hline
      & 0722860101 & PL & 2.139(4) & -- & -- & -- & 375.54/127 & NH\\
      && \textbf{EPLP} & 0.199(13) & -- & 0.52(3) & -- & 124.70/126 & 253.46 ($<0.001$)\\
      && BPL & 2.002(14) & 2.234(9) & 0.96(5) & -- & 150.88/125 & 93.06 ($<0.001$)\\\hline
      & 0722860701 & PL & 2.131(3) & -- & -- & -- & 399.55/131 & NH\\
      && \textbf{EPLP} & 0.152(10) & -- & 0.44(3) & -- & 139.27/130 & 242.96 ($<0.001$)\\
      && BPL & 1.978(17) & 2.188(6) & 0.77(3) & -- & 162.20/129 & 94.38 ($<0.001$)\\\hline
      & 0722860201 & PL & 2.188(4) & -- & -- & -- & 346.74/128 & NH\\
      && \textbf{EPLP} & 0.171(13) & -- & 0.33(3) & -- & 148.99/127 & 168.56 ($<0.001$)\\
      && BPL & 2.131(7) & 2.352(18) & 1.66(9) & -- & 163.11/126 & 70.93 ($<0.001$)\\\hline
      & 0722860301 & PL & 2.182(4) & -- & -- & -- & 358.56/127 & NH\\
      && \textbf{EPLP} & 0.201(14) & -- & 0.40(3) & -- & 150.85/126 & 173.49 ($<0.001$)\\
      && BPL & 2.040(15) & 2.280(10) & 0.94(4) & -- & 153.00/125 & 83.97 ($<0.001$)\\\hline
      & 0722860401 & PL & 2.110(4) & -- & -- & -- & 279.33/128 & NH\\
      && \textbf{EPLP} & 0.114(11) & -- & 0.40(5) & -- & 170.83/127 & 80.66 ($<0.001$)\\
      && BPL & 2.005(17) & 2.153(7) & 0.83(6) & -- & 186.70/126 & 31.26 ($<0.001$)\\
      \hline\hline
      OJ 287 & 0300480201 & \textbf{PL} & 1.616(22) & -- & -- & -- & 81.20/79 & NH\\
      && EPLP & 0.125(67) & -- & 40.25(47.22) & -- & 98.79/78 & --\\
      && BPL & 1.806(94) & 1.539(38) & 0.97(23) & -- & 72.29/77 & 4.75 (1.14$\times10^{-2}$)\\
      && BBEL & 0.074(344) & -- & 28.11(455.014) & 1.66(32) & 74.39/76 & 2.32 (8.21$\times10^{-2}$)\\\hline
      & 0300480301 & PL & 1.936(10) & -- & -- & -- & 322.67/115 & NH\\
      && EPLP & 0.012(31) & -- & 99.31(1176.1) & -- & 336.58/114 & --\\
      && \textbf{BPL} & 2.441(57) & 1.782(16) & 0.79(4) & -- & 126.85/113 & 87.22 ($<0.001$)\\
      && BBEL & --0.404(27) & -- & 1.10(3) & 199.4(--) & 122.63/112 & 60.90 ($<0.001$)\\\hline
      & 0401060201 & PL & 1.797(8) & -- & -- & -- & 127.07/120 & NH\\
      && EPLP & 0.053(25) & -- & 99.88(204.32) & -- & 154.33/119 & --\\
      && \textbf{BPL} & 1.856(17) & 1.703(26) & 1.59(25) & -- & 101.49/118 & 14.87 ($<0.001$)\\
      && BBEL & --0.109(81) & -- & 0.17(20) & 24.14(2004.4) & 99.73/117 & 10.69 ($<0.001$)\\\hline
      & 0502630201 & \textbf{PL} & 1.750(8) & -- & -- & -- & 111.02/119 & NH\\
      && EPLP & 0.066(25) & -- & 99.77(164.29) & -- & 123.87/118 & --\\
      && BPL & 1.750(8) & 0.214(--) & 32.98(--) & -- & 111.02/117 & --\\
      && BBEL & 0.099(122) & -- & 9.62(31.09) & 1.67(27) & 108.61/116 & 0.86 (4.65$\times10^{-1}$)\\\hline
      & 0679380701 & PL & 1.797(6) & -- & -- & -- & 207.22/126 & NH\\
      && EPLP & 0.053(19) & -- & 99.90(158.42) & -- & 259.74/125 & --\\
      && \textbf{BPL} & 1.888(17) & 1.706(16) & 1.33(13) & -- & 142.28/124 & 28.30 ($<0.001$)\\
      && BBEL & 0.026(92) & -- & 55.89(766.00) & 1.63(13) & 141.85/123 & 18.89 ($<0.001$)\\\hline
      & 0761500201 & PL & 2.187(5) & -- & -- & -- & 648.53/128 & NH\\
      && \textbf{EPLP} & --0.303(13) & -- & 2.42(8) & -- & 143.71/127 & 446.12 ($<0.001$)\\
      && BPL & 2.313(9) & 1.933(15) & 1.49(5) & -- & 153.63/126 & 202.95 ($<0.001$)\\
      && BBEL & --0.229(54) & -- & 3.65(1.22) & 1.65(29) & 140.53/125 & 150.62 ($<0.001$)\\\hline
    \end{tabular}
\end{table*}

\begin{table*}
    \contcaption{Spectral properties of studied AGNs.}
    \begin{tabular}{|c|c|c|c|c|c|c|c|c|}
      Object & Obs. ID & Model & $\Gamma$/$\beta$/$\Gamma_1$ & $\Gamma_2$ & $E_{p}/E_{b}$ (keV) & kT (keV) & $\chi^2$/dof & F-test (prob.)\\\hline\hline
      & 0830190501 & \textbf{PL} & 2.070(9) & -- & -- & -- & 139.45/113 & NH\\
      && EPLP & --0.018(25) & -- & 98.15(587.21) & -- & 139.10/112 & 0.28 (5.97$\times10^{-1}$)\\
      && BPL & 2.083(14) & 2.028(41) & 1.92(1.05) & -- & 137.11/111 & 0.95 (3.91$\times10^{-1}$)\\
      && BBEL & 0.171(102) & -- & 0.42(14) & 1.67(32) & 134.85/110 & 1.25 (2.95$\times10^{-1}$)\\
      \hline
      & 0854591201 & PL & 2.810(6) & -- & -- & -- & 171.86/112 & NH\\
      && EPLP & 0.068(20) & -- & 0.00(0) & -- & 194.45/111 & --\\
      && BPL & 2.828(7) & 2.613(54) & 2.21(30) & -- & 139.77/110 & 12.63 ($<0.001$)\\
      && \textbf{BBEL} & 0.383(62) & -- & 0.06(2) & 1.19(7) & 107.67/109 & 21.66 ($<0.001$)\\
      \hline\hline
      Mrk 501 & 0113060201 & PL & 2.238(5) & -- & -- & -- & 135.86/127 & NH\\
      && EPLP & 0.062(14) & -- & 0.01(2) & -- & 126.03/126 & 9.83 (2.14$\times10^{-3}$)\\
      && \textbf{BPL} & 2.063(68) & 2.258(6) & 0.60(6) & -- & 106.22/125 & 17.44 ($<0.001$)\\\hline
      & 0113060401 & PL & 2.238(5) & -- & -- & -- & 137.03/125 & NH\\
      && EPLP & 0.036(16) & -- & 0.001(2) & -- & 144.18/124 & --\\
      && \textbf{BPL} & 1.422(929) & 2.246(6) & 0.42(5) & -- & 124.58/123 & 6.15 (2.86$\times10^{-3}$)\\\hline
      & 0652570101 & PL & 2.430(2) & -- & -- & -- & 879.10/132 & NH\\
      && \textbf{EPLP} & 0.154(6) & -- & 0.05(1) & -- & 196.37/131 & 455.46 ($<0.001$)\\
      && BPL & 2.288(9) & 2.492(4) & 0.79(2) & -- & 211.85/130 & 204.73 ($<0.001$)\\\hline
      & 0652570201 & PL & 2.452(2) & -- & -- & -- & 975.02/131 & NH\\
      && \textbf{EPLP} & 0.184(7) & -- & 0.07(1) & -- & 157.90/130 & 672.74 ($<0.001$)\\
      && BPL & 2.322(7) & 2.536(5) & 0.89(2) & -- & 209.34/129 & 235.92 ($<0.001$)\\\hline
      & 0652570301 & PL & 2.088(2) & -- & -- & -- & 565.49/133 & NH\\
      && EPLP & 0.007(6) & -- & 0.00(0) & -- & 607.36/132 & --\\
      && \textbf{BPL} & 2.139(4) & 1.985(7) & 1.53(6) & -- & 201.31/131 & 118.49 ($<0.001$) \\\hline
      & 0652570401 & \textbf{PL} & 2.069(2) & -- & -- & -- & 208.83/133 & NH\\
      && EPLP & 0.009(6) & -- & 0.00(0) & -- & 215.15/132 & --\\
      && BPL & 2.076(4) & 2.053(7) & 1.62(42) & -- & 199.32/131 & 3.13 (4.72$\times10^{-2}$)\\
      \hline\hline
      S5 0716+714 & 0150495601 & PL & 2.419(6) & -- & -- & -- & 6392.94/126 & NH\\
      && \textbf{EPLP} & --1.126(12) & -- & 1.83(1) & -- & 257.53/125 & 2978.01 ($<0.001$)\\
      && BPL & 2.883(10) & 1.507(13) & 1.43(1) & -- & 402.38/124 & 923.05 ($<0.001$)\\\hline
      & 0502271401 & PL & 2.277(7) & -- & -- & -- & 500.93/123 & NH\\
      && \textbf{EPLP} & --0.397(18) & -- & 2.50(10) & -- & 93.92/122 & 528.70 ($<0.001$)\\
      && BPL & 2.435(13) & 1.947(20) & 1.40(5) & -- & 89.60/121 & 277.74 ($<0.001$)\\\hline
    \end{tabular}
\end{table*}

\section{Results}\label{sec:res}
We present the results of our analysis below. Along with light curves, HR-plots, and spectral fits of the mentioned observations, we carried out the countrate histograms for studied sources which showed an interesting relation between bumps of bimodal distribution and the best-fit spectral models of the corresponding observations (see Appendix for the plots discussed below).

\subsection{RBS 2070 (H 2356-309)}
We looked into nine XMM-Newton observations scattered on a timescale between June 2005 and December 2013. In the light curves (see Fig. \ref{fig:LC_RBS}, left panels), the countrate points appear to be scattered, showing no coherent variability pattern, except observation 0504370701, where a clear harder-when-brighter trend is visible during a slight increase in countrate. Likewise, no clear trend in the countrate-HR plots can be observed (see Fig. \ref{fig:LC_RBS}, middle panels). Six observations are well-fitted with EPLP model ($E_{\rm p}\approx$ 0.3--0.5 keV, $\beta\approx$ 0.1--0.2), two more observations are well-fitted with different BPL models with $E_{b}\approx$ 0.6 and 1.8 keV and one more -- with PL model ($\Gamma\approx$ 2.25; see Fig. \ref{fig:LC_RBS}, right panels). According to \cite{2001AIPC..599..586C} and \cite{2005ExA....20...31G}, the X-ray spectrum of this source could be characterized by a broken power-law with a synchrotron peak at 1.8 keV using BeppoSAX observations. \cite{2008A&A...478..395M} reported the results of their fitting of one of the XMM-Newton observations (0304080601) using the log-parabolic model. They stated that the value of peak energy equals $0.78(17)$ keV. 

The countrate histogram of the source as presented in Fig. \ref{fig:hist} (top left panel) shows a bimodal distribution with a second peak formed mainly by two observations (0504370701 and 0693500101) which are best fitted with BPL model. The observations contributing to the first peak, however, are best fitted with the EPLP and PL models.

\subsection{OJ 287}
Our study was conducted using eight XMM-Newton observations which have been held from April 2005 to April 2020. The source showed no reasonable variability pattern both in light curves and countrate-HR planes (see Fig. \ref{fig:LC_OJ}, left and middle panels). Only one observation is well-fitted with BBEL model ($E_{\rm p}\approx$ 0.1 keV, $\beta\approx$ 0.4, $kT\approx$ 1.2 keV; see Fig. \ref{fig:LC_OJ}, right panels), three more are fitted with BPL model ($E_{\rm b}\approx$ 0.8, 1.3 and 1.6 keV), for three more the best-fit model is PL ($\Gamma\approx$ 1.62, 1.75 and 2.07), and the remaining one observation is well-fitted with EPLP model ($\beta\approx$ --0.3, $E_{\rm p}\approx$ and 2.4 keV).

For this source, we report a complex and peculiar distribution (see Fig. \ref{fig:hist}, top right panel). There are four small bumps in total. It is possible that first three peaks would form a single peak if we have more observations of this source. Nevertheless, the fourth bump is prominently forming another peak and contains only one observation, which is best fitted with BBEL model (0854591201) with the peak energy 0.06 keV. It is worth noting that the third bump also contains only one observation which is best fitted with BPL model (0679380701; break energy -- 1.3 keV). Spectra models from other observations do not show any correlation in peak alignment.

Several multiwavelength analyses of the source, including the XMM-Newton data, were performed to study the source \citep{2018MNRAS.480.1999G, 2021MNRAS.504.5575K, 2021ApJ...923...51K, 2020MNRAS.498L..35K}. \cite{2020ApJ...890...47P} performed a spectral and temporal study of OJ 287 based on the 2015 and 2018 XMM-Newton observations (0761500201 and 0830190501). They state that the best-fit model in their research was the log-parabolic model. Furthermore, the appearance of the soft excess during 2015 and its absence in 2018 was consistent with the presence of an accretion-disk signature. Based on these considerations, the authors believe that the soft X-ray excess and UV emission appeared to be primarily as a result of reflection phenomena.

\subsection{Mrk 501}
We examined six XMM-Newton observations which were made between July 2002 and February 2011. Our study does not show any systematic variability in light curves or countrate-HR plots (see Fig. \ref{fig:LC_Mrk}, left and middle panels). Of the three spectral models, three observations were fitted with the BPL model with $E_{\rm b}\approx$ 0.4 keV, 0.6 keV and 1.5 keV, two were well-represented by EPLP model ($E_{\rm p}\approx$ 60 eV, $\beta\approx$ 0.17), and one observation was best-fit by PL model with $\Gamma\approx$ 2.07 (see Fig. \ref{fig:LC_Mrk}, right panels).

\cite{2018A&A...619A..93B} obtained a clear harder-when-brighter trend on countrate-HR (see Section 4) plot of one of the NuSTAR observations of this source. However, the spectra they studied were well-fitted by different models showing no systematic dependency. \cite{2019ApJ...885....8W} systematically studied the $E_{\rm p}-L_{\rm p}$ and $E_{\rm p}-1/b$ relations (see Section 4.2.2 for description of these spectral parameters) for 14 BL Lac objects, including Mrk 501, using the 3–25 keV RXTE/PCA and 0.3–10 keV Swift/XRT data. They pointed out that there is a positive correlation between $E_{\rm p}$ and $1/\beta$ in the case of Mrk 501. However, they posit that this correlation can not be explained by either the energy-dependent acceleration probability scenario or the stochastic acceleration process.

The countrate histogram is best fitted with a bimodal distribution (see Fig. \ref{fig:hist}, bottom left panel). The second peak is formed by three observations which are best fitted by BPL and PL models (0113060201, 0113060401 and 0652570401). Going to the fainter observations, the next one is the fourth BPL model observation. Together with observations which are best-fitted by EPLP, it forms the first peak.

\subsection{S5 0716+714}
We looked into two XMM-Newton observations separated by three years and a half (April 2004 and September 2007). The light curve of both observations showed significant variability (fractional variability $\sim$ 23--34\%). The first observation (0150495601) showed a complex harder-when-brighter trend, yet the second one (0502271401) followed a completely opposite trend -- harder-when-fainter (see Fig. \ref{fig:LC_S5}, left and middle panels). The spectra are best-fitted with different EPLP models ($E_{\rm p}\approx$ 1.83 and 2.50, $\beta\approx$ --1.1 and --0.4; see Fig. \ref{fig:LC_S5}, right panels). That, in combination with the Gaussian distribution of countrate (see Fig. \ref{fig:hist}, bottom right panel), leads to no conclusive peculiarities.

\cite{2016MNRAS.462.1508G} analyzed only one observation of the source, which was taken by XMM-Newton during the source's high state and found strong intraday variability in it. Similarly, \cite{2020Galax...8...66K} report that in high states of S5 0716+714, the observations show a complex temporal evolution of intensity vs. HR and also that this source shows anti-correlation between countrate evolution and HR.

\section{Discussion}\label{sec:dsc}
In this section, we present our interpretation and discussion on the obtained results in terms of the standard model of blazars, i.e., black hole powered central engine and the extended radio jets providing grounds for particle acceleration and energy dissipation events.

\subsection{X-ray Variability in Blazars}
X-ray observations reveal important processes that occur in the innermost regions of the central engine of an AGN. There usually are three main components of X-ray emission from the AGNs: soft X-ray excess, neutral iron line and the Compton hump. Typically, the spectra of blazars show pure power-law shapes which are clear of any emission or absorption properties. This is caused by the Doppler boosted jet emission which dominates over the coronal emission. At the same time, the origin of hard X-ray variability in blazars over various timescales could be the up-scattering of the soft photon fields positioned in different geometrical parts of an AGN. In such manner, any modulation in the photon field, high-energy electron population, and magnetic field can generate X-ray variability, which can transmit along the jets.

As discussed in \cite{2018A&A...619A..93B}, the hard X-ray emission in HSPs is probably due to the high energy tail of the synchrotron emission from the large-scale jets. The variable emission then can be related to the particle acceleration and synchrotron emission by the electrons of the highest energy. In such a case, the minimum variability timescales can be directly connected to the particle acceleration and cooling timescales. To estimate the synchrotron cooling timescale, i.e. cooling due to synchrotron emission, we define $t_{\rm cool}$ in seconds as energy of an electron divided by synchrotron power loss, $t_{\rm cool} = \gamma m_ec^2/P_{\rm syn}$. Also, for the velocity $\beta=v/c$

\begin{equation}
    t_{\rm cool} = \dfrac{3}{4}\dfrac{m_ec}{\sigma_T U_b \gamma\beta^2}\approx 7.74\cdot10^8\cdot\gamma^{-1}B^{-2}.
\end{equation}

Similarly, the inverse-Compton (IC) cooling timescale of synchrotron self-Compton scenario depends on the energy density of the external photon field and the electron energy. This timescale can be expressed as

\begin{equation}
    t_{IC}\approx\dfrac{3}{4}\dfrac{m_ec}{\sigma_T\cdot U\gamma},
\end{equation}
where $U$ represents photon density which can be $U_{\rm syn}$ and $U_{\rm ext}$ for synchrotron self-Compton and external Compton, respectively.

Figure \ref{fig:index_Fvar}, presenting the distribution of the fractional variability over the power-law indexes $\Gamma_X$, points out that the ISPs tend to be more variable in their steeper spectral states while the HSPs show the opposite relation. This result can be explained in terms of the broadband SED of blazars. In case of ISPs the observed X-ray band is closer to the synchrotron peak energy. Thus, the X-ray emission is largely contributed by the population of electrons with energies near to the peak energy of the distribution. In other words, when the ISPs become steeper, there is an excess of particles with energies comparable to the peak energy, resulting in larger variability. On the other hand, in case of HSPs the observed X-ray emission is contributed by the particles which represent the lower end of the power-law distribution. For HSPs in their steeper states this energy region is contributed by lower energy particles which require cooling on longer timescales and thereby appear less variable in intra-day timescales.

\subsection{Countrate Distribution}
Study of countrate distribution in blazars can reveal the nature of the processes contributing to the variable emission. In the scenario where the observed variability could be contributed by the combination of a large number of stochastic events, e. g. shocks and magnetic reconnection events in the turbulent regions of either the jets and the accretion disks, countrate distribution can be a key in addressing issues like whether the nature of the processes is additive or multiplicative. In general, PDFs that are represented by normal distribution are considered to be produced in linear additive process; whereas log-normal distribution with a heavy tail could be signatures of non-linear multiplicative process \citep[see][]{Uttley2005}. Moreover, in the X-ray band, a linear relationship between the long-term countrate and short-term variability has been found to be characteristic of many binary black hole systems. More recently, log-normal countrate distribution with linear RMS-countrate relation was also observed in the $\gamma$-ray observations of a number of blazars including S5 0716+714 \citep[see][]{Bhatta2020}.

In the accretion disk based model, the flux and countrate variability is partially fed by the fluctuations in the disk, while the latter can occur at different radii and thereby be governed by viscosity fluctuations according to the local viscous timescales. Consequently, these fluctuations modulate the mass accretion rates at larger distances from the black hole (\citealt{Lyubarskii1997}; see also \citealt{2006MNRAS.367..801A}). In the case of strong disk-jet connection, such behaviours can propagate along the jet and thereby can be observed in the MWL emission from the jets \citep{Giebels2009}. In another jet based model, \cite{Biteau2012} showed that the emission from isotropically oriented mini-jets scattered along the jet could result in power-law distribution with the linear RMS-flux relation. We note that while the countrate distribution of the blazars S5 0716+714, OJ 287, and Mrk 501, was consistent with log-normal PDF, the countrate distribution of the remaining source, RBS 2070 exhibited a clear bi-modal log-normal PDF (see Fig. \ref{fig:hist}). The observed difference between the normal and log-normal fits for the studied sources is small and moreover, we have analyzed only four sources. Therefore, the possibility of the log-normal distribution can not be ruled out conclusively.

Such a bi-modal flux distribution was also reported in the Rossi X-ray Timing Explorer observations of the TeV blazars Mrk 421 and Mrk 501 \citep{Khatoon2020}. The observation implies a more complex scenario in which the observed X-ray emission could be contributed by more than one emission zone. It is also possible that the sources possess two distinct dynamics states, e. g. hard and soft states, resulting in the observed bi-modal feature in the flux histogram. Out of our sample sources, Mrk 501, S5 0716+714 and OJ 287 (the latter of which is known to show signs of disk-jet relation; \cite{2016ApJ...819L..37V}) exhibit the log-normal countrate distribution. Histograms of other sample sources tend to be normal.

\subsection{Power Spectral Density}
The PSD analysis performed extensively on the MWL light curves of blazars suggests that the underlying PSD model is consistent with a simple power law \citep[see][]{Bhatta2020,Isobe2015,Shimizu2013,Nakagawa2013,Chatterjee2008} The PSD analysis of the sample observations was carried out by estimating the DFP of the individual observations and subsequently, PSD model of single power-law was fitted (see Appendix). 

In general, the slopes listed in the 7th column of Table \ref{tab:var} show that the sources OJ 287, Mrk 501, and RBS 2070 all have comparable slopes within the range --0.3 and 1.4 (with the exception of OBSID 0113060401 of Mrk 501 that has --1.4). It is seen that the only significantly different source in this respect is S5 0716+714 with slopes greater than 1.8. The outlier nature of S5 0716+714 is also evident in the Figures \ref{fig:flux_DFT} and \ref{fig:index_Fvar}. This behavior could be caused by the fact that this source shows evidence of strong variability possibly linked with its small jet viewing angle $\sim 3^\circ$ \citep{2017ApJ...846...98J}. Moreover, the slopes of the periodograms of the source OJ 287 are consistent with the results obtained in the previous study \cite[see][]{2020Galax...8...58K}.

Furthermore, it is important to note that the analysis shows the variable nature of the PSD slope of the sample sources. A similar variable PSD slope was also reported in the intra-day Suzaku observations of the TeV Blazar PKS 2155-304 \citep{2021ApJ...909..103Z}. The changing PSD slopes in the light curves of a source can be ascribed to the non-stationary emission processes on intra-day timescales, and thereby key to the broader issue (whether the overall blazar variability are driven by stationary or non-stationary processes). Variable PSD slopes can arise due to transient non-stationarity in the underlying processes, such as local instabilities in the turbulent region \citep[see e. g.][]{Calafut2015,Marscher2014}. Similar local non-stationarity in the form of variable PDF was also detected in the $\gamma$-ray observations of blazars \citep{2021MNRAS.508.1446D}.

\subsection{Spectral Fitting and Countrate -- Hardness Ratio Relation}
We have plotted the countrate against the hardness ratio for the sources to study their relation. Yet, we did not note any clear correlation that could be generalized for all observations. Only in two cases we detected clear evidence of harder-when-brighter within the observation period, and in one more observation we noted a softer-when-brighter trend. However, no signs of hysteresis loops in the countrate-HR plane could be observed.

The nature of this correlation in blazars is yet to be unveiled. For optical range, there are two different tendencies. The bluer-when-brighter behaviour is mostly related to BL Lacs in the intra-day timescales. The redder-when-brighter trend is frequently observed in FSRQs \citep{Ikejiri2011}. For $\gamma$-ray range, blazars are seen to act similarly: the spectrum either hardens with the source intensity, or softens with the countrate enhancements.

The softer-when-brighter trend seen in the observation of S5 0716+714 could be due to excess of soft photons during active states. The observed feature can be explained in the context of synchrotron emission where cooling timescale is inversely proportional to the energy of the photons ($t_{\rm cool}\sim1/E_{\rm ph}$). As a result, higher energy photons cool faster leaving an excess of lower energy (soft) photons, which take a longer time to cool. In general, it is possible that the countrate-HR diagram can yield structures that are more complex than harder-when-brighter or softer-when-brighter trends \citep[see][]{2016MNRAS.458.2350W}.

In case of HSP, the observed X-ray emission mainly results from the synchrotron processes. Shock waves that propagate along the jet in blazars can inject a power-law electron distribution via Fermi acceleration processes. The accelerated particles lose their energy as they plunge into the enhanced magnetic field at the shock front and thereby release a large output of synchrotron emission. The spectral shape of such emission largely mimics the power-law shape of the electron distribution with a relation $\alpha=(p-1)/2$, where $\alpha$ and $p$ represent the power-law index of the emission and the particle distribution, respectively. Similarly, during the particle injection the emission spectral shape may evolve over time. In particular, if the acceleration time-scale is significantly shorter than the cooling time-scales, not all the particles have enough time to cool. In such scenario, we might expect a break in the emission spectrum \citep[see Fig. 1 in][]{1998A&A...333..452K}. It is also possible that a power-law spectrum of particle distribution can yield a curved emission spectrum in the presence of a number of competing processes, such as stochastic acceleration, synchrotron, and inverse-Compton cooling. \cite{Massaro2004} argued that such curvature in the blazar spectra could be naturally produced by the distribution of the relativistic particles with a curvature at the time of the injection.

In our study, the best-fit models are EPLP (11 observations), BPL (8 observations), PL (5 observations) and BBEL (1 observation). Overall, our best-fit models are in agreement with the literature results mentioned in the previous section (see Table \ref{tab:spec}). However, there are differences in the model parameters, which can be explained by the difference in the methodology of the researches. The negative spectral power indexes in our sample are mainly in range from --0.3 to 1, except S5 0716+714 where we detect $-\beta_p>1.7$ for both observations. Additionally, observations of S5 0716+714 are highly variable, while the other sources show a very low variability ($F_{\mathrm{var}}<6\%$).

\section{Conclusions}\label{sec:con}
We analyzed 25 XMM-Newton observations of four blazars, including 2 ISPs and 2 HSPs. The following are the findings from our analyses.

\begin{itemize}
    \item Most of the sources show low level or no variability at all. At the same time, S5 0716+714 showed high fractional variability with the highest variability amplitudes in the sample.
    \item We did not detect any correlation between the soft and hard energy emission that may be inferred for all observations.
    \item Nearly half of the observations are best fitted by the EPLP model, approximately one third -- with the BPL model. PL and BBEL models turned out to be not favorable by sample sources.
    \item We note that the highest countrate observations are best fitted with the BPL model. Moreover, we identified a close correlation of the mean countrate states and the photon indices.
    \item We found that ISPs favor more variable behaviour in their steeper spectral states, while HSPs are less variable in their steeper spectral states.
\end{itemize}

In general, the observed results are in agreement with the standard model of blazars, that is, supermassive black holes powering the relativistic jets that are closely aligned to the line of sight. The power-law shape of the X-ray spectra are signatures of non-thermal emission from the jets. Moreover, the variability of the X-ray emission are typical of any jet-accretion systems. However, the moderate variability reported here suggests that the emission regions could be larger than the ones referred from the intra-day variability timescales -- with the exception of S5 0716+714 which exhibits significant MWL variability. The spectral shape of the non-thermal emission is consistent with the competing timescales (particle and cooling timescales) in synchrotron and/or inverse-Compton processes. In particular, the ISPs and HSPs show a distinction in the observed variability. Such distinction is possibly caused by the fact that in ISPs the particles responsible for the X-ray band emission lie closer to the peak energy. On the other hand, in the HSPs the emission formed by population of lower energy particles cools more slowly resulting in longer variability timescale. Finally, the relation observed between the countrate states and the best-fit spectral models could be explored further involving a larger sample of blazars or a larger sample of data.

\section*{Data availability}
The data underlying this work will be shared on reasonable request to the corresponding author.

\section*{Acknowledgements}
GB appreciates the financial support by the Narodowe Centrum Nauki (NCN) grant UMO-2017/26/D/ST9/01178. RP acknowledges the institutional support of the Silesian University in Opava and the grant SGS/12/2019. RP was also supported by the Student Grant Foundation of the Silesian University in Opava, Grant No. SGF/4/2020, which has been carried out within the EU OPSRE project entitled “Improving the quality of the internal grant scheme of the Silesian University in Opava”, reg. number: CZ.02.2.69/0.0/0.0/19\_073/0016951. We thank the anonymous reviewer for his/her constructive comments that helped us improve the quality of the paper significantly.

\bibliographystyle{mnras}

\clearpage
\appendix
\section{Light curves, hardness ratio plots, spectral fits and discrete Fourier periodograms of XMM-Newton sources}

\setcounter{figure}{0}

\begin{figure*}\label{app:rbs}
	\centering
	\caption{LCs, HR plots and spectral fits derived from observations of RBS 2070.}
	\label{fig:LC_RBS}
	\begin{minipage}{.3\textwidth} 
		\centering 
		\includegraphics[width=.99\linewidth]{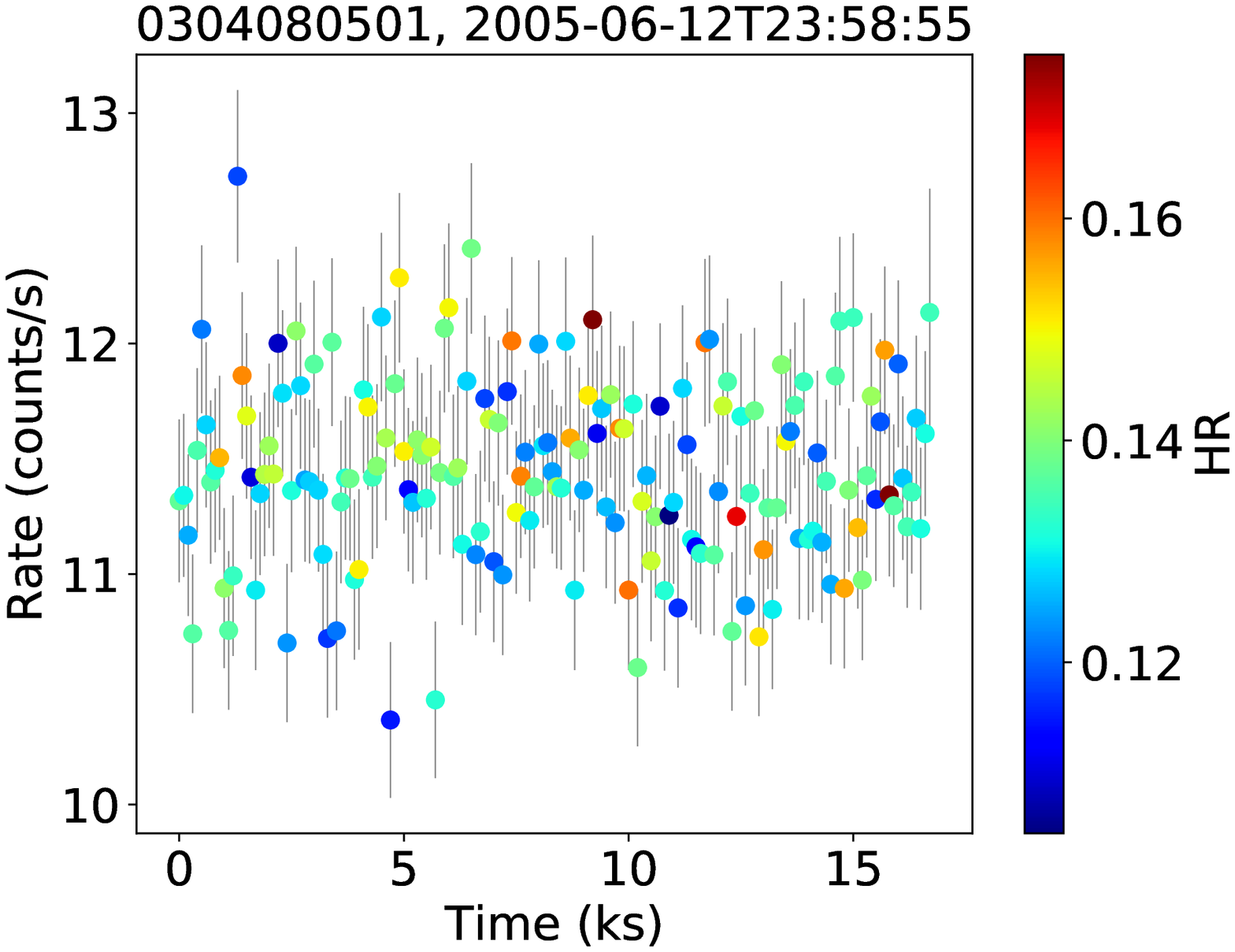}
	\end{minipage}
	\begin{minipage}{.3\textwidth} 
		\centering 
		\includegraphics[width=.99\linewidth]{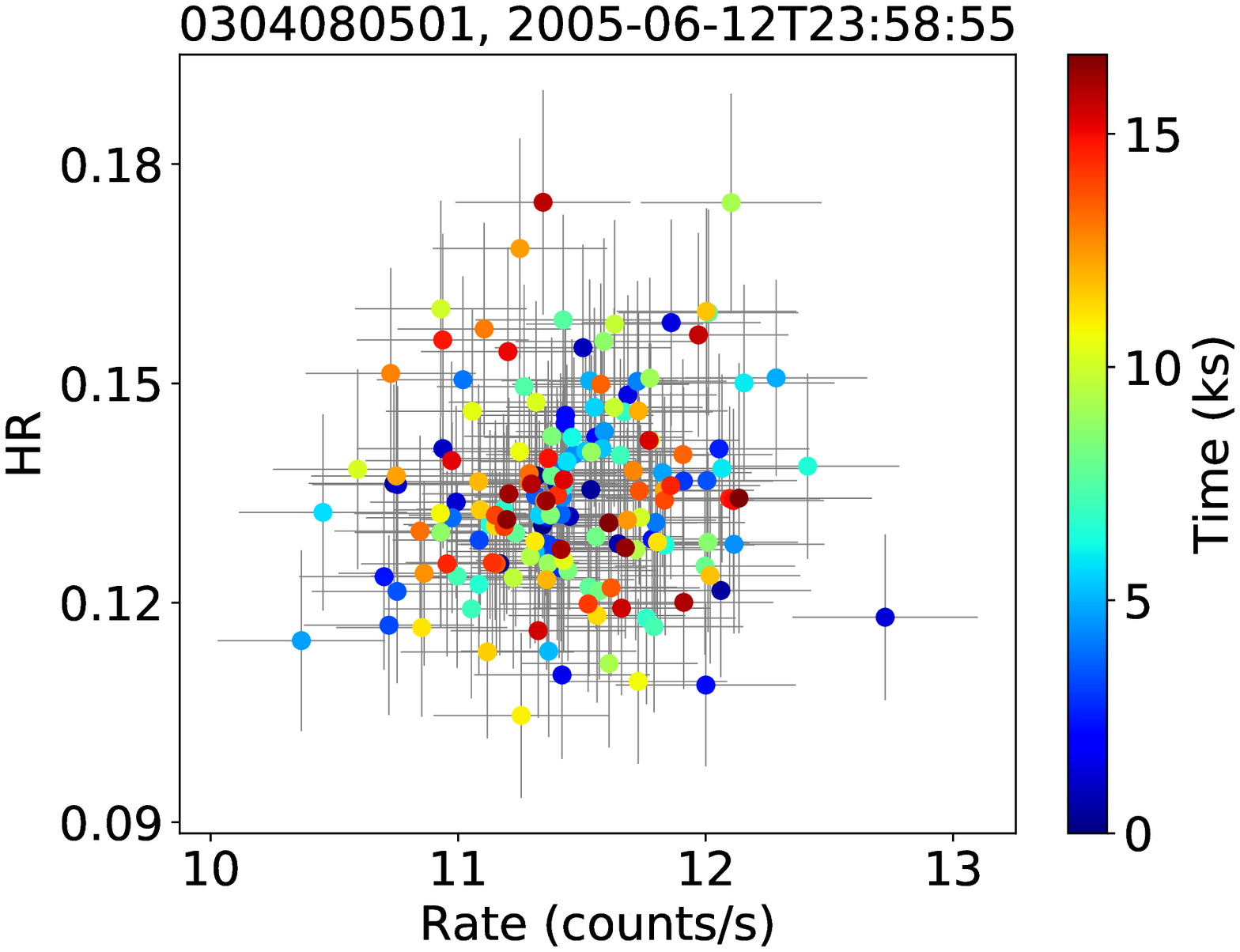}
	\end{minipage}
	\begin{minipage}{.3\textwidth} 
		\centering 
		\includegraphics[height=.99\linewidth, angle=-90]{PlotsPersonal/0304080501_PL.eps}
	\end{minipage}
	\begin{minipage}{.3\textwidth} 
		\centering 
		\includegraphics[width=.99\linewidth]{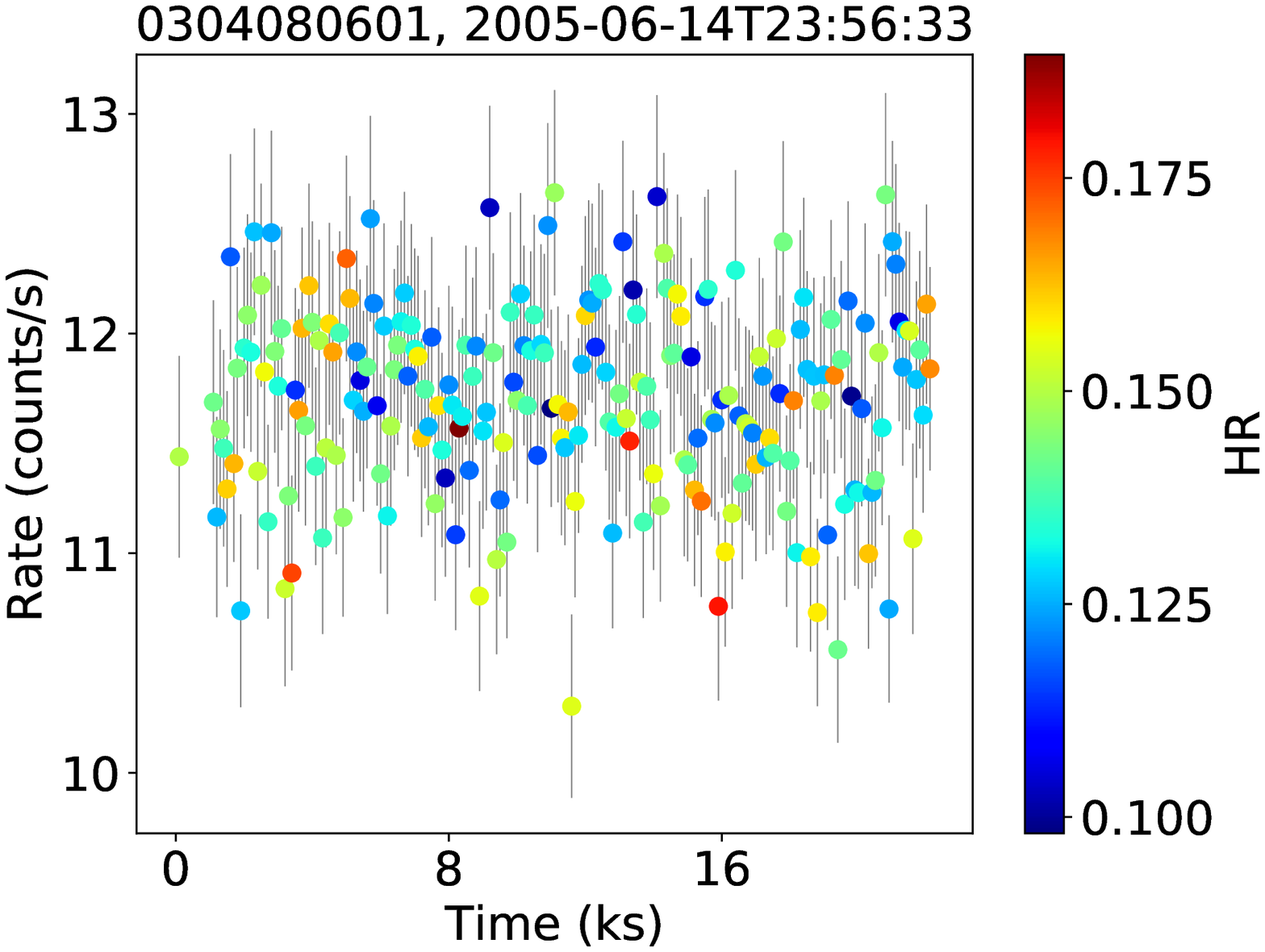}
	\end{minipage}
	\begin{minipage}{.3\textwidth} 
		\centering 
		\includegraphics[width=.99\linewidth]{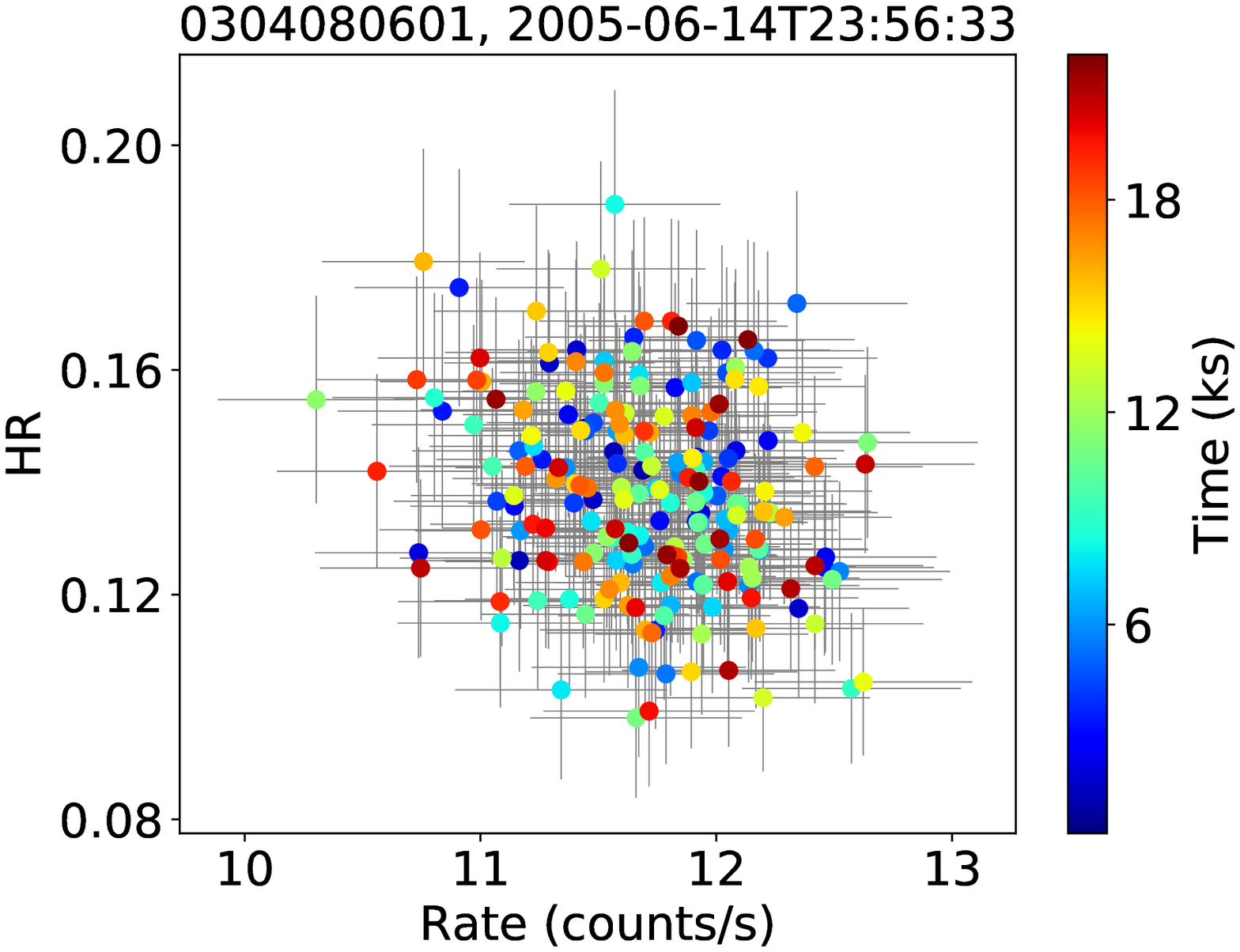}
	\end{minipage}
	\begin{minipage}{.3\textwidth} 
		\centering 
		\includegraphics[height=.99\linewidth, angle=-90]{PlotsPersonal/0304080601_EPLP.eps}
	\end{minipage}
	\begin{minipage}{.3\textwidth} 
		\centering 
		\includegraphics[width=.99\linewidth]{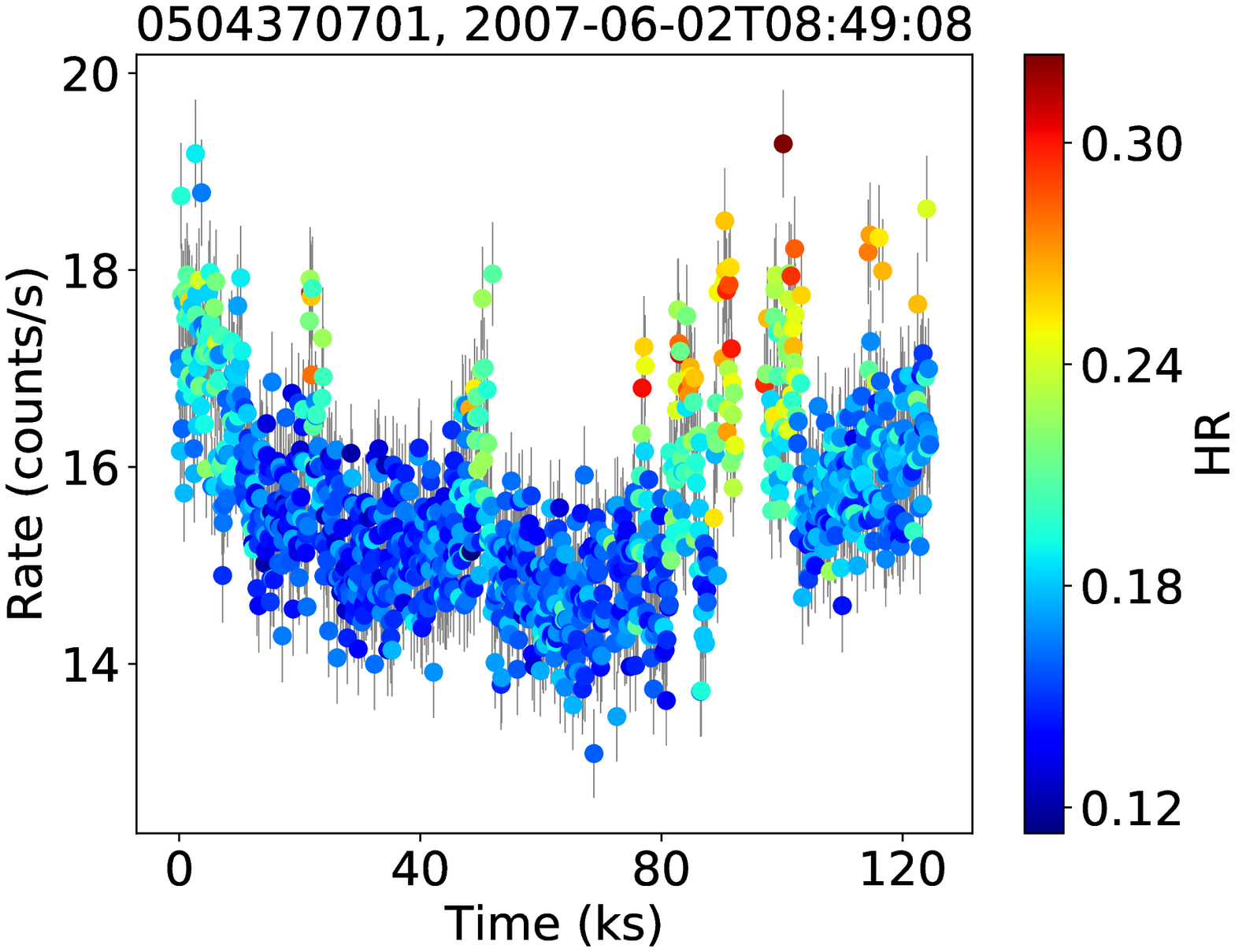}
	\end{minipage}
	\begin{minipage}{.3\textwidth} 
		\centering 
		\includegraphics[width=.99\linewidth]{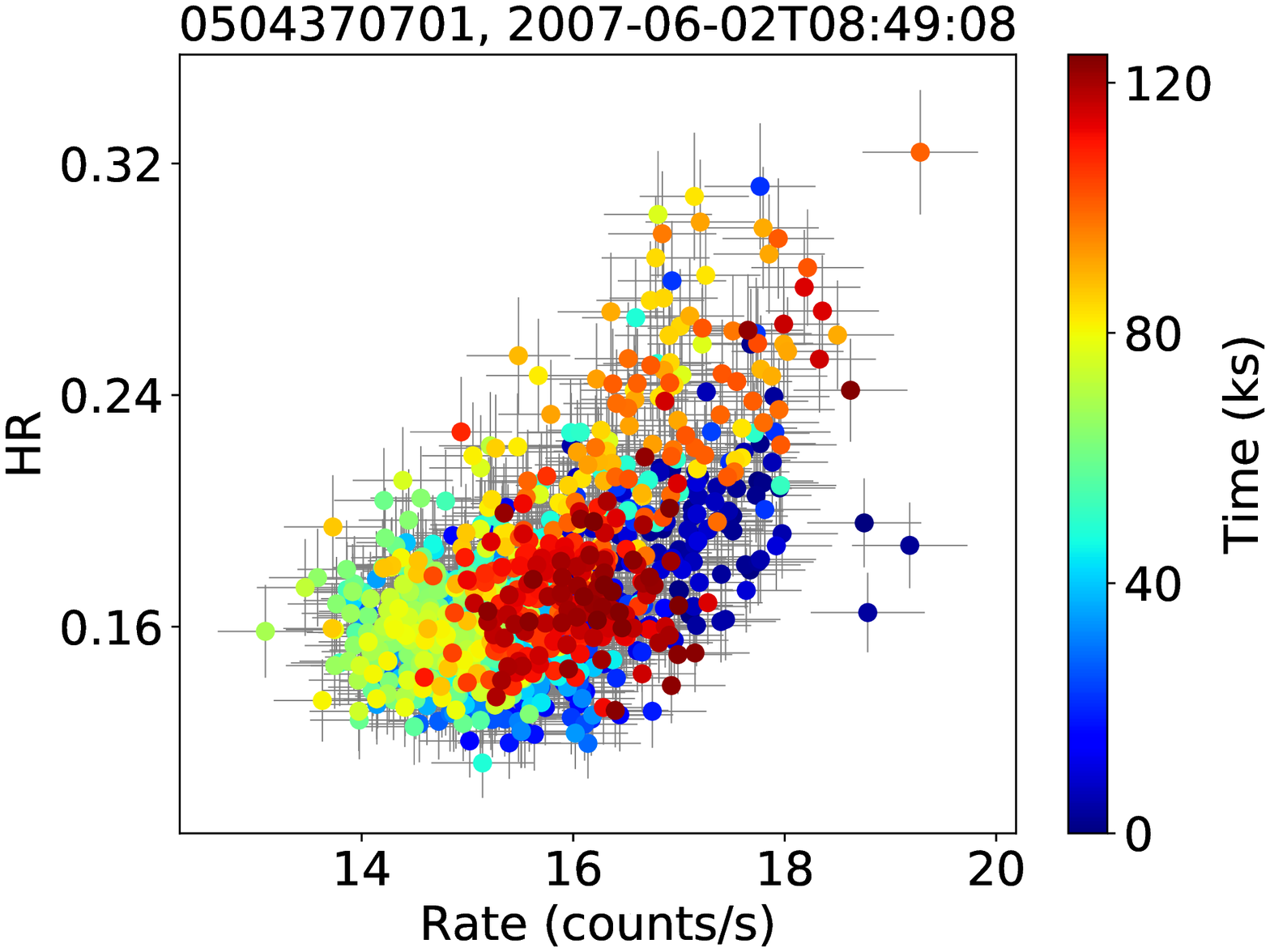}
	\end{minipage}
	\begin{minipage}{.3\textwidth} 
		\centering 
		\includegraphics[height=.99\linewidth, angle=-90]{PlotsPersonal/0504370701_BPL.eps}
	\end{minipage}
	\begin{minipage}{.3\textwidth} 
		\centering 
		\includegraphics[width=.99\linewidth]{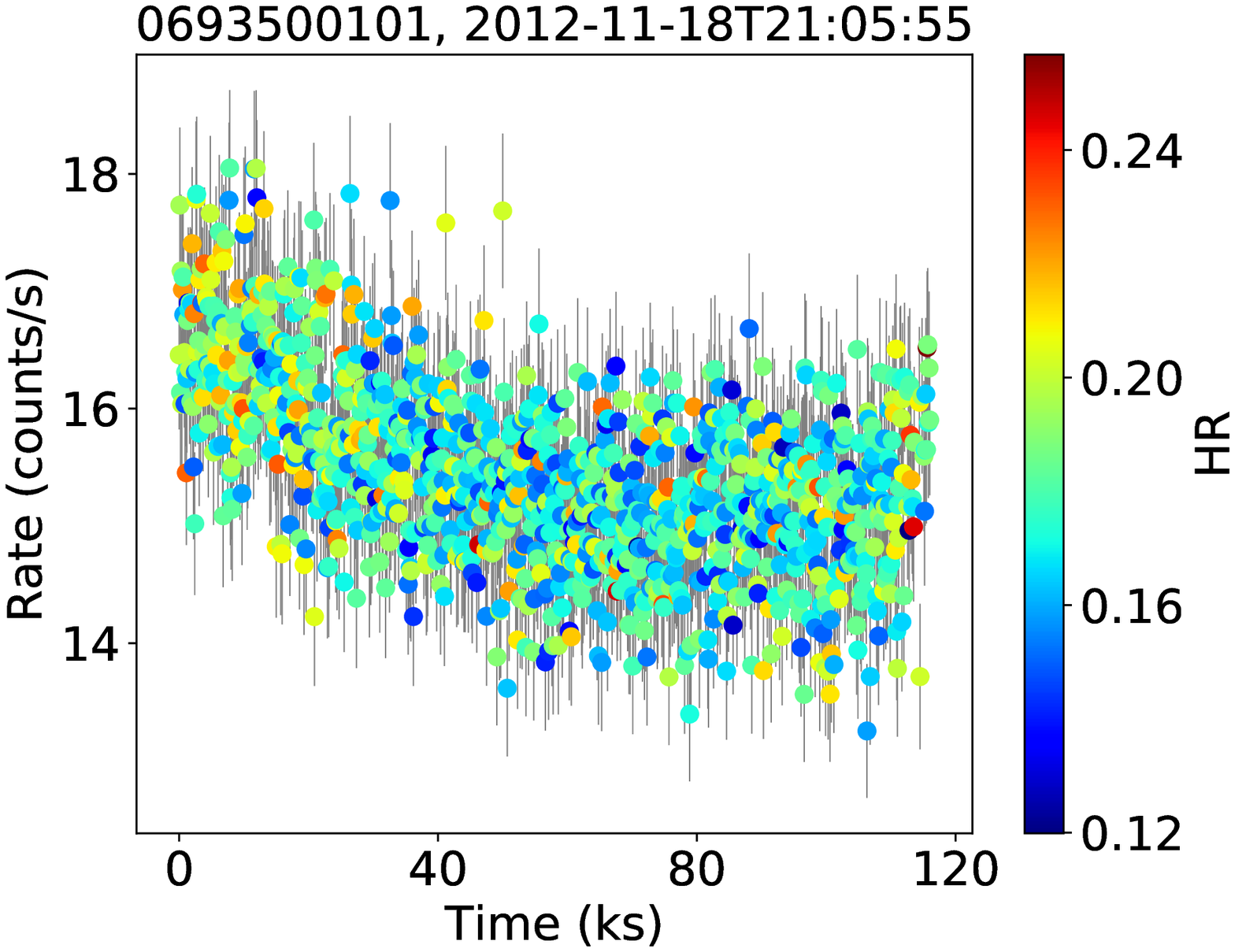}
	\end{minipage}
	\begin{minipage}{.3\textwidth} 
		\centering 
		\includegraphics[width=.99\linewidth]{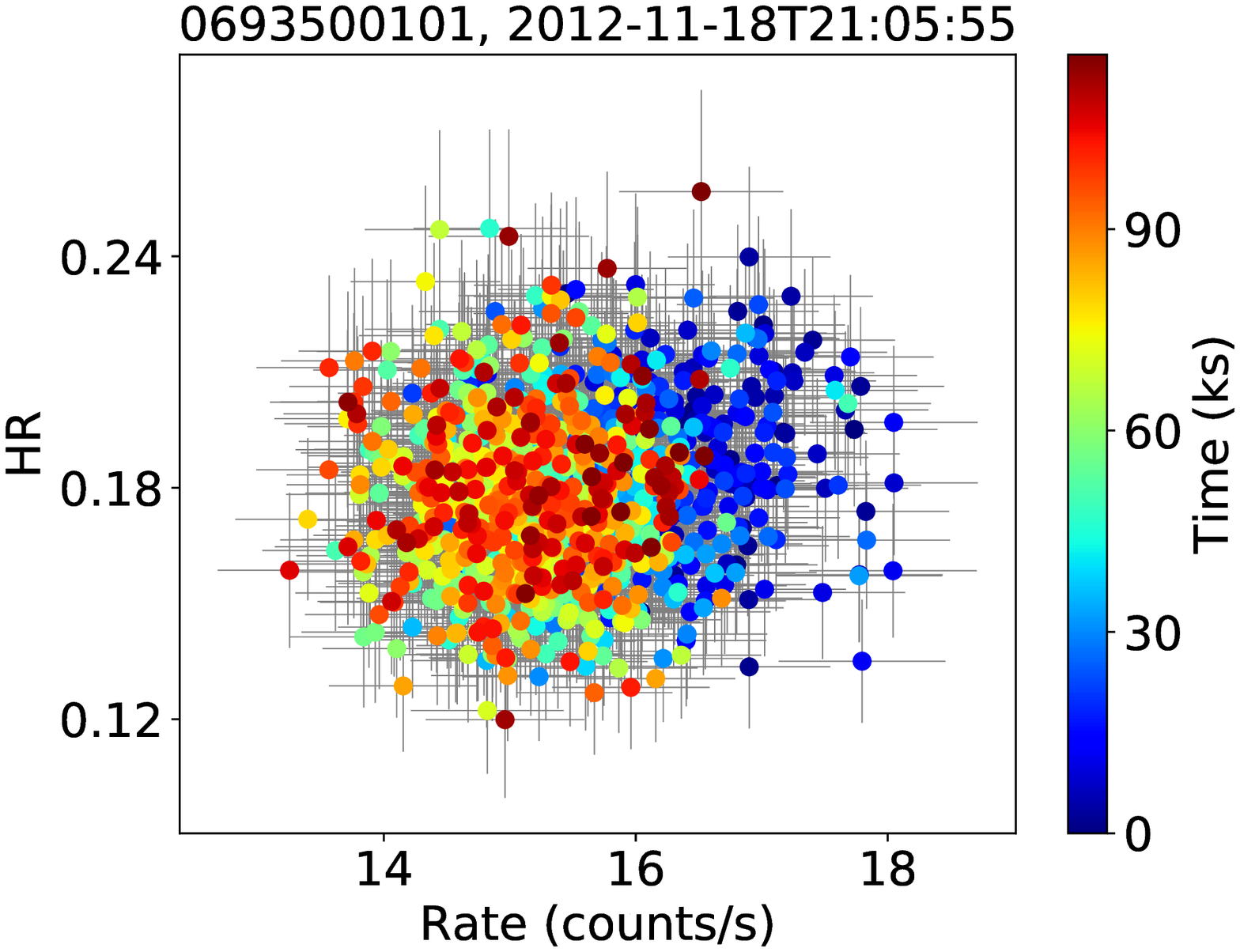}
	\end{minipage}
	\begin{minipage}{.3\textwidth} 
		\centering 
		\includegraphics[height=.99\linewidth, angle=-90]{PlotsPersonal/0693500101_BPL.eps}
	\end{minipage}
	\begin{minipage}{.3\textwidth} 
		\centering 
		\includegraphics[width=.99\linewidth]{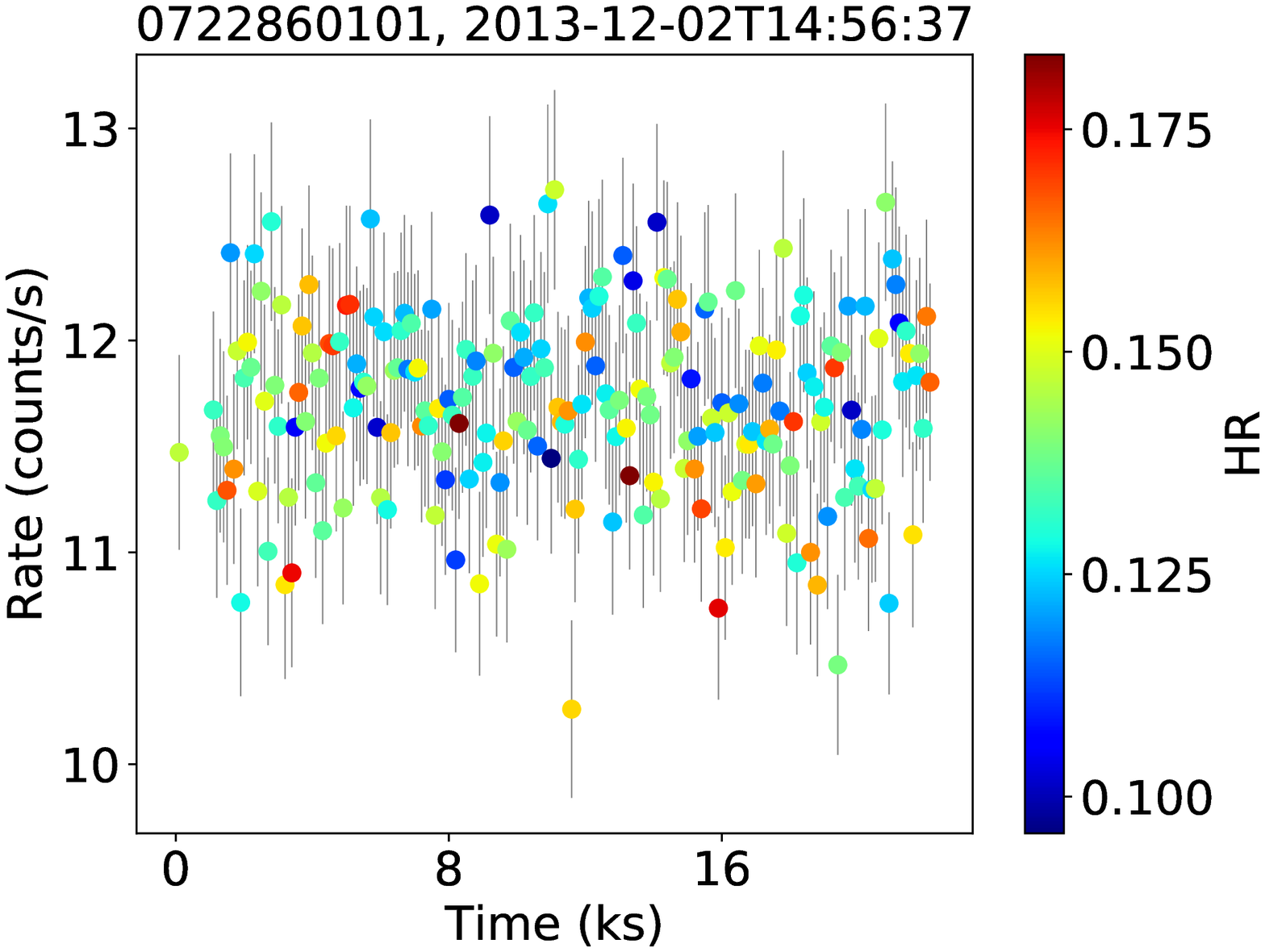}
	\end{minipage}
	\begin{minipage}{.3\textwidth} 
		\centering 
		\includegraphics[width=.99\linewidth]{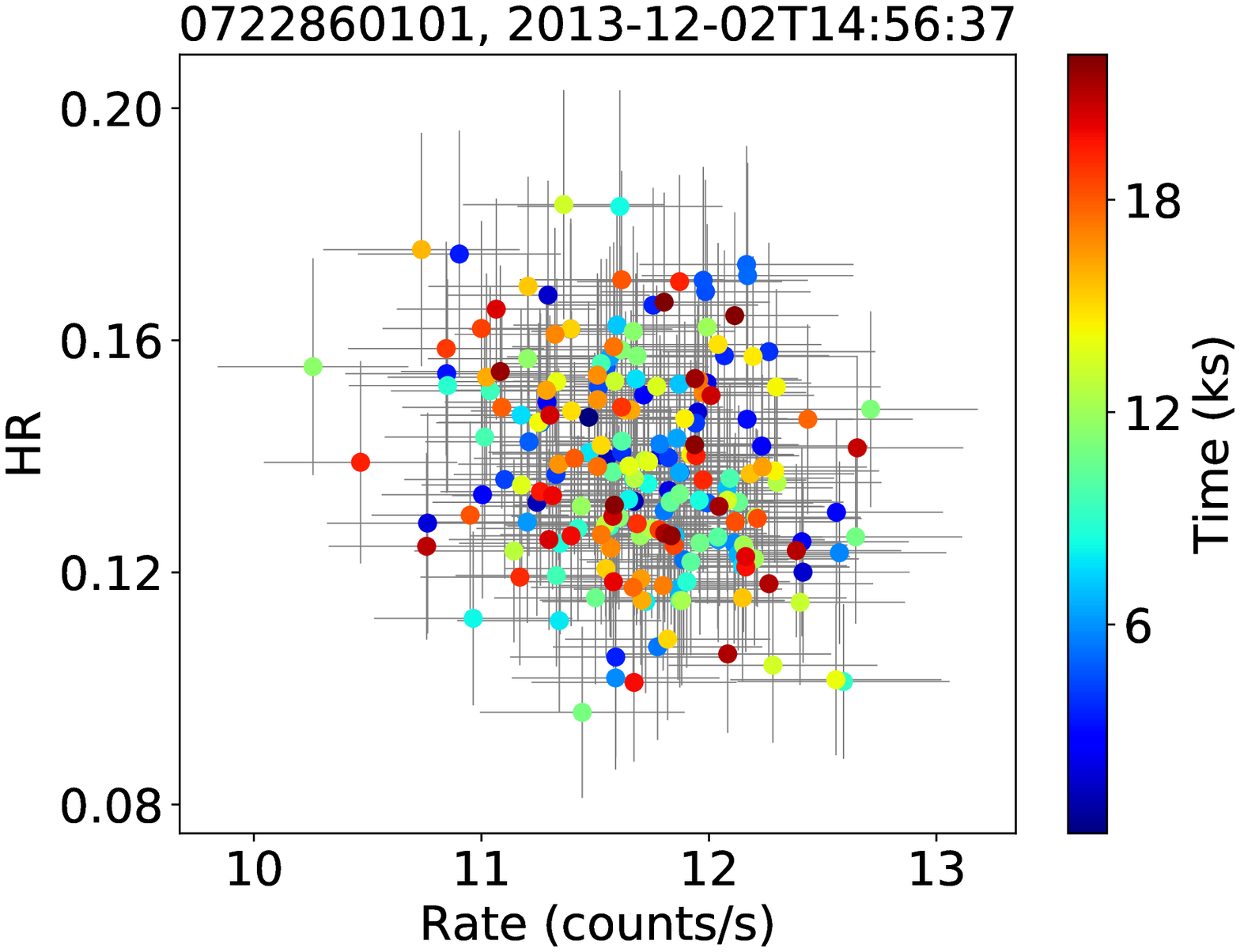}
	\end{minipage}
	\begin{minipage}{.3\textwidth} 
		\centering 
		\includegraphics[height=.99\linewidth, angle=-90]{PlotsPersonal/0722860101_EPLP.eps}
	\end{minipage}
\end{figure*}

\begin{figure*}
    \centering
	\begin{minipage}{.3\textwidth} 
		\centering 
		\includegraphics[width=.99\linewidth]{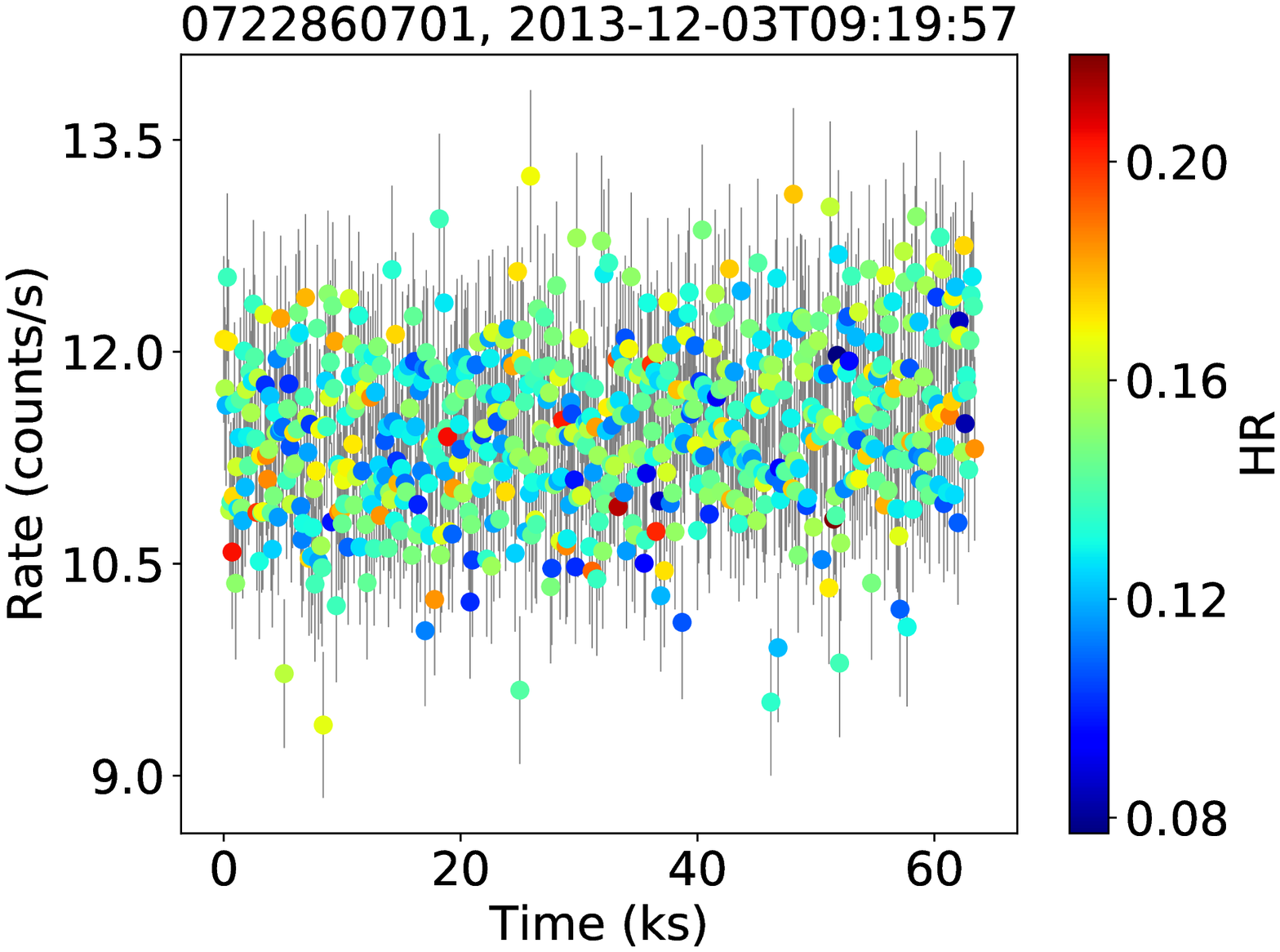}
	\end{minipage}
	\begin{minipage}{.3\textwidth} 
		\centering 
		\includegraphics[width=.99\linewidth]{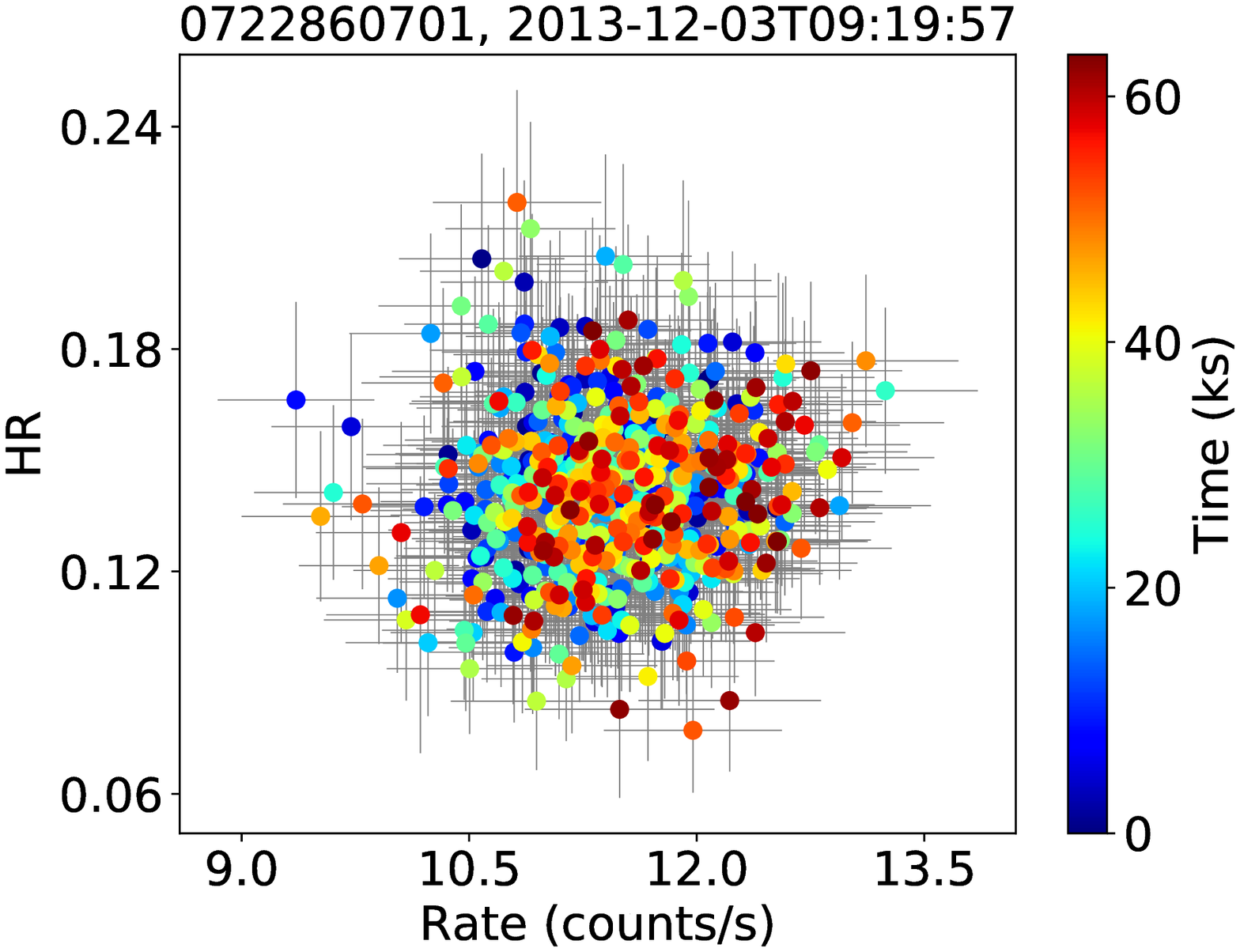}
	\end{minipage}
	\begin{minipage}{.3\textwidth} 
		\centering 
		\includegraphics[height=.99\linewidth, angle=-90]{PlotsPersonal/0722860701_EPLP.eps}
	\end{minipage}
	\begin{minipage}{.3\textwidth} 
		\centering 
		\includegraphics[width=.99\linewidth]{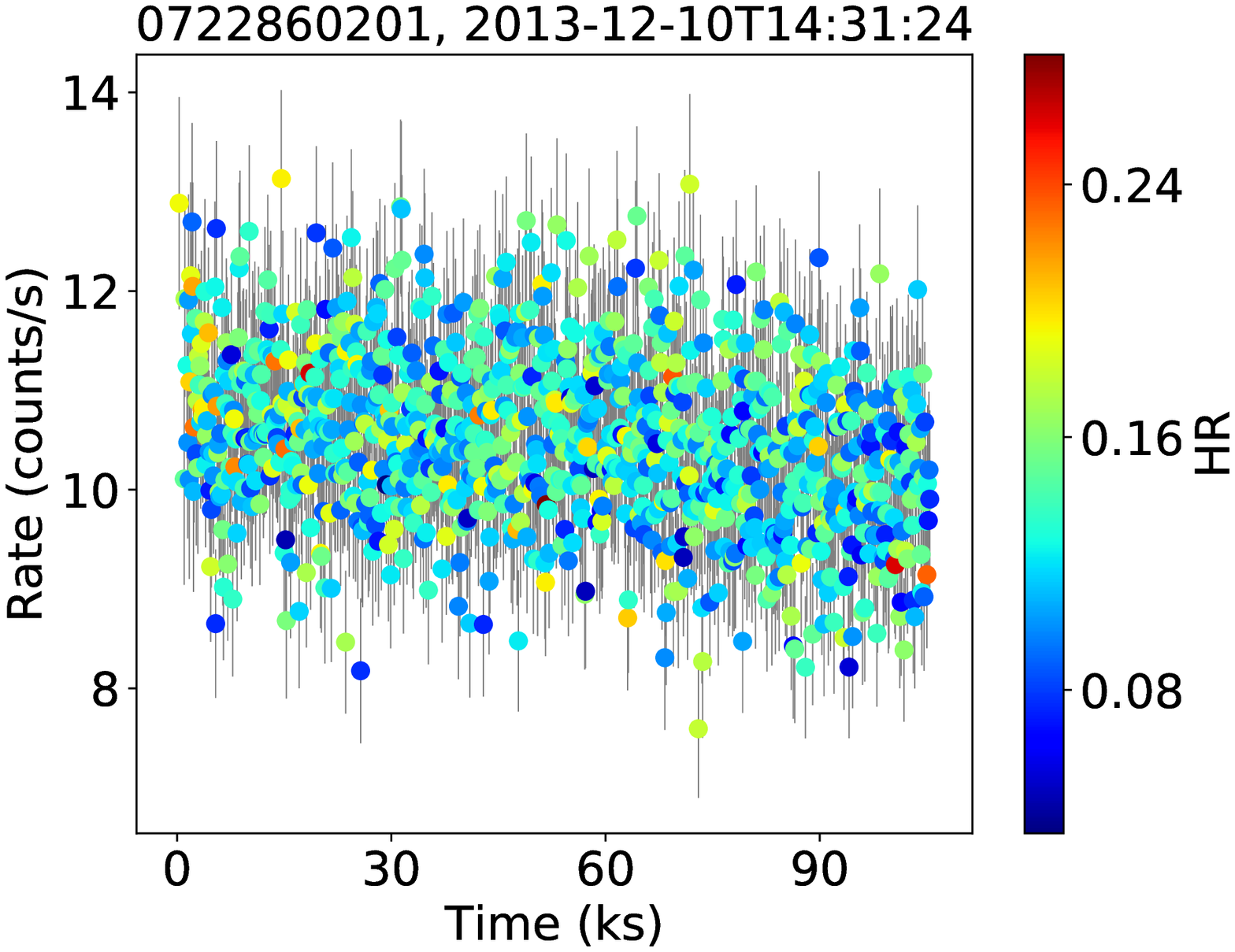}
	\end{minipage}
	\begin{minipage}{.3\textwidth} 
		\centering 
		\includegraphics[width=.99\linewidth]{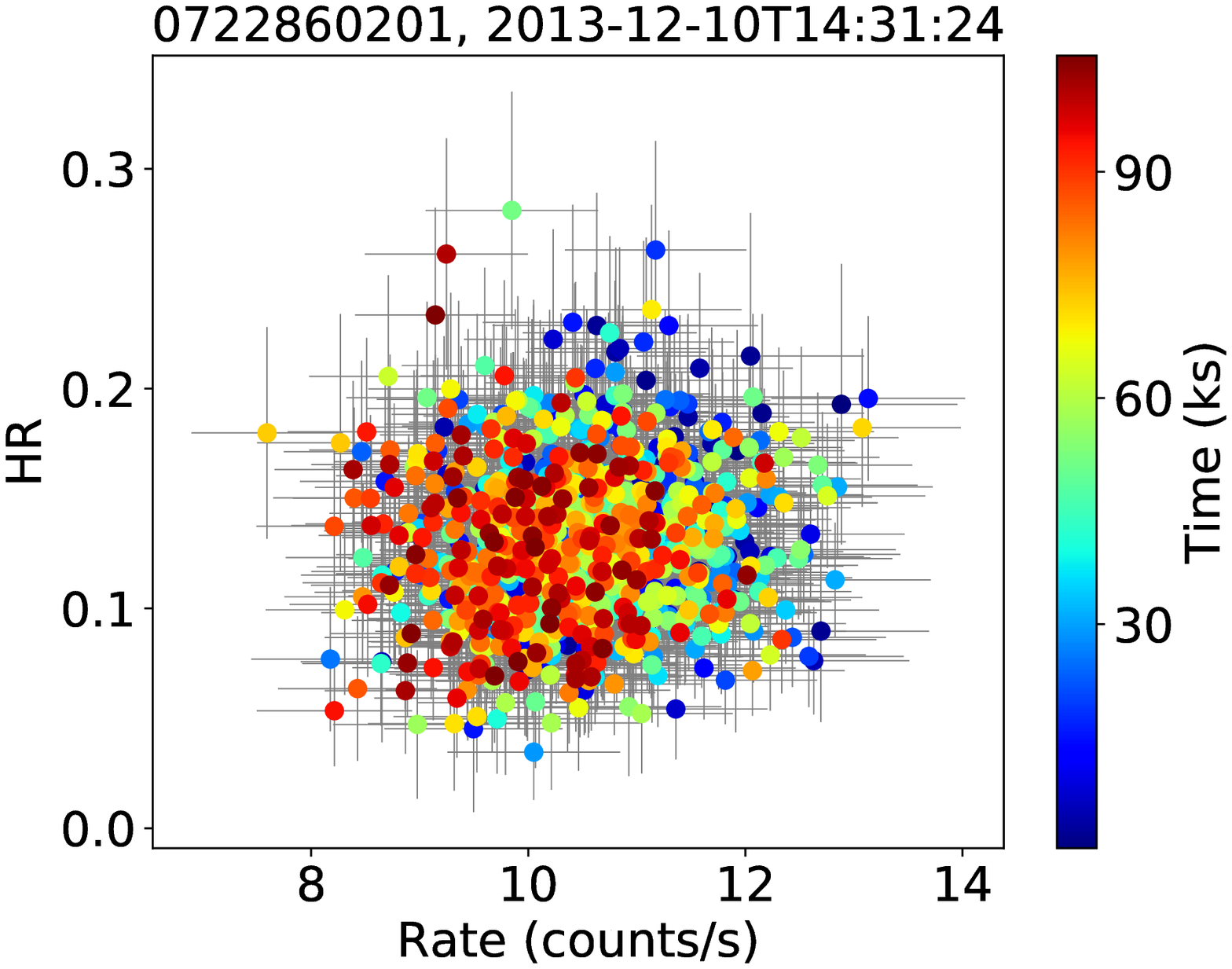}
	\end{minipage}
	\begin{minipage}{.3\textwidth} 
		\centering 
		\includegraphics[height=.99\linewidth, angle=-90]{PlotsPersonal/0722860201_EPLP.eps}
	\end{minipage}
	\begin{minipage}{.3\textwidth} 
		\centering 
		\includegraphics[width=.99\linewidth]{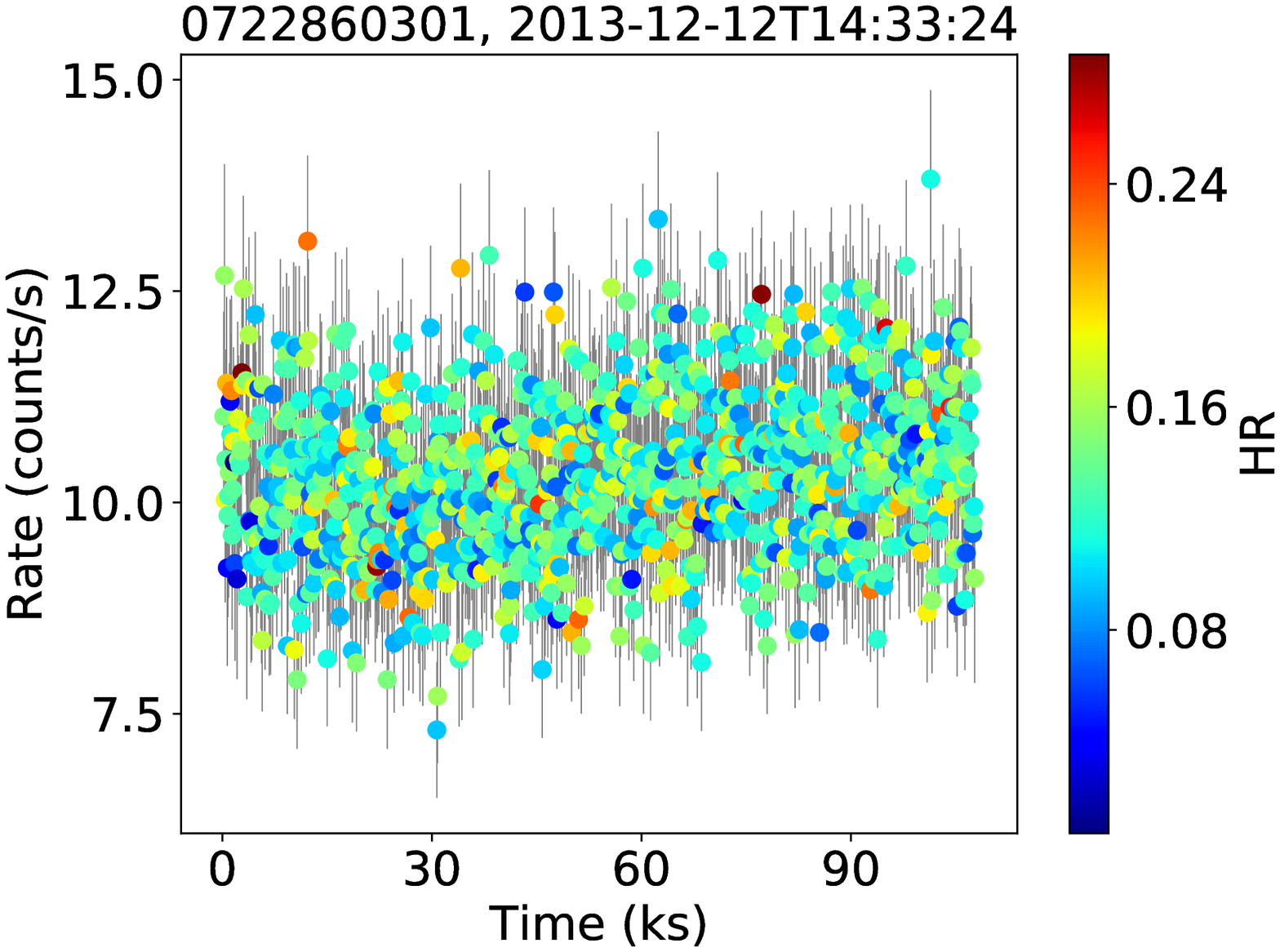}
	\end{minipage}
	\begin{minipage}{.3\textwidth} 
		\centering 
		\includegraphics[width=.99\linewidth]{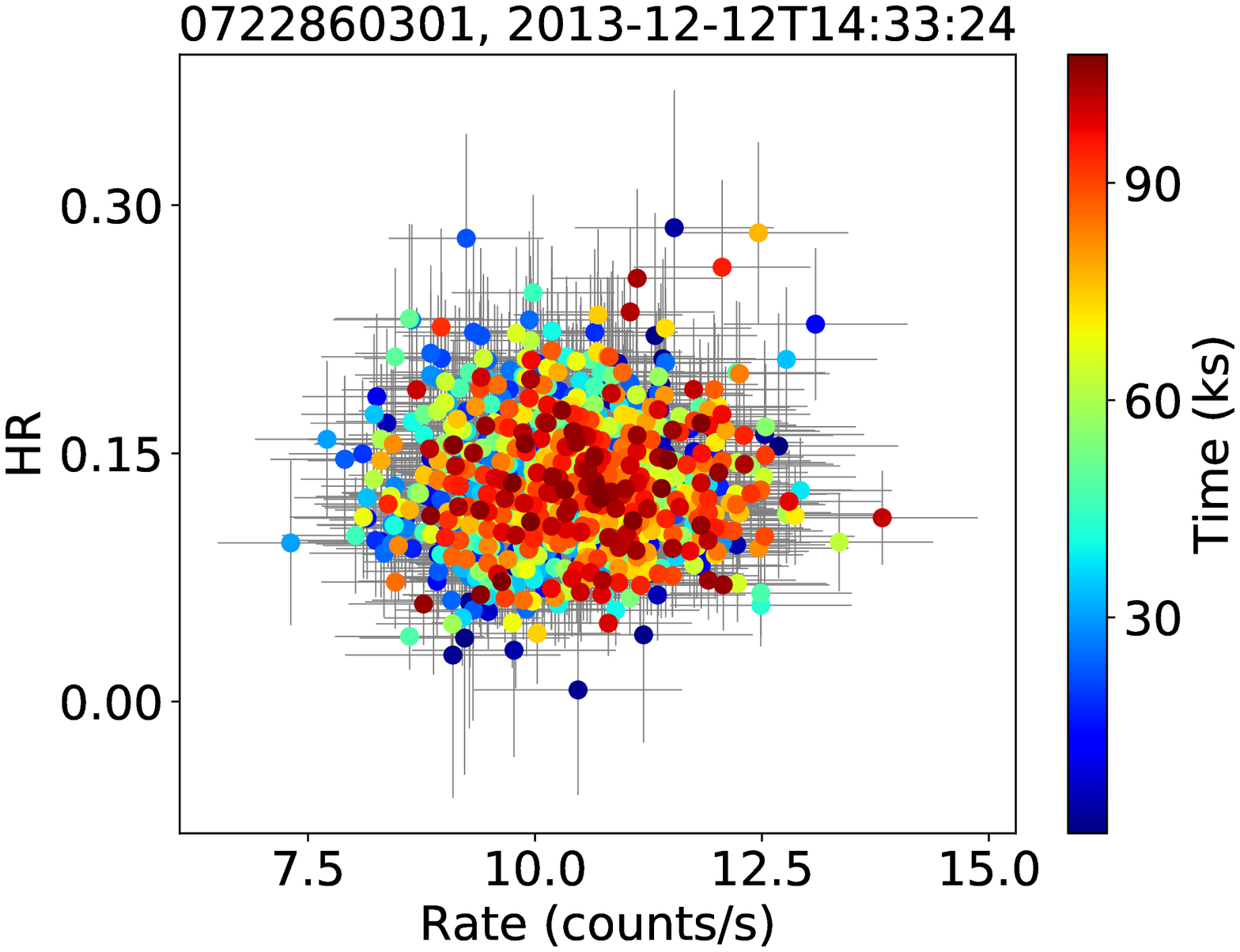}
	\end{minipage}
	\begin{minipage}{.3\textwidth} 
		\centering 
		\includegraphics[height=.99\linewidth, angle=-90]{PlotsPersonal/0722860301_EPLP.eps}
	\end{minipage}
	\begin{minipage}{.3\textwidth} 
		\centering 
		\includegraphics[width=.99\linewidth]{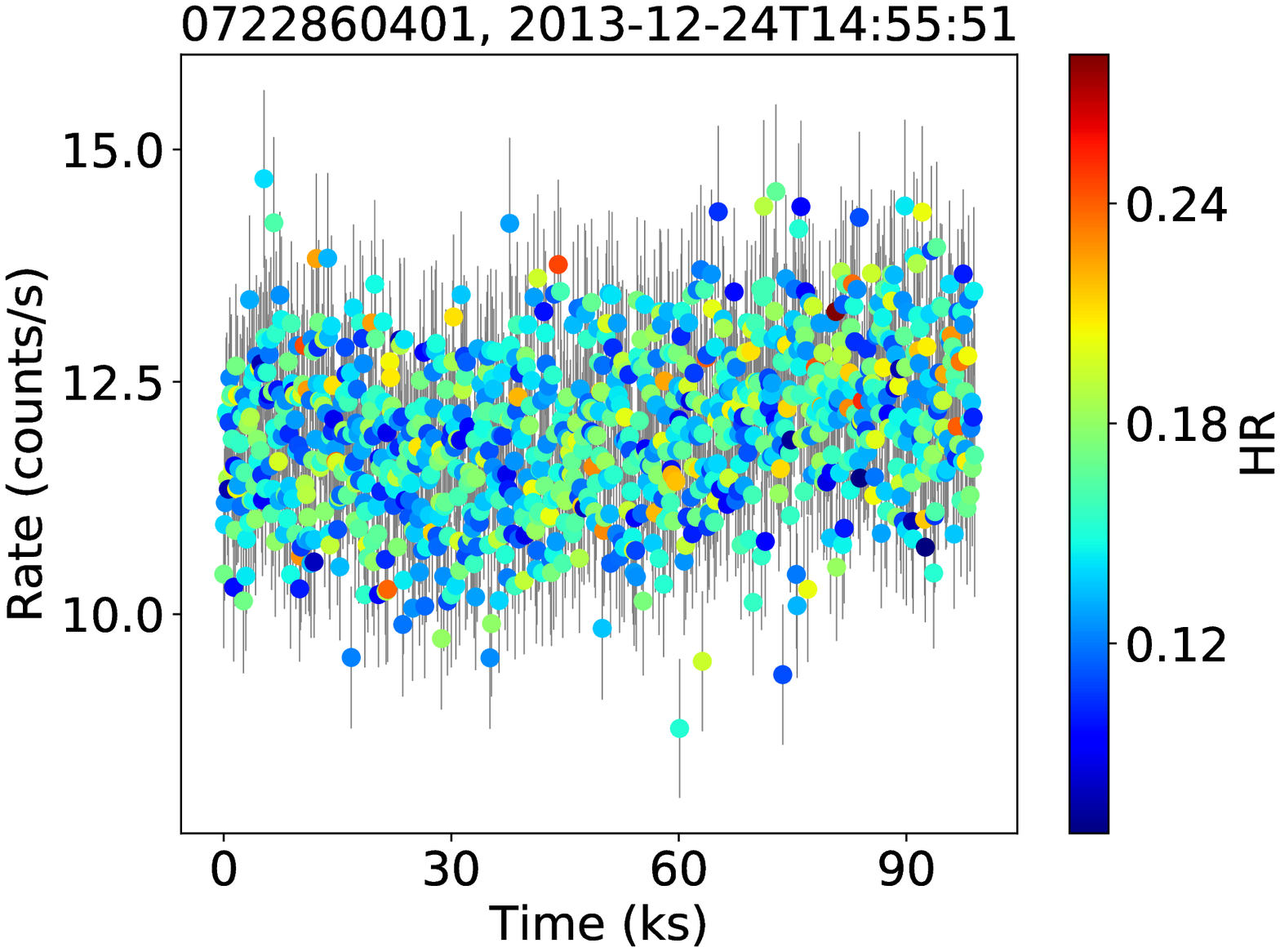}
	\end{minipage}
	\begin{minipage}{.3\textwidth} 
		\centering 
		\includegraphics[width=.99\linewidth]{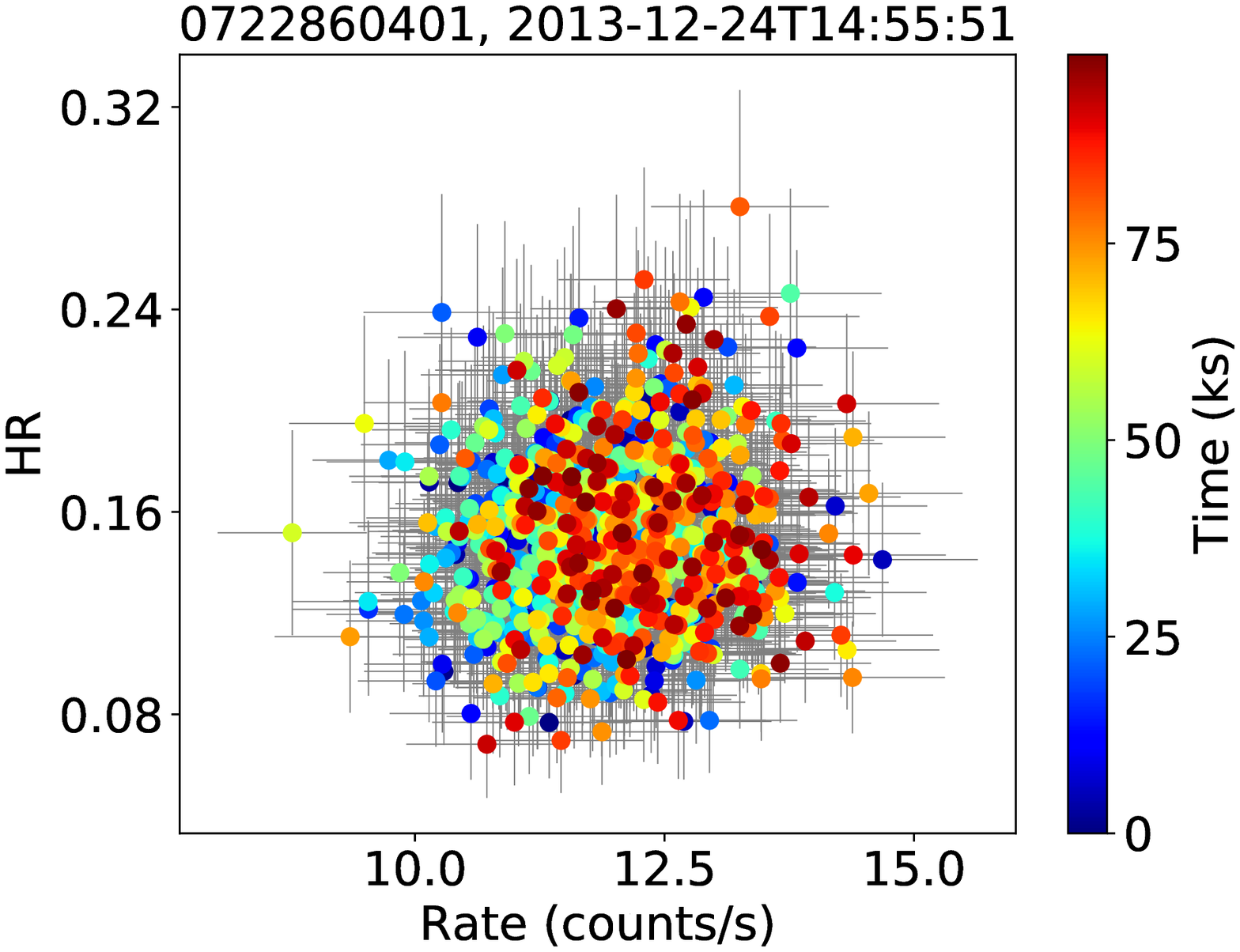}
	\end{minipage}
	\begin{minipage}{.3\textwidth} 
		\centering 
		\includegraphics[height=.99\linewidth, angle=-90]{PlotsPersonal/0722860401_EPLP.eps}
	\end{minipage}
\end{figure*}

\begin{figure*}\label{app:mrk}
	\centering
	\caption{LCs, HR plots and spectral fits derived from observations of Mrk 501.}
	\label{fig:LC_Mrk}
	\begin{minipage}{.3\textwidth} 
		\centering 
		\includegraphics[width=.99\linewidth]{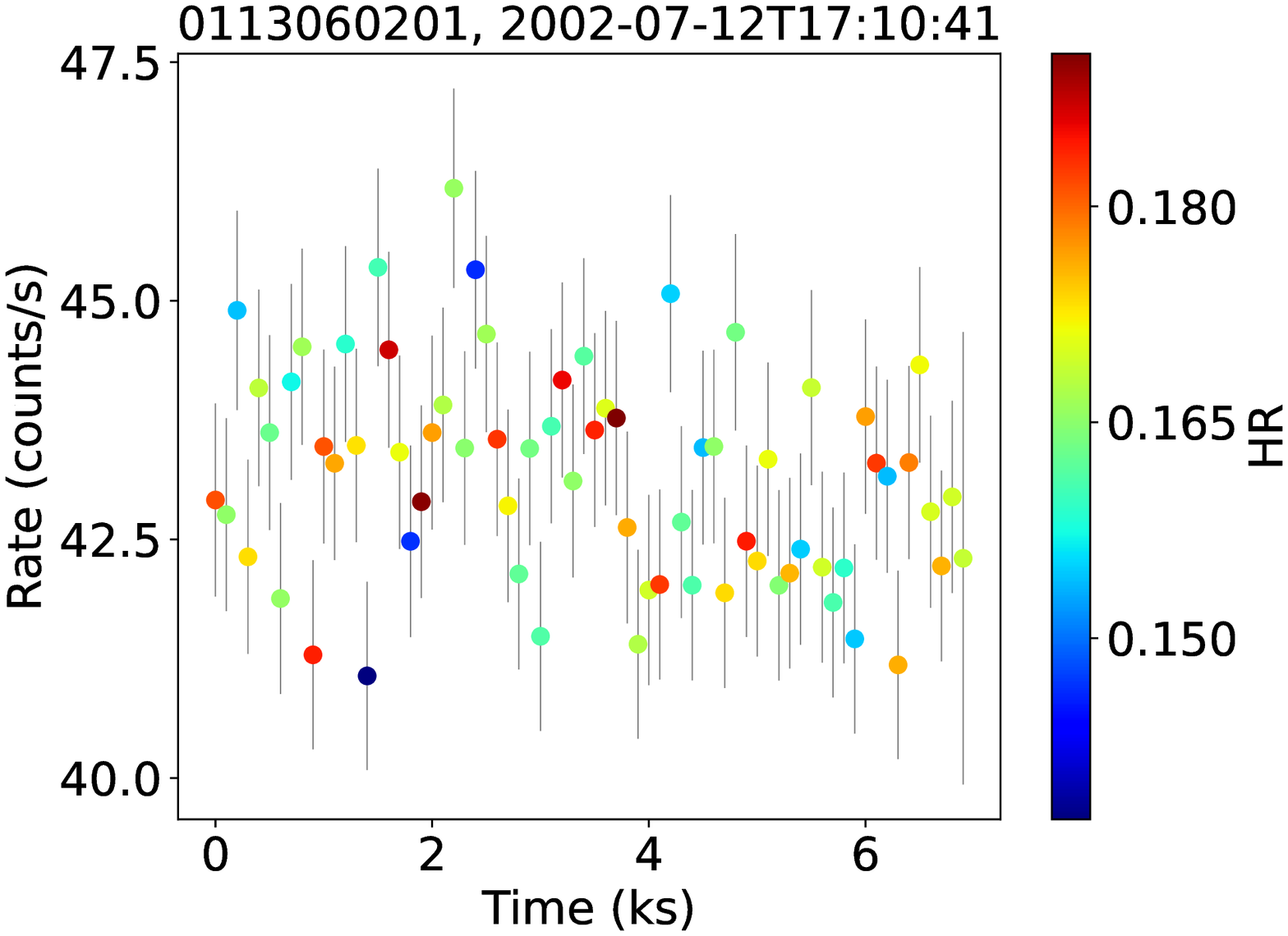}
	\end{minipage}
	\begin{minipage}{.3\textwidth} 
		\centering 
		\includegraphics[width=.99\linewidth]{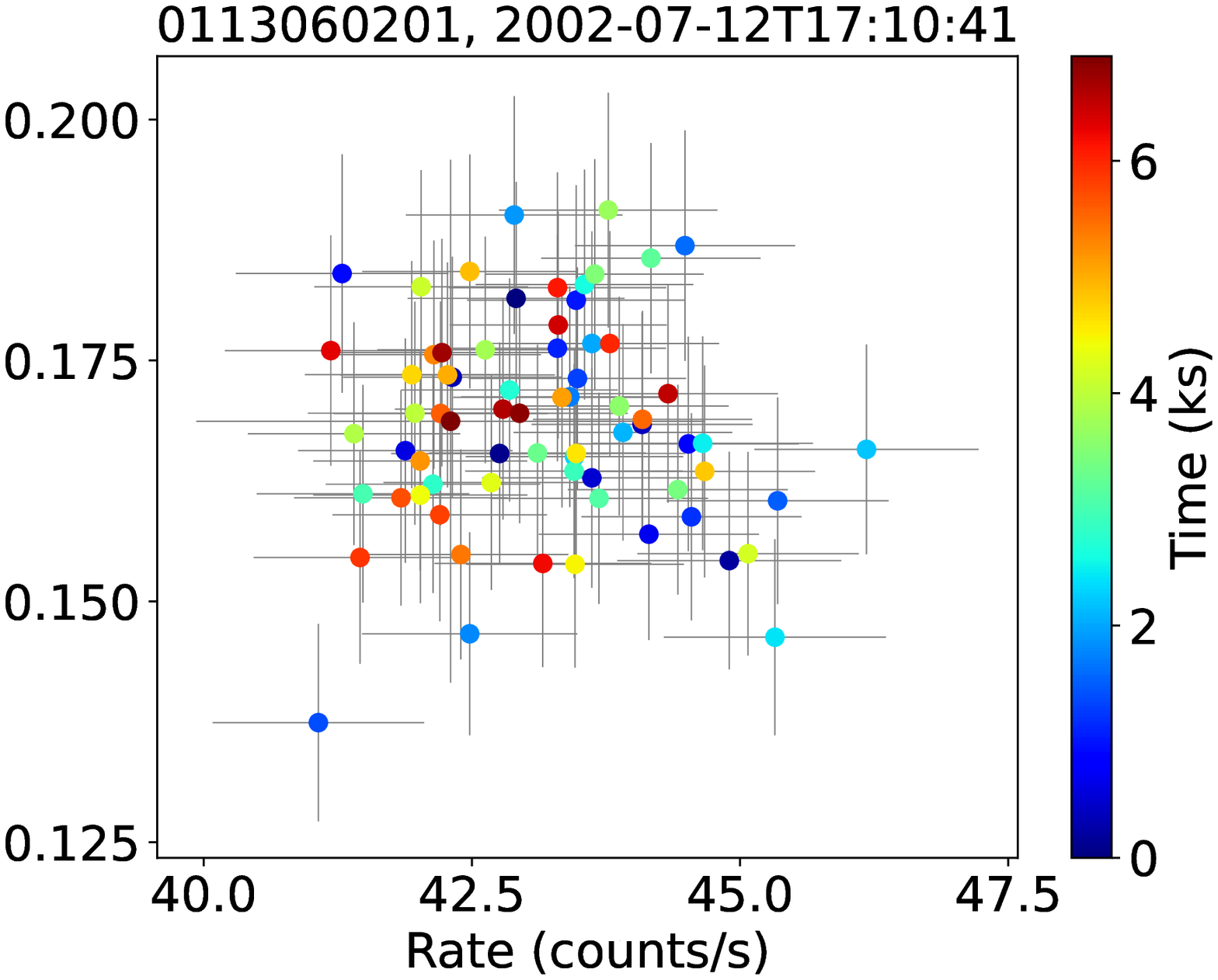}
	\end{minipage}
	\begin{minipage}{.3\textwidth} 
		\centering 
		\includegraphics[height=.99\linewidth, angle=-90]{PlotsPersonal/0113060201_BPL.eps}
	\end{minipage}
\end{figure*}

\begin{figure*}
	\centering
	\begin{minipage}{.3\textwidth} 
		\centering 
		\includegraphics[width=.99\linewidth]{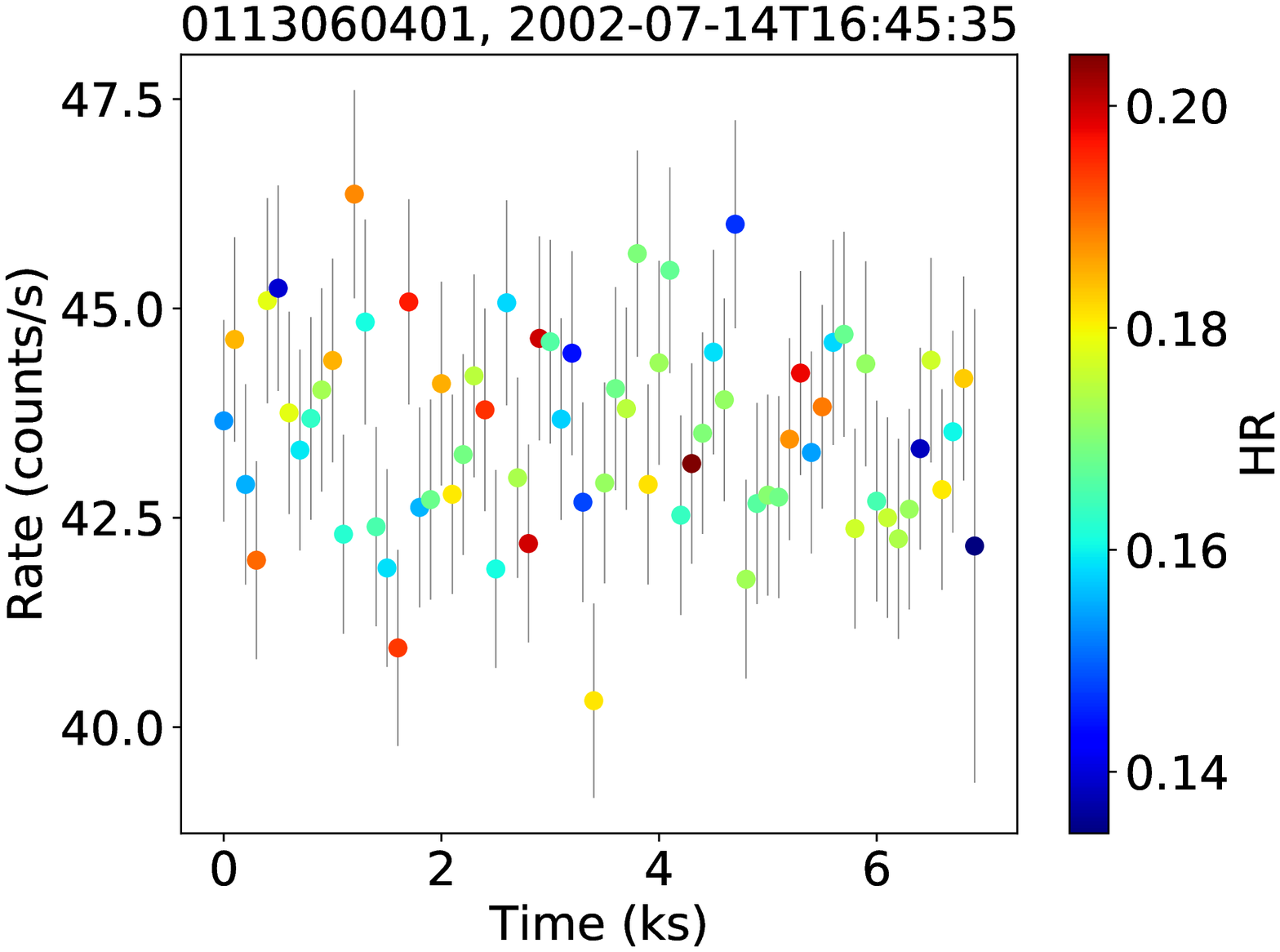}
	\end{minipage}
	\begin{minipage}{.3\textwidth} 
		\centering 
		\includegraphics[width=.99\linewidth]{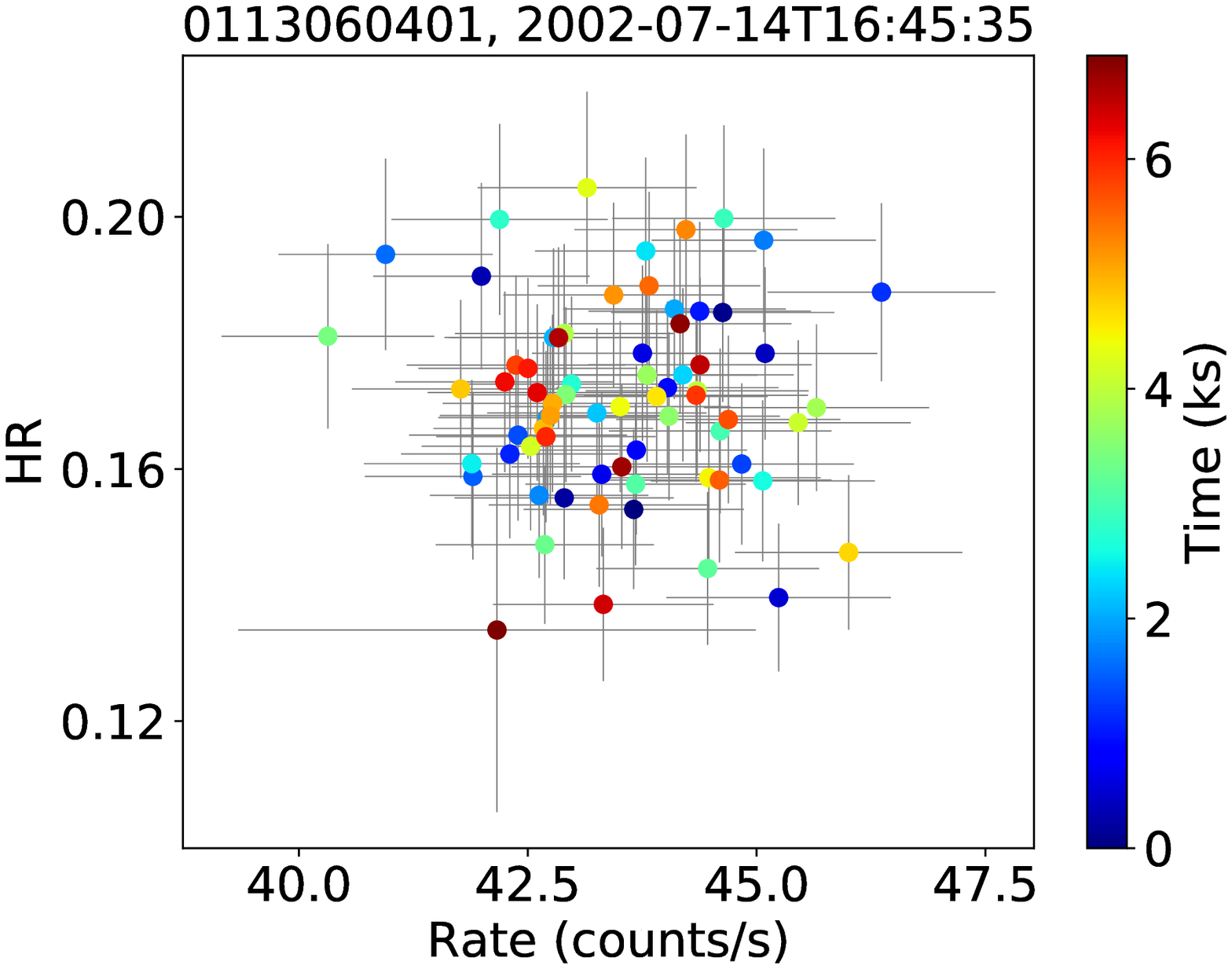}
	\end{minipage}
	\begin{minipage}{.3\textwidth} 
		\centering 
		\includegraphics[height=.99\linewidth, angle=-90]{PlotsPersonal/0113060401_BPL.eps}
	\end{minipage}
	\begin{minipage}{.3\textwidth} 
		\centering 
		\includegraphics[width=.99\linewidth]{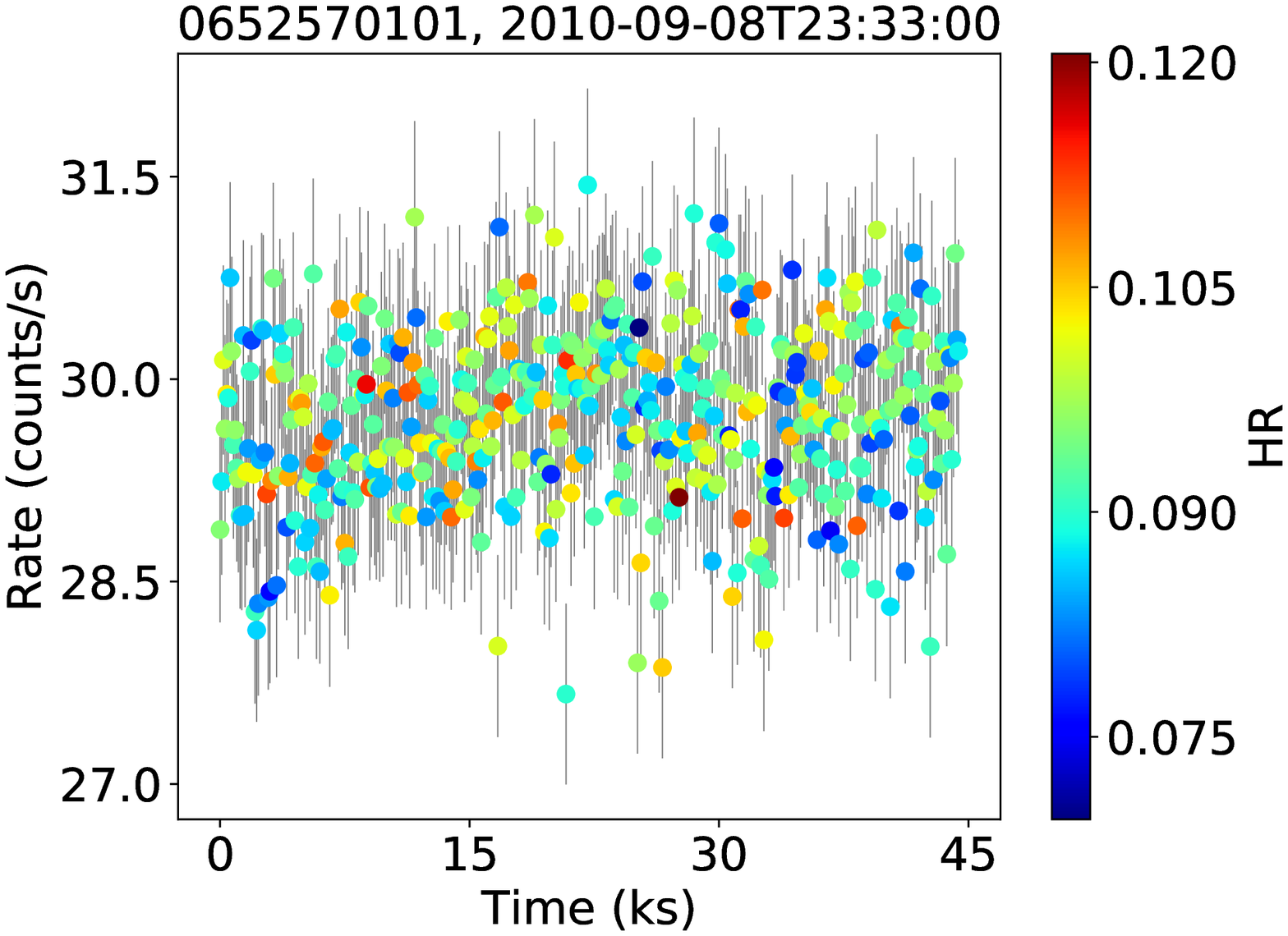}
	\end{minipage}
	\begin{minipage}{.3\textwidth} 
		\centering 
		\includegraphics[width=.99\linewidth]{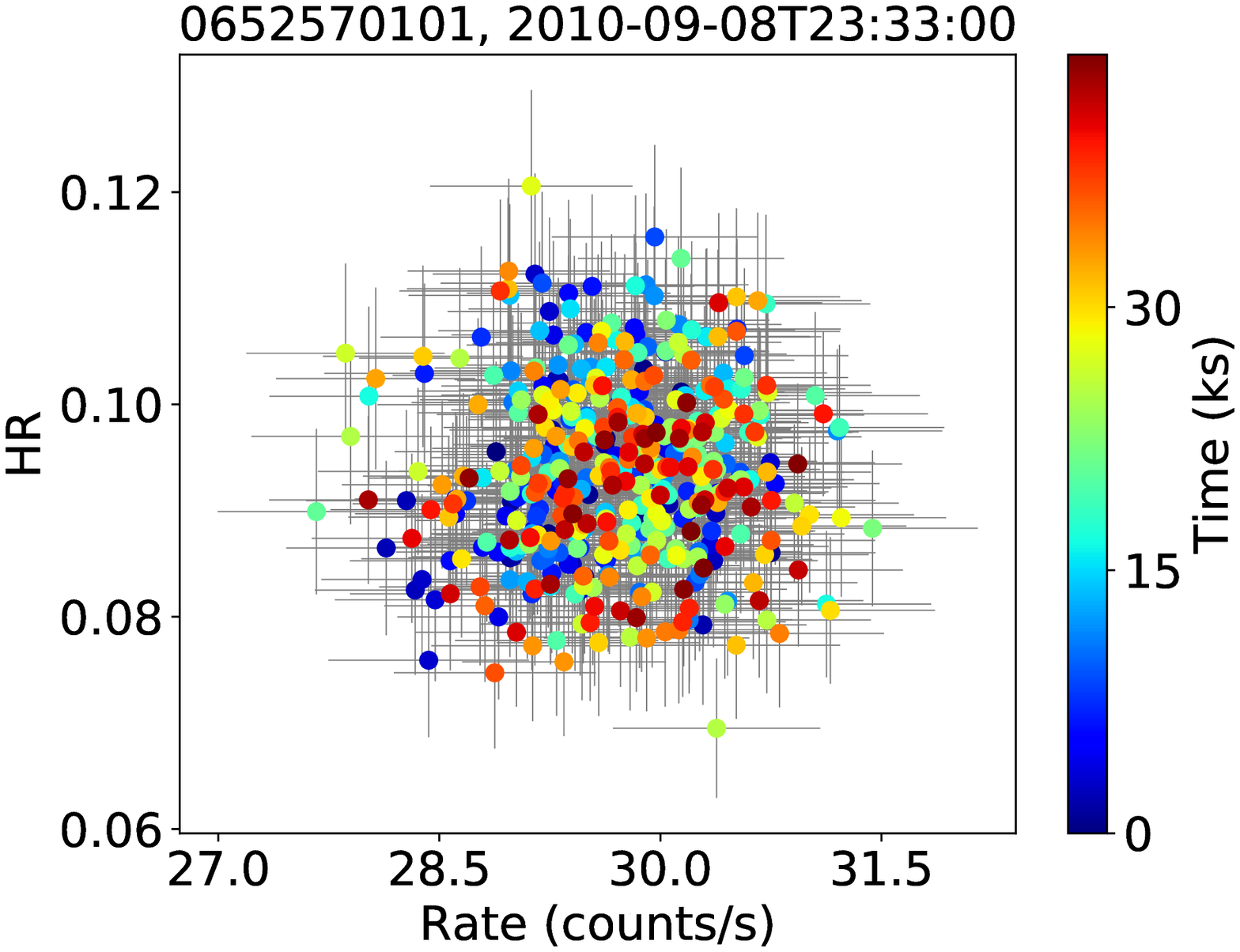}
	\end{minipage}
	\begin{minipage}{.3\textwidth} 
		\centering 
		\includegraphics[height=.99\linewidth, angle=-90]{PlotsPersonal/0652570101_EPLP.eps}
	\end{minipage}
	\begin{minipage}{.3\textwidth} 
		\centering 
		\includegraphics[width=.99\linewidth]{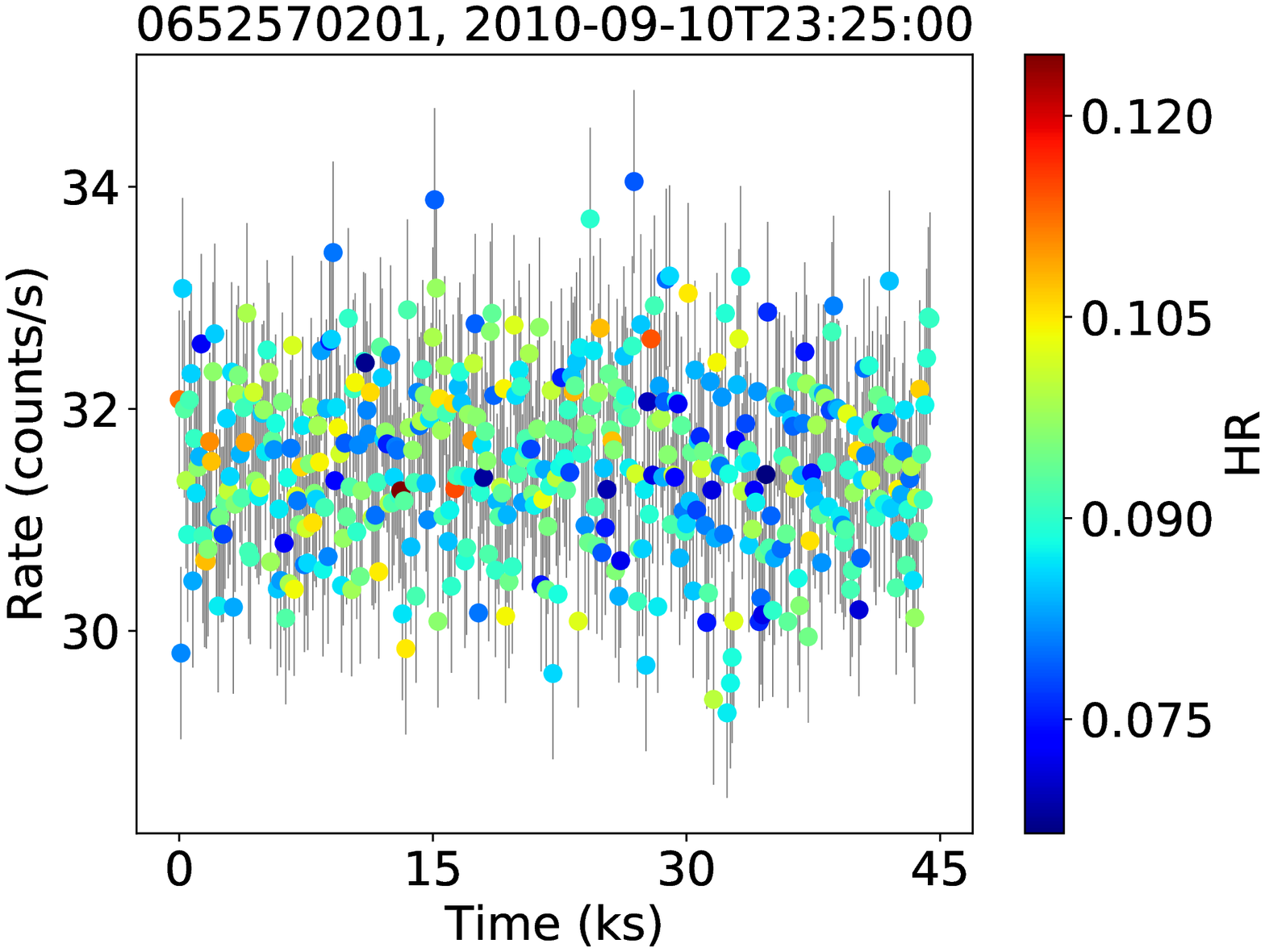}
	\end{minipage}
	\begin{minipage}{.3\textwidth} 
		\centering 
		\includegraphics[width=.99\linewidth]{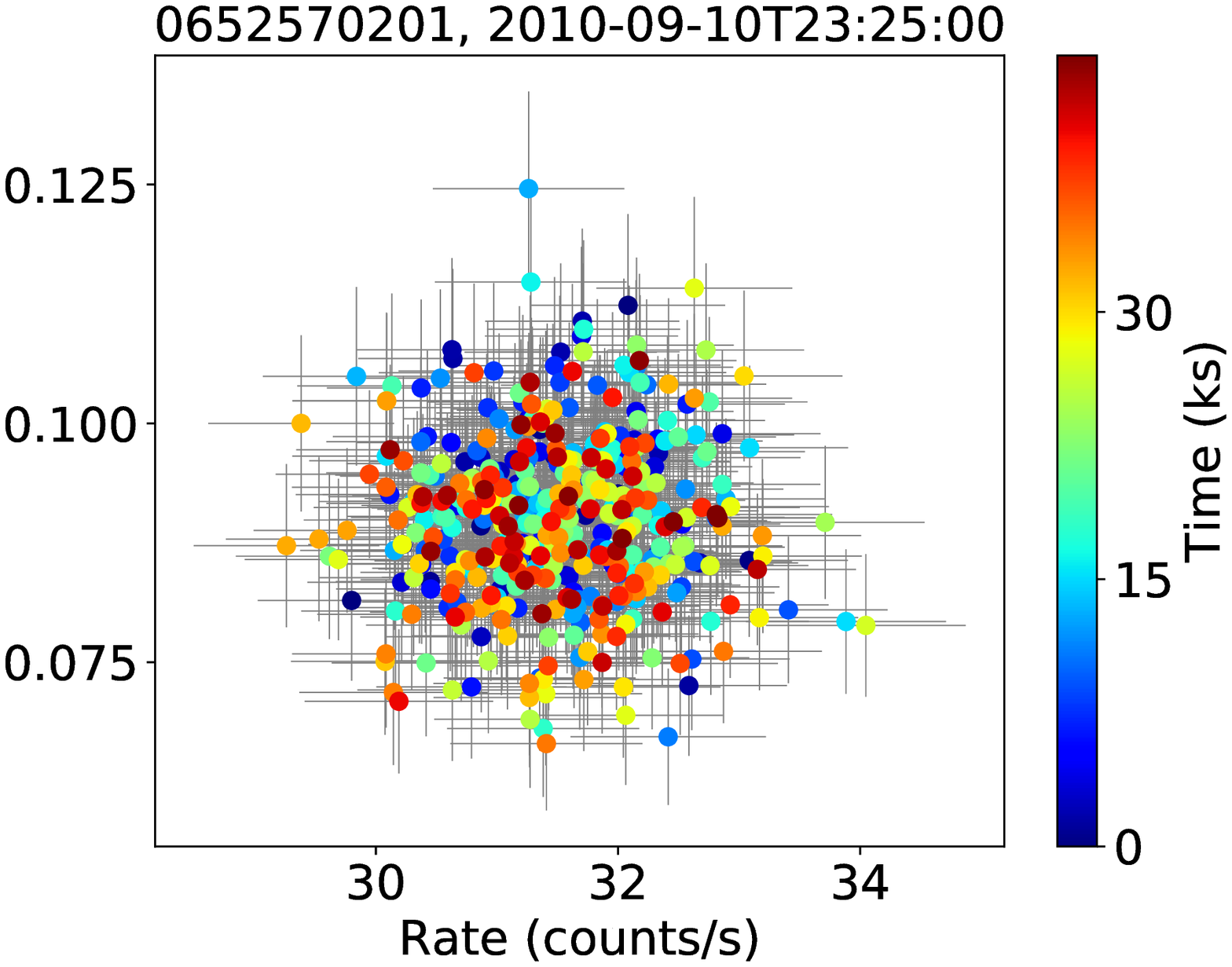}
	\end{minipage}
	\begin{minipage}{.3\textwidth}
		\centering 
		\includegraphics[height=.99\linewidth, angle=-90]{PlotsPersonal/0652570201_EPLP.eps}
	\end{minipage}
	\begin{minipage}{.3\textwidth}
		\centering 
		\includegraphics[width=.99\linewidth]{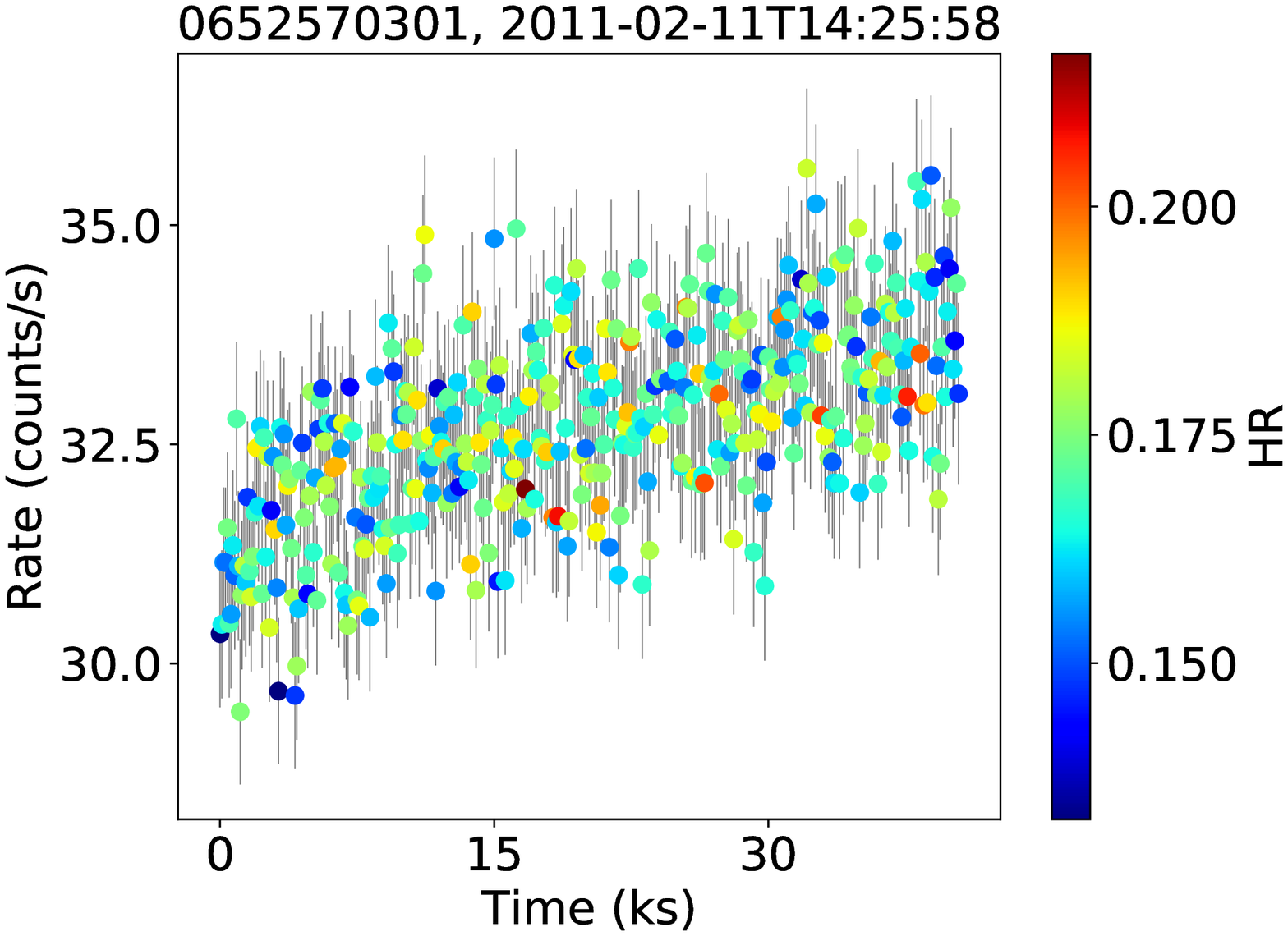}
	\end{minipage}
	\begin{minipage}{.3\textwidth} 
		\centering 
		\includegraphics[width=.99\linewidth]{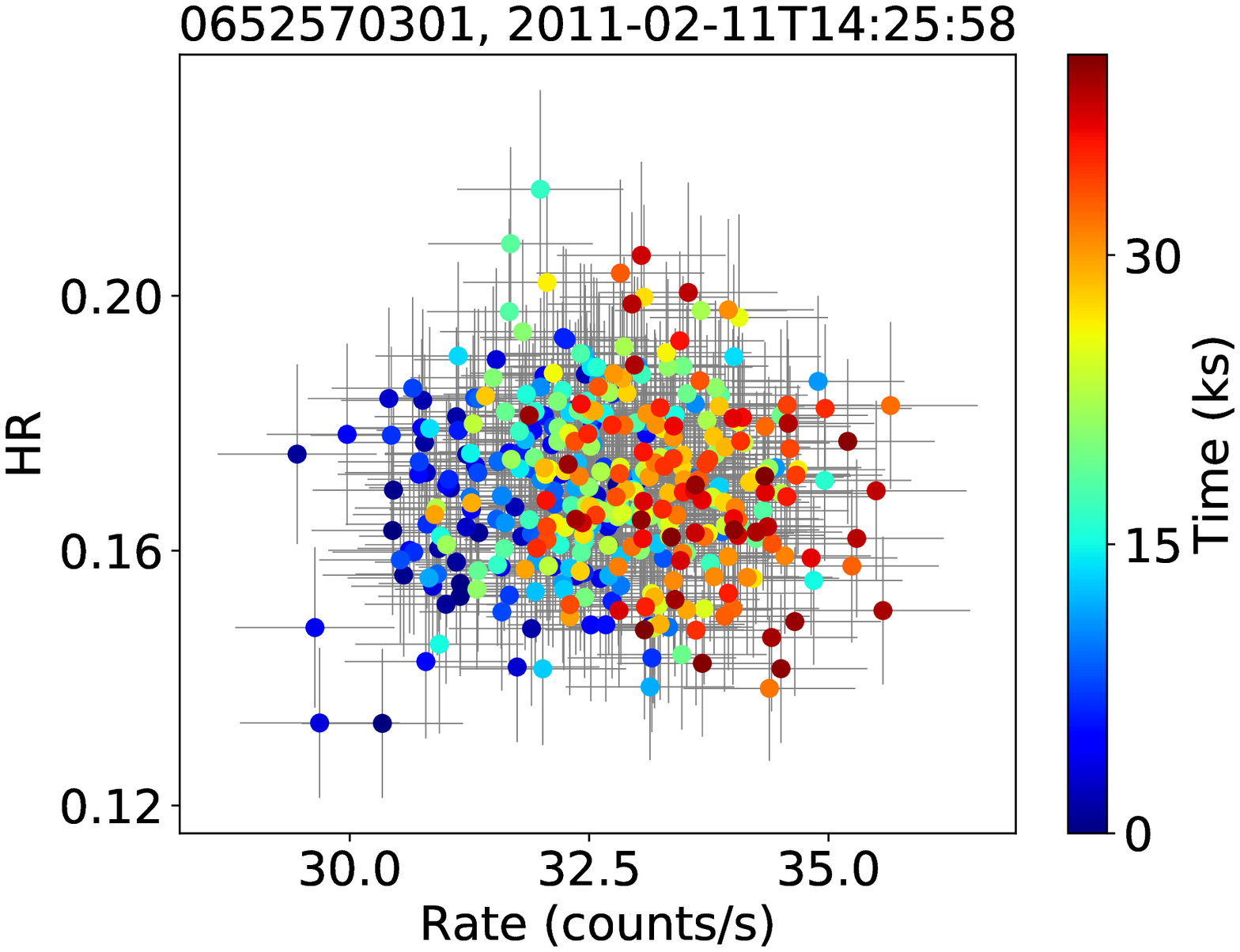}
	\end{minipage}
	\begin{minipage}{.3\textwidth} 
		\centering 
		\includegraphics[height=.99\linewidth, angle=-90]{PlotsPersonal/0652570301_BPL.eps}
	\end{minipage}
	\begin{minipage}{.3\textwidth} 
		\centering 
		\includegraphics[width=.99\linewidth]{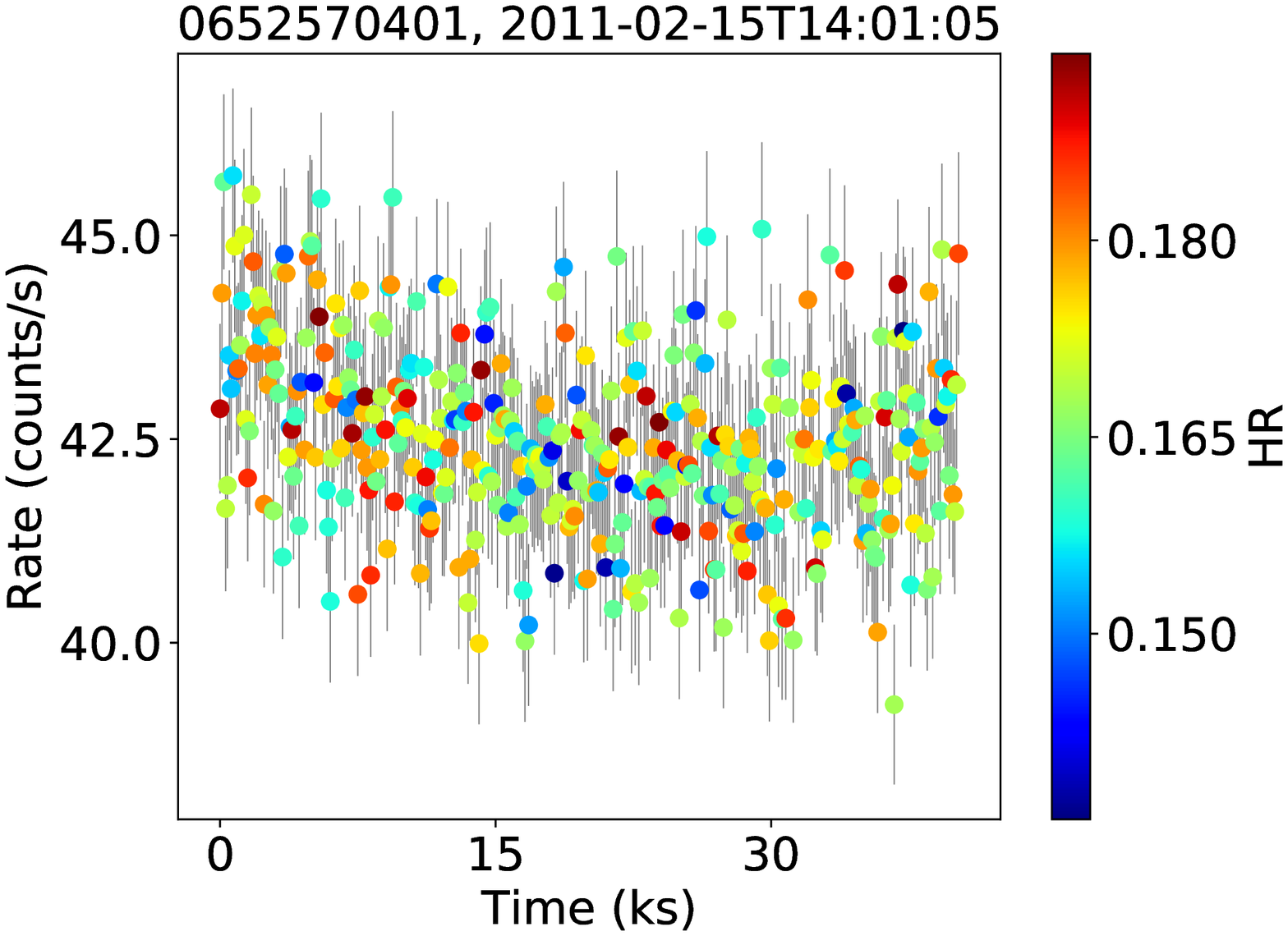}
	\end{minipage}
	\begin{minipage}{.3\textwidth} 
		\centering 
		\includegraphics[width=.99\linewidth]{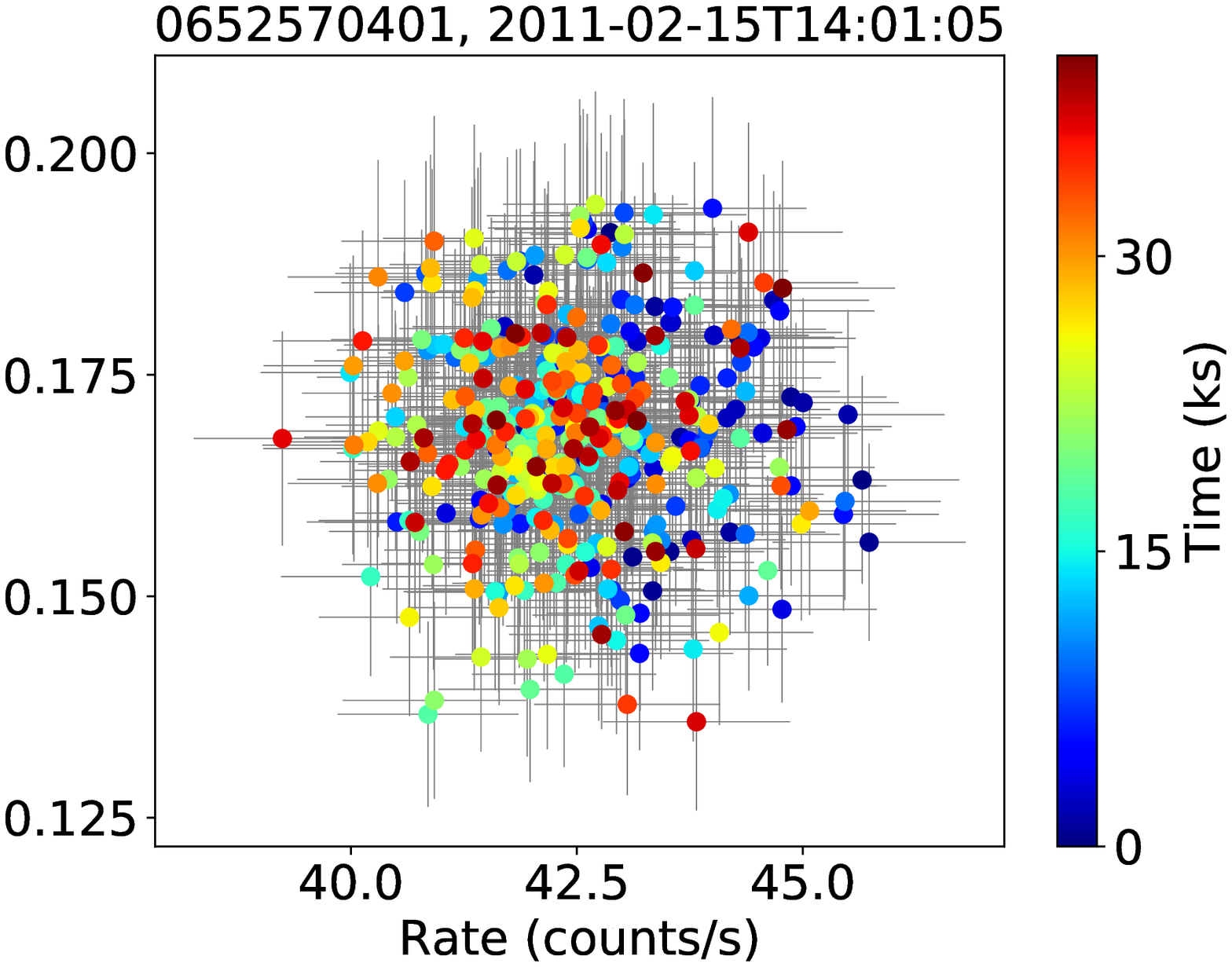}
	\end{minipage}
	\begin{minipage}{.3\textwidth} 
		\centering 
		\includegraphics[height=.99\linewidth, angle=-90]{PlotsPersonal/0652570401_PL.eps}
	\end{minipage}
\end{figure*}

\begin{figure*}\label{app:oj}
	\centering
	\caption{LCs, HR plots and spectral fits derived from observations of OJ 287.}
	\label{fig:LC_OJ}
	\begin{minipage}{.3\textwidth} 
		\centering 
		\includegraphics[width=.99\linewidth]{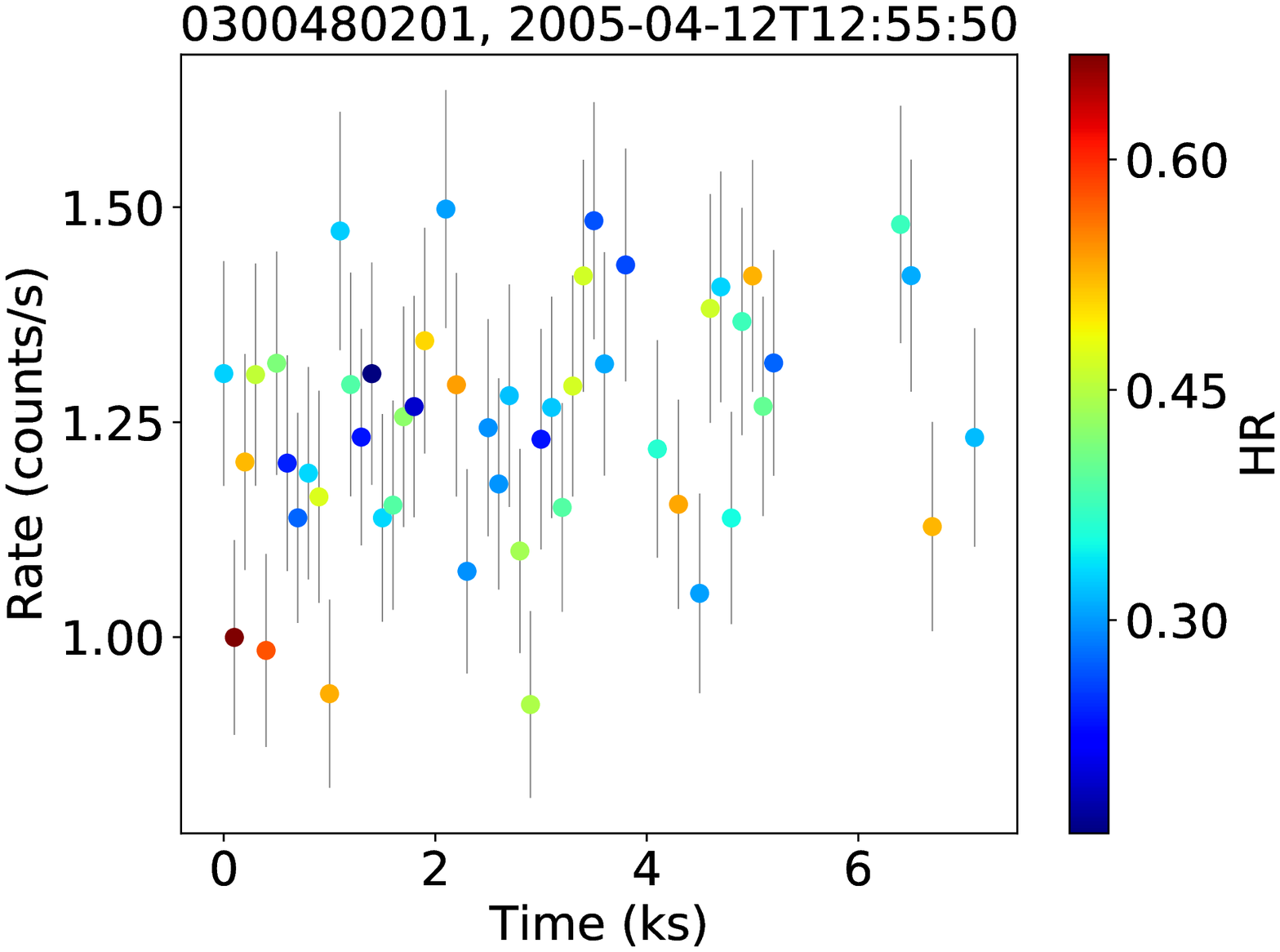}
	\end{minipage}
	\begin{minipage}{.3\textwidth} 
		\centering 
		\includegraphics[width=.99\linewidth]{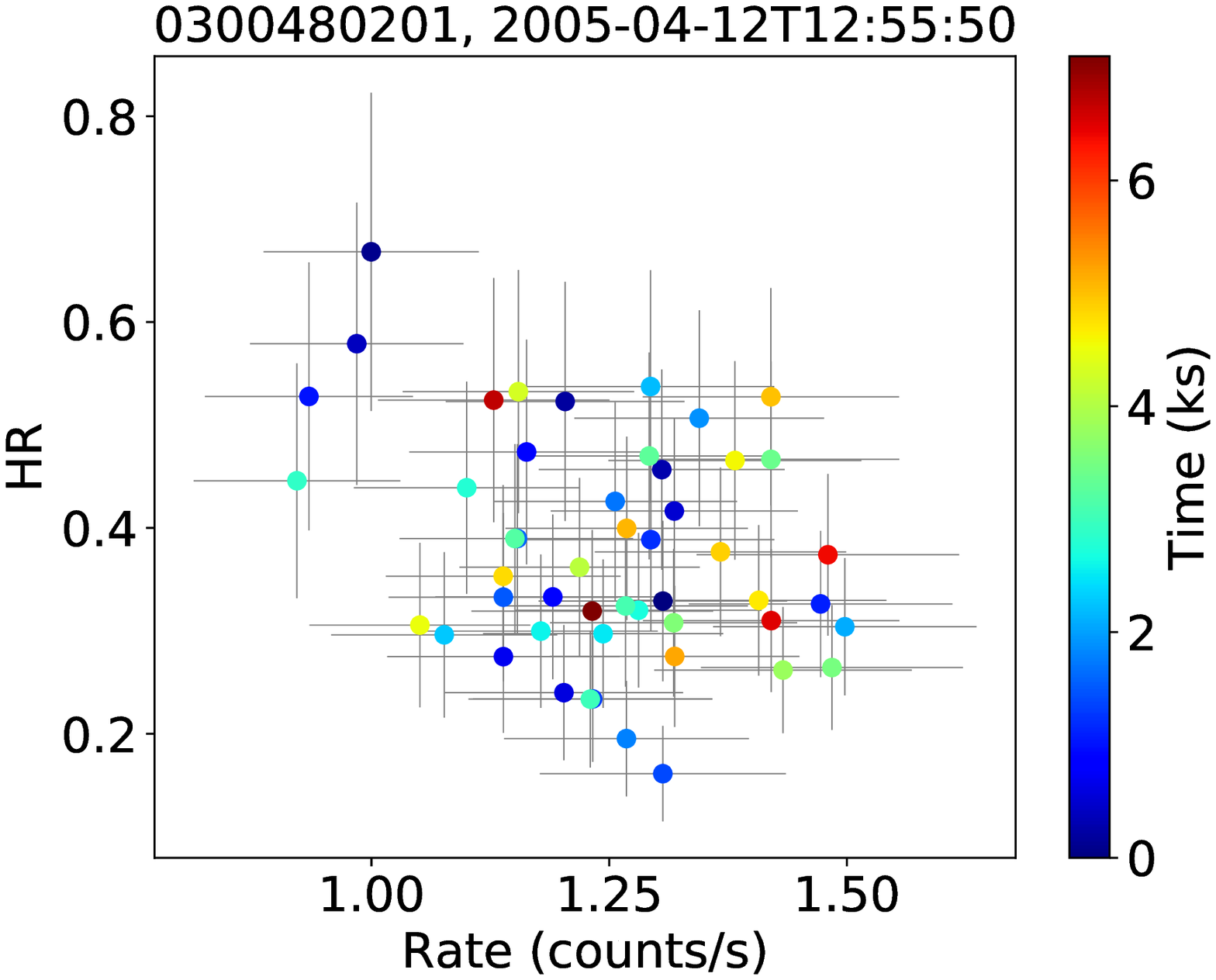}
	\end{minipage}
	\begin{minipage}{.3\textwidth} 
		\centering 
		\includegraphics[height=.99\linewidth, angle=-90]{PlotsPersonal/0300480201_PL.eps}
	\end{minipage}
	\begin{minipage}{.3\textwidth} 
		\centering 
		\includegraphics[width=.99\linewidth]{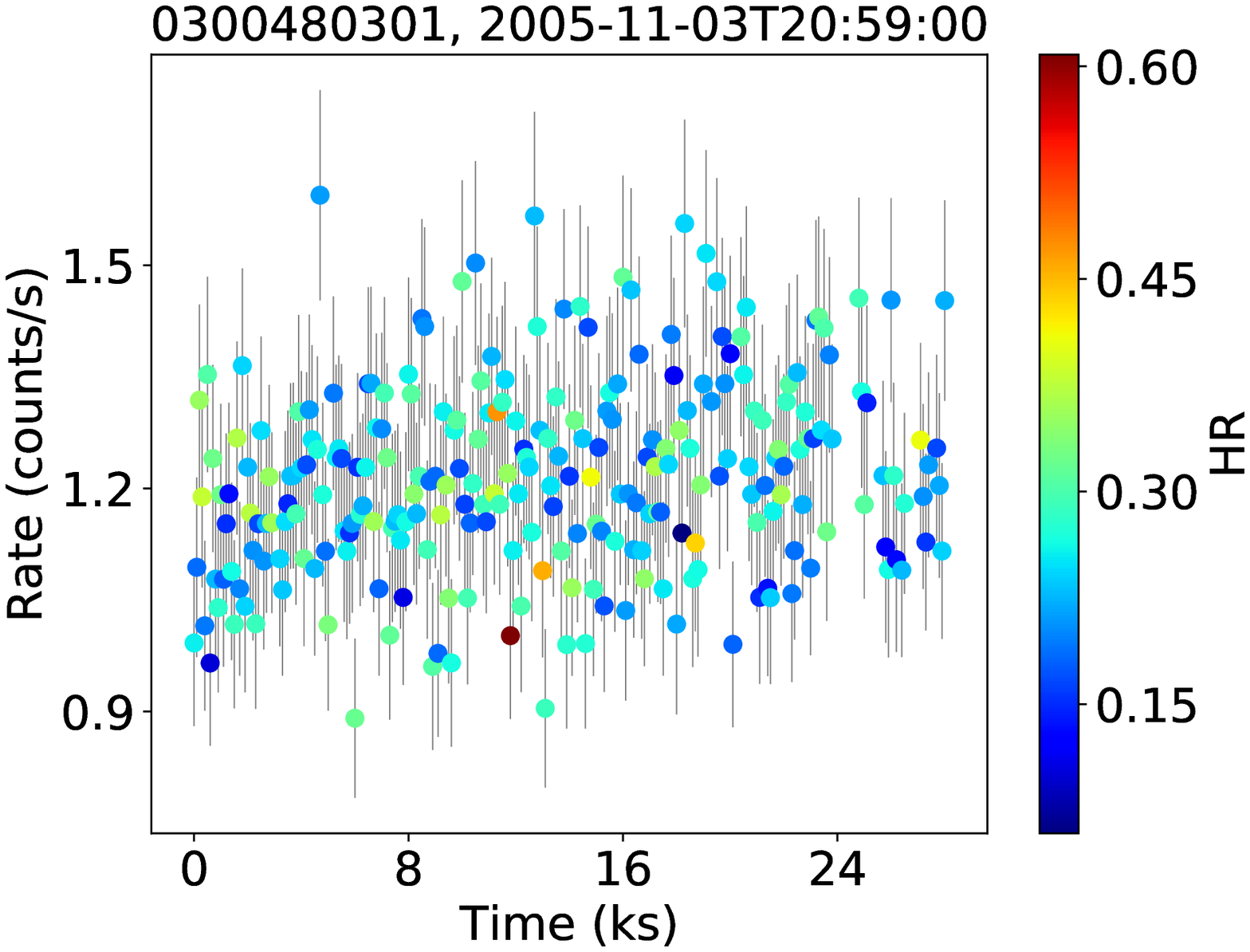}
	\end{minipage}
	\begin{minipage}{.3\textwidth} 
		\centering 
		\includegraphics[width=.99\linewidth]{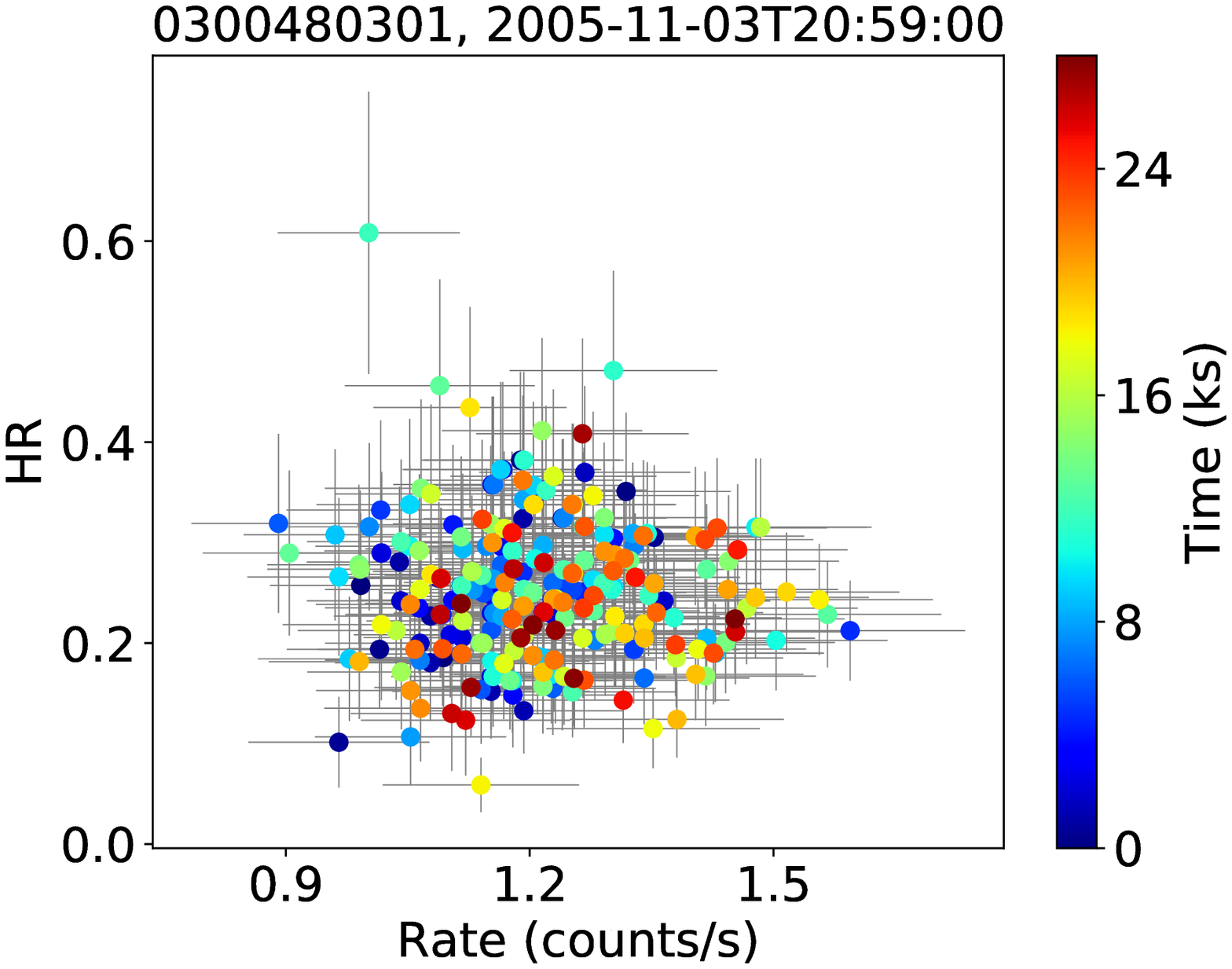}
	\end{minipage}
	\begin{minipage}{.3\textwidth} 
		\centering 
		\includegraphics[height=.99\linewidth, angle=-90]{PlotsPersonal/0300480301_BPL.eps}
	\end{minipage}
	\begin{minipage}{.3\textwidth} 
		\centering 
		\includegraphics[width=.99\linewidth]{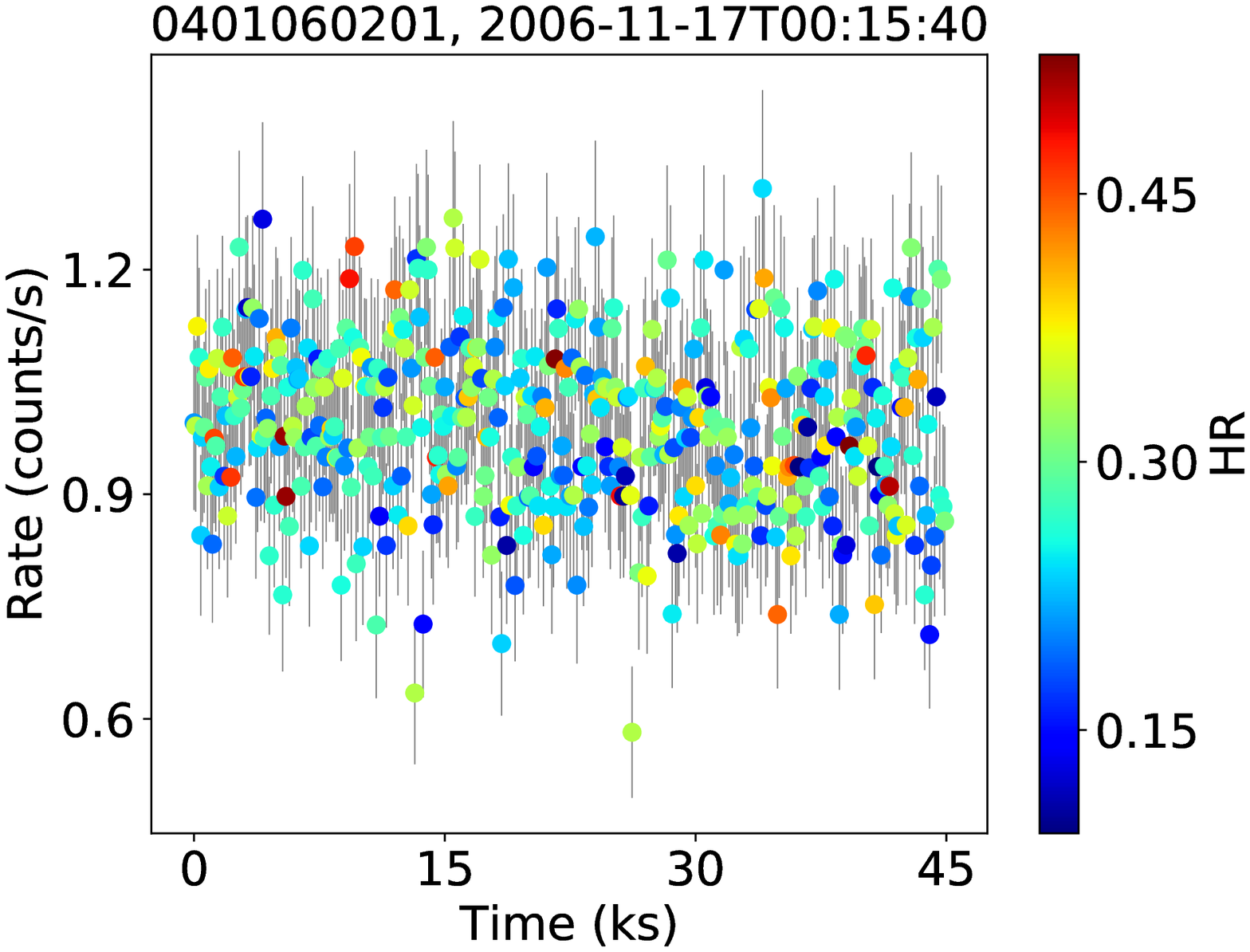}
	\end{minipage}
	\begin{minipage}{.3\textwidth} 
		\centering 
		\includegraphics[width=.99\linewidth]{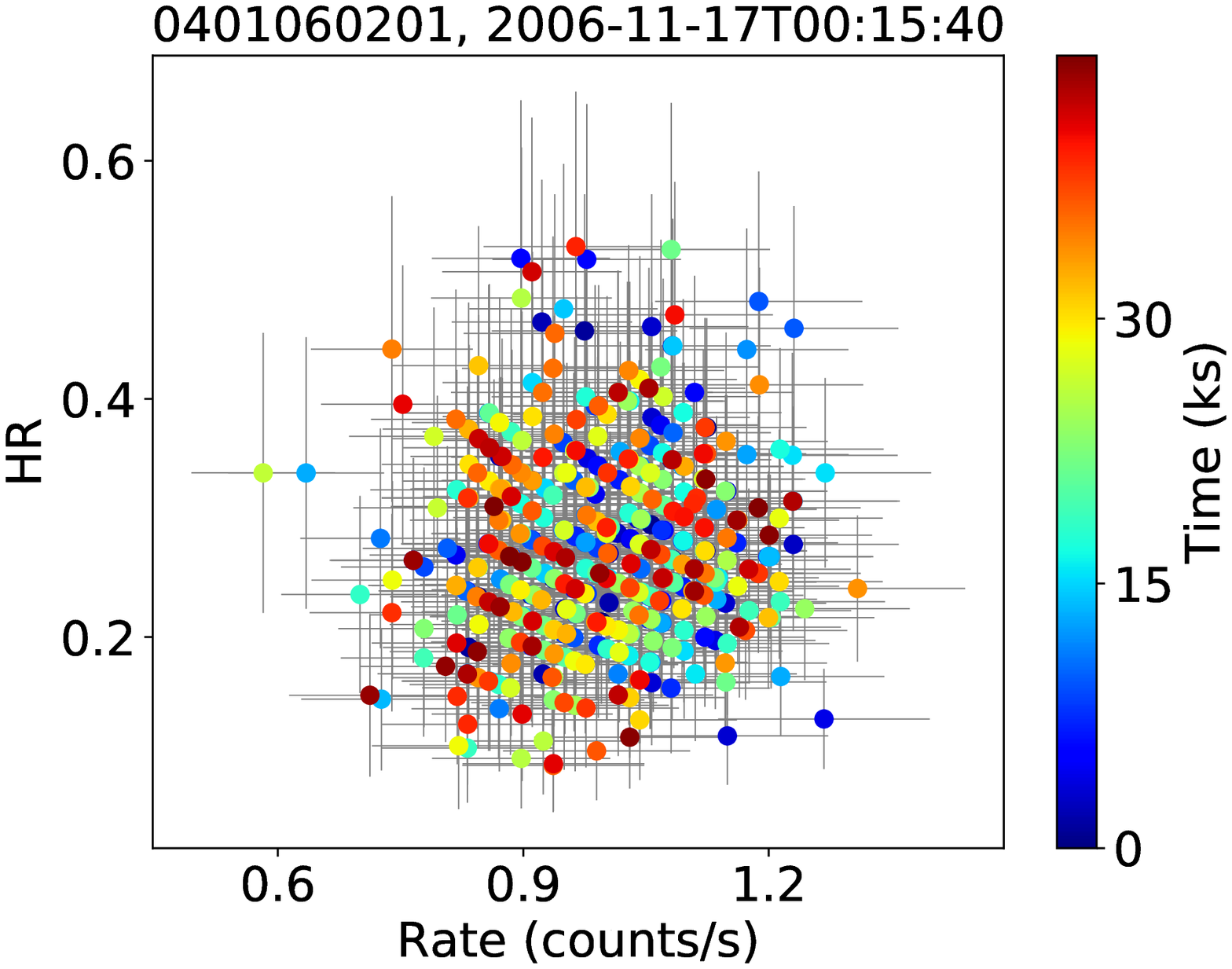}
	\end{minipage}
	\begin{minipage}{.3\textwidth} 
		\centering 
		\includegraphics[height=.99\linewidth, angle=-90]{PlotsPersonal/0401060201_BPL.eps}
	\end{minipage}
	\begin{minipage}{.3\textwidth} 
		\centering 
		\includegraphics[width=.99\linewidth]{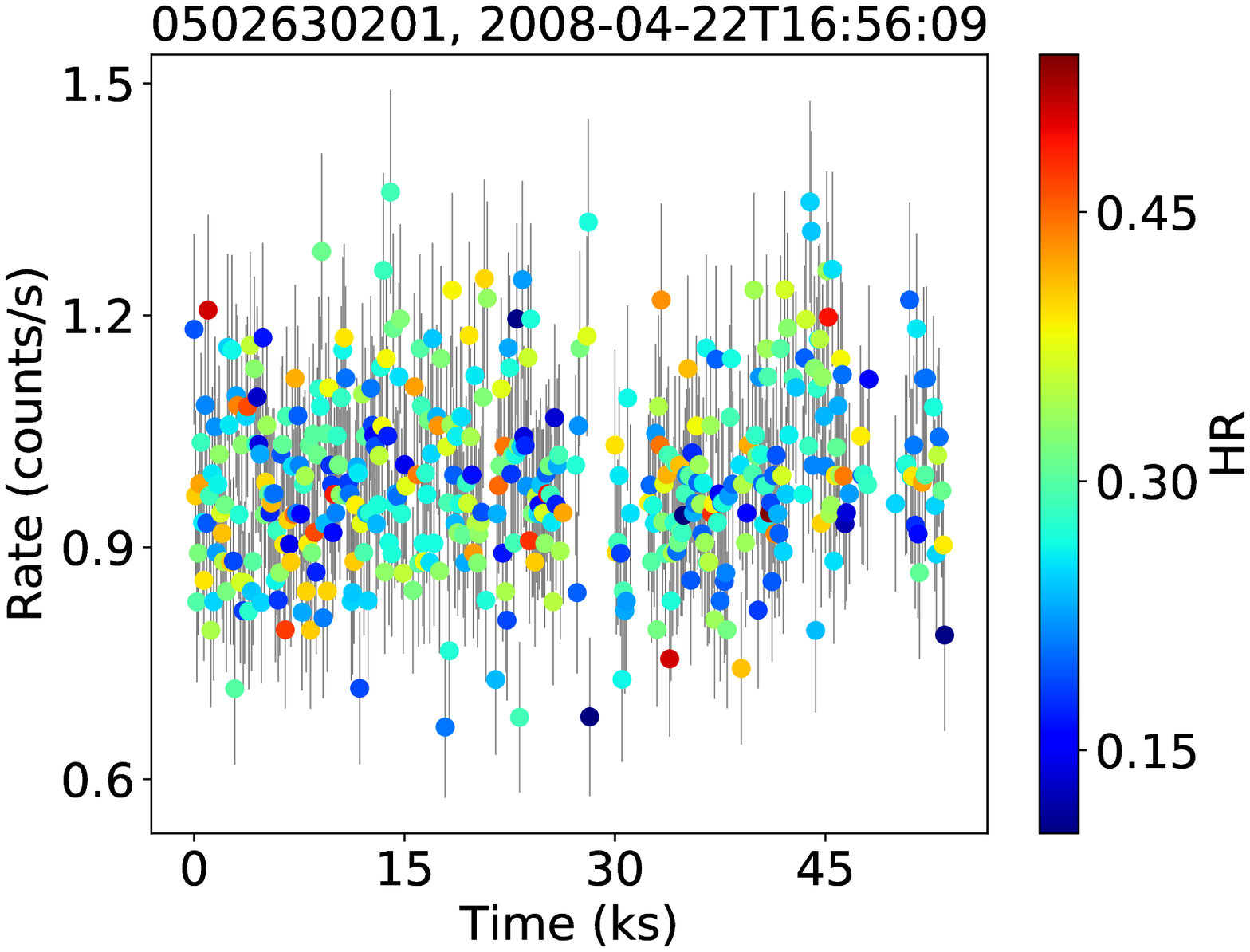}
	\end{minipage}
	\begin{minipage}{.3\textwidth} 
		\centering 
		\includegraphics[width=.99\linewidth]{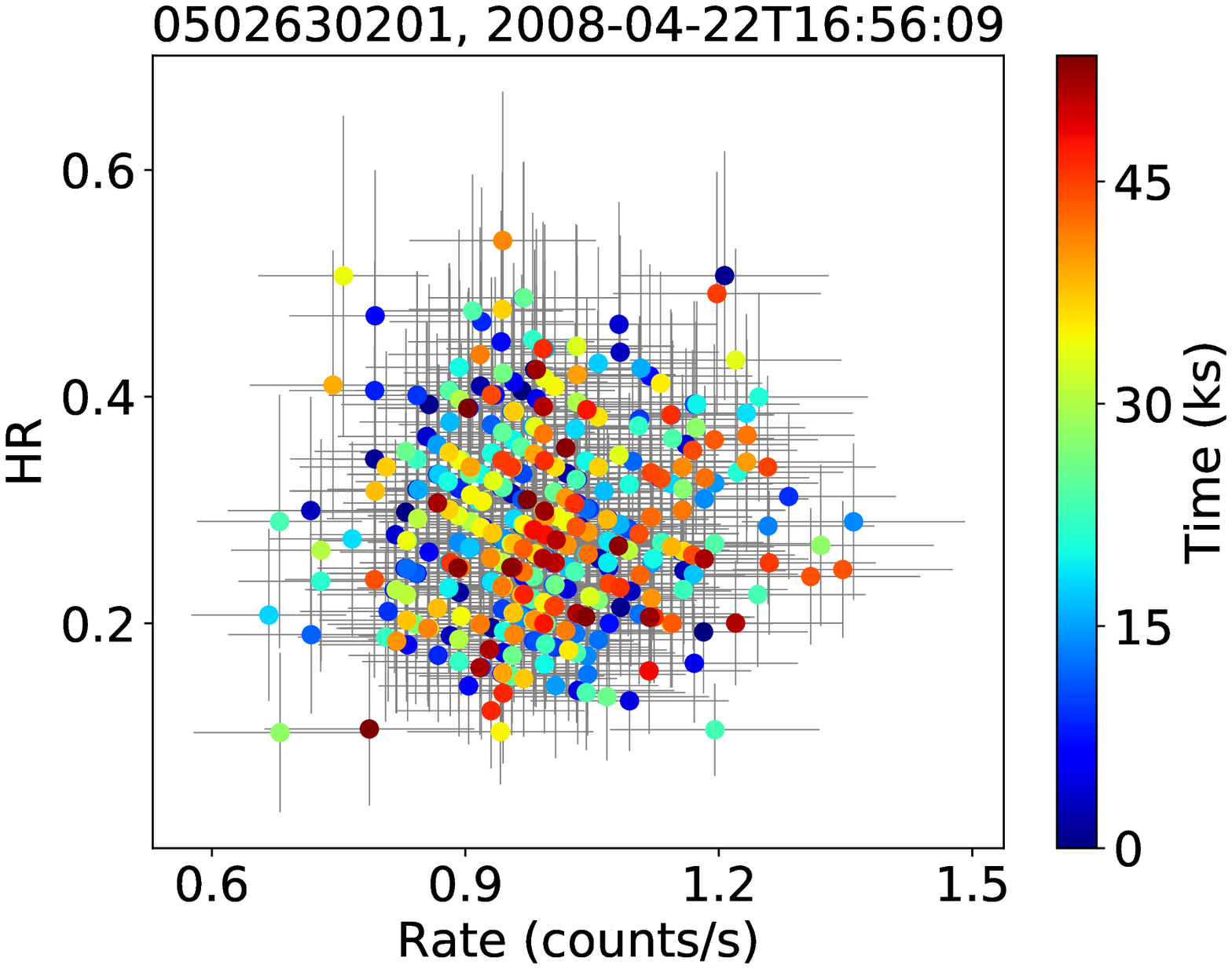}
	\end{minipage}
	\begin{minipage}{.3\textwidth} 
		\centering 
		\includegraphics[height=.99\linewidth, angle=-90]{PlotsPersonal/0502630201_PL.eps}
	\end{minipage}
	\begin{minipage}{.3\textwidth} 
		\centering 
		\includegraphics[width=.99\linewidth]{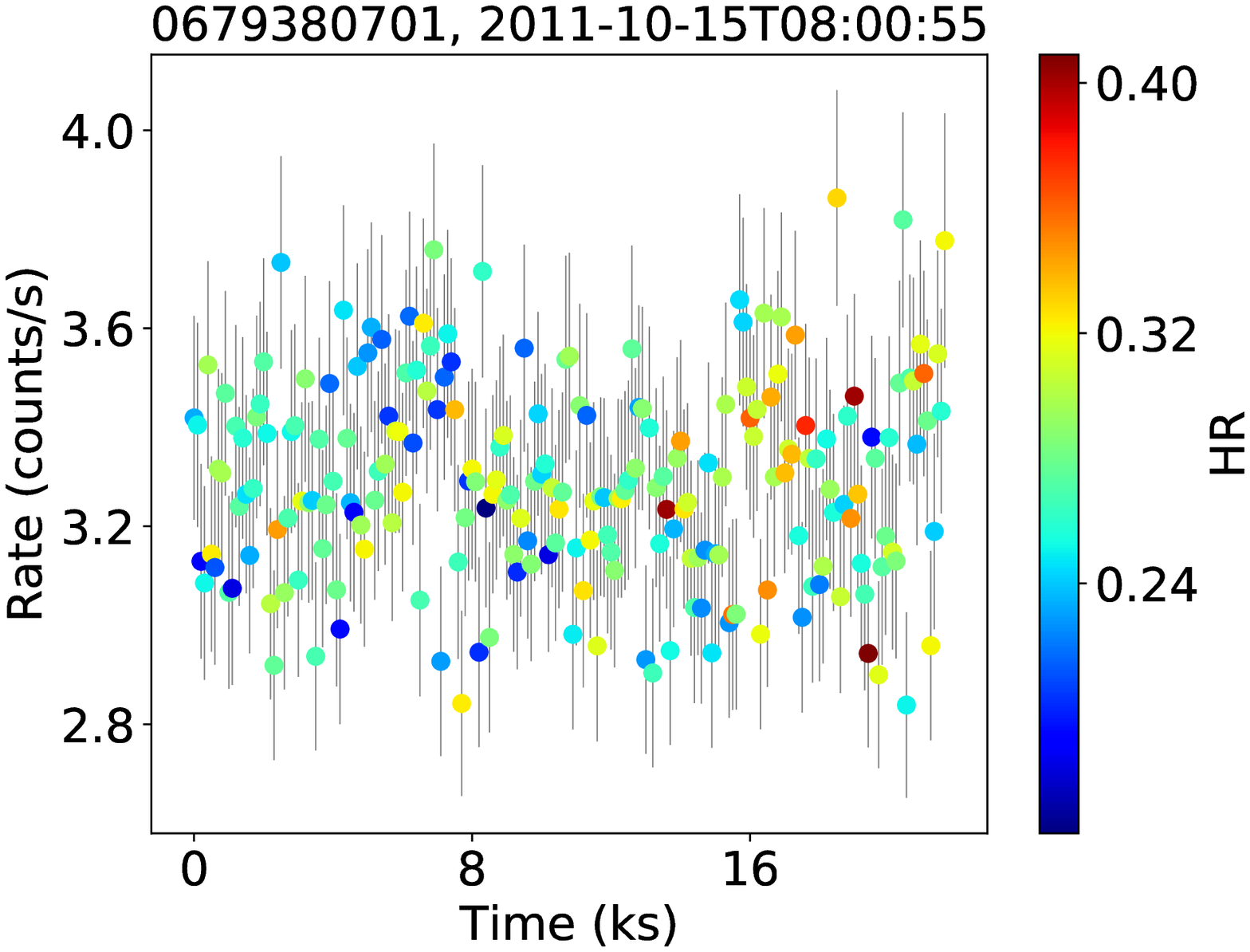}
	\end{minipage}
	\begin{minipage}{.3\textwidth} 
		\centering 
		\includegraphics[width=.99\linewidth]{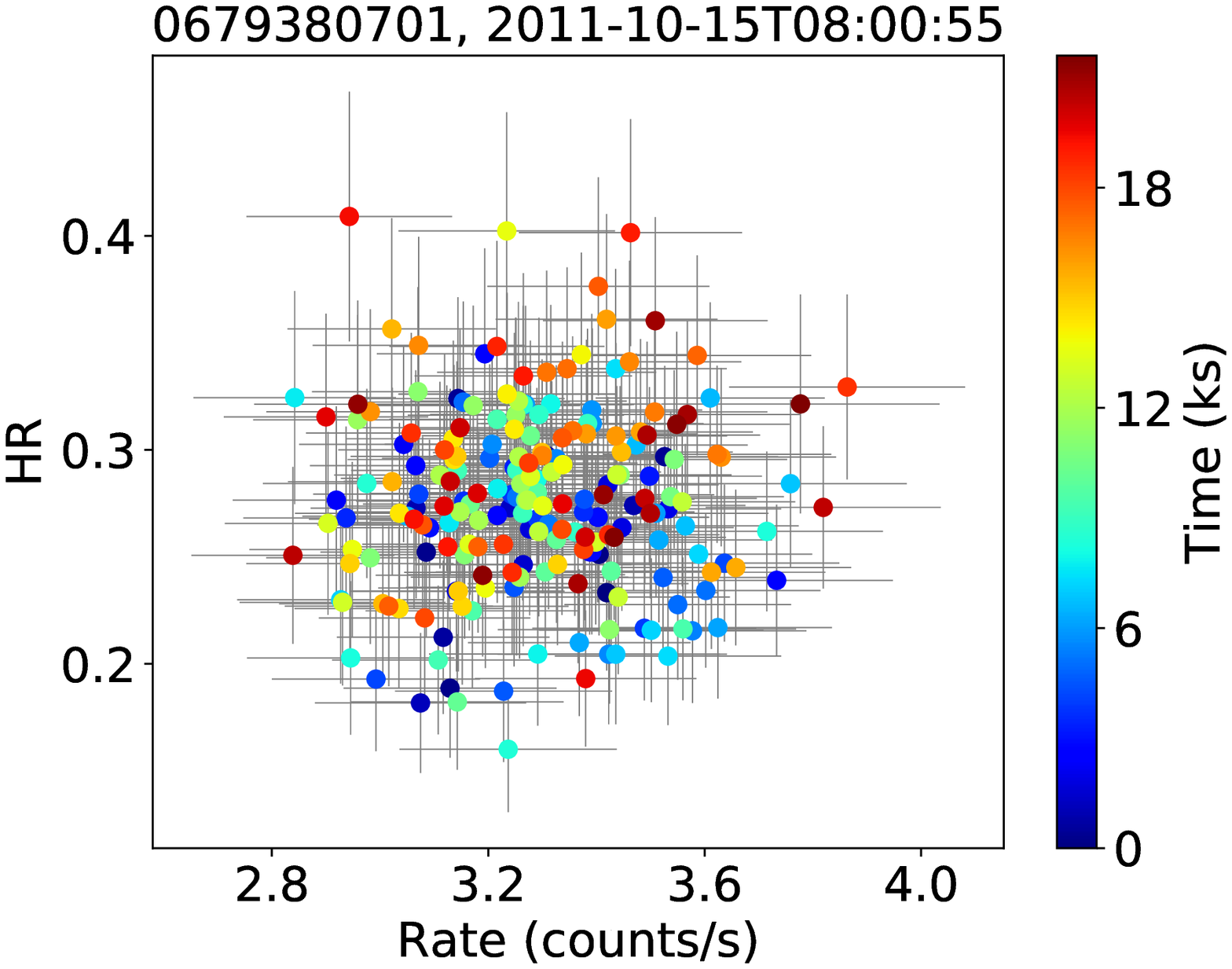}
	\end{minipage}
	\begin{minipage}{.3\textwidth} 
		\centering 
		\includegraphics[height=.99\linewidth, angle=-90]{PlotsPersonal/0679380701_BPL.eps}
	\end{minipage}
\end{figure*}

\begin{figure*}
	\centering
	\begin{minipage}{.3\textwidth} 
		\centering 
		\includegraphics[width=.99\linewidth]{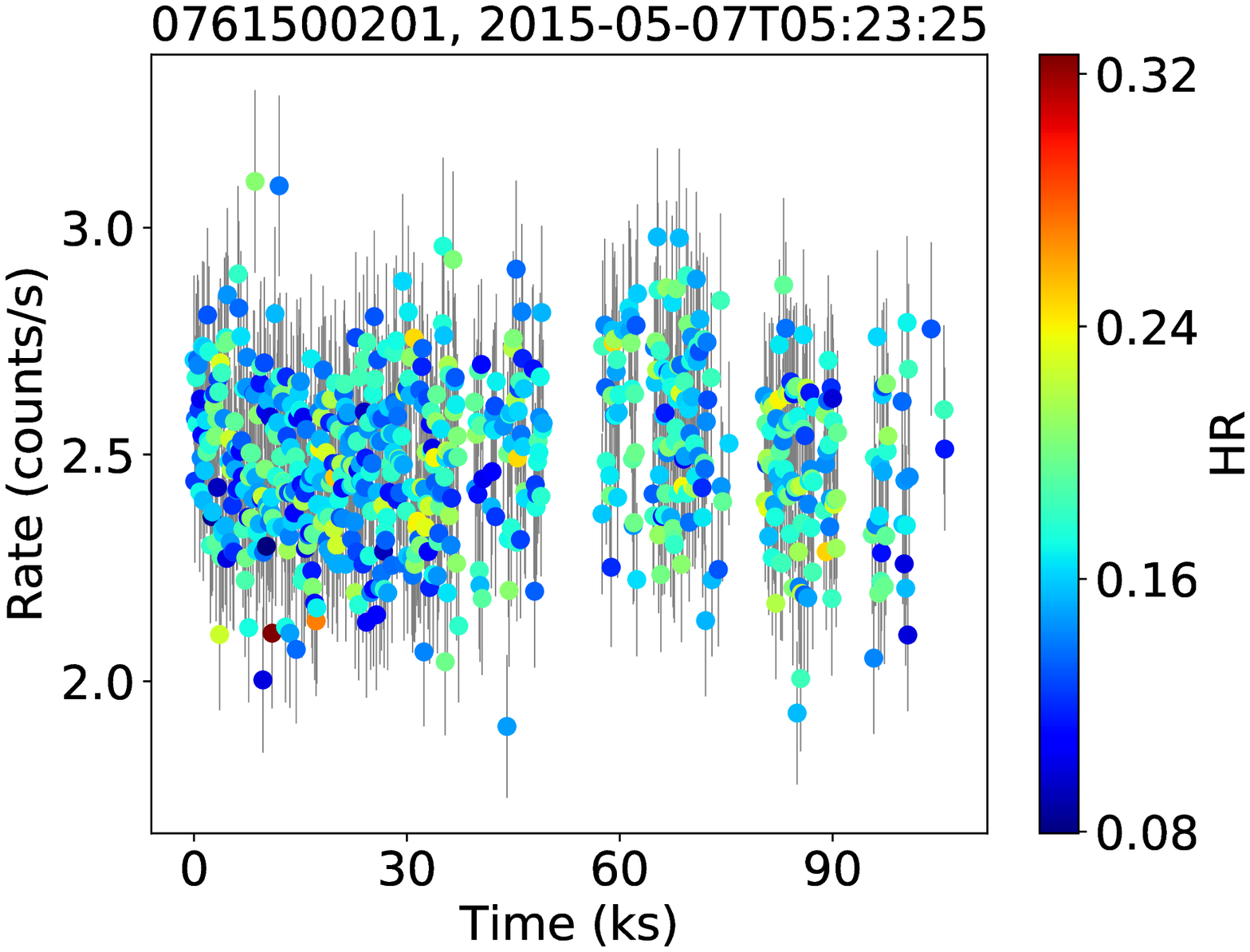}
	\end{minipage}
	\begin{minipage}{.3\textwidth} 
		\centering 
		\includegraphics[width=.99\linewidth]{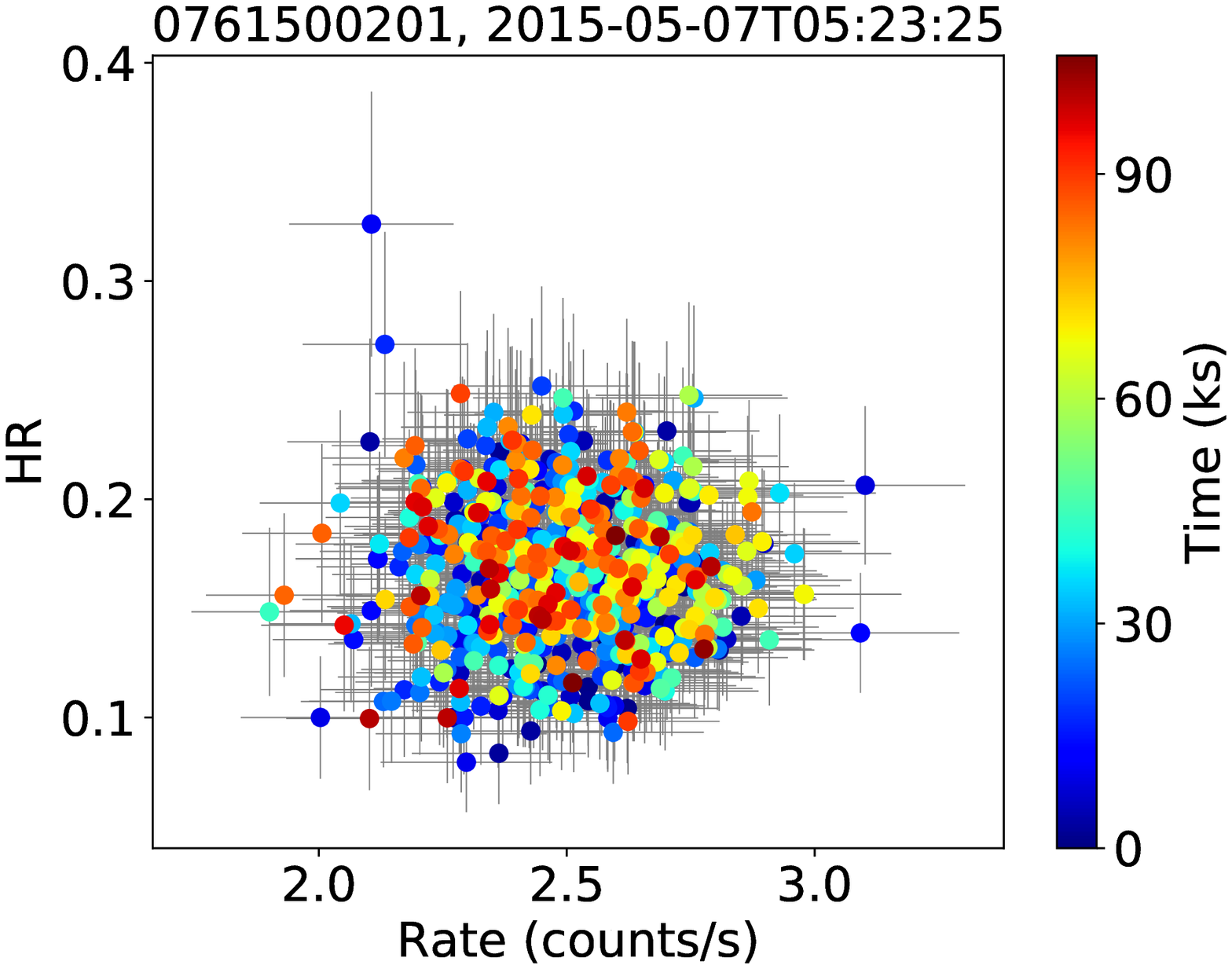}
	\end{minipage}
	\begin{minipage}{.3\textwidth} 
		\centering 
		\includegraphics[height=.99\linewidth, angle=-90]{PlotsPersonal/0761500201_EPLP.eps}
	\end{minipage}
	\begin{minipage}{.3\textwidth} 
		\centering 
		\includegraphics[width=.99\linewidth]{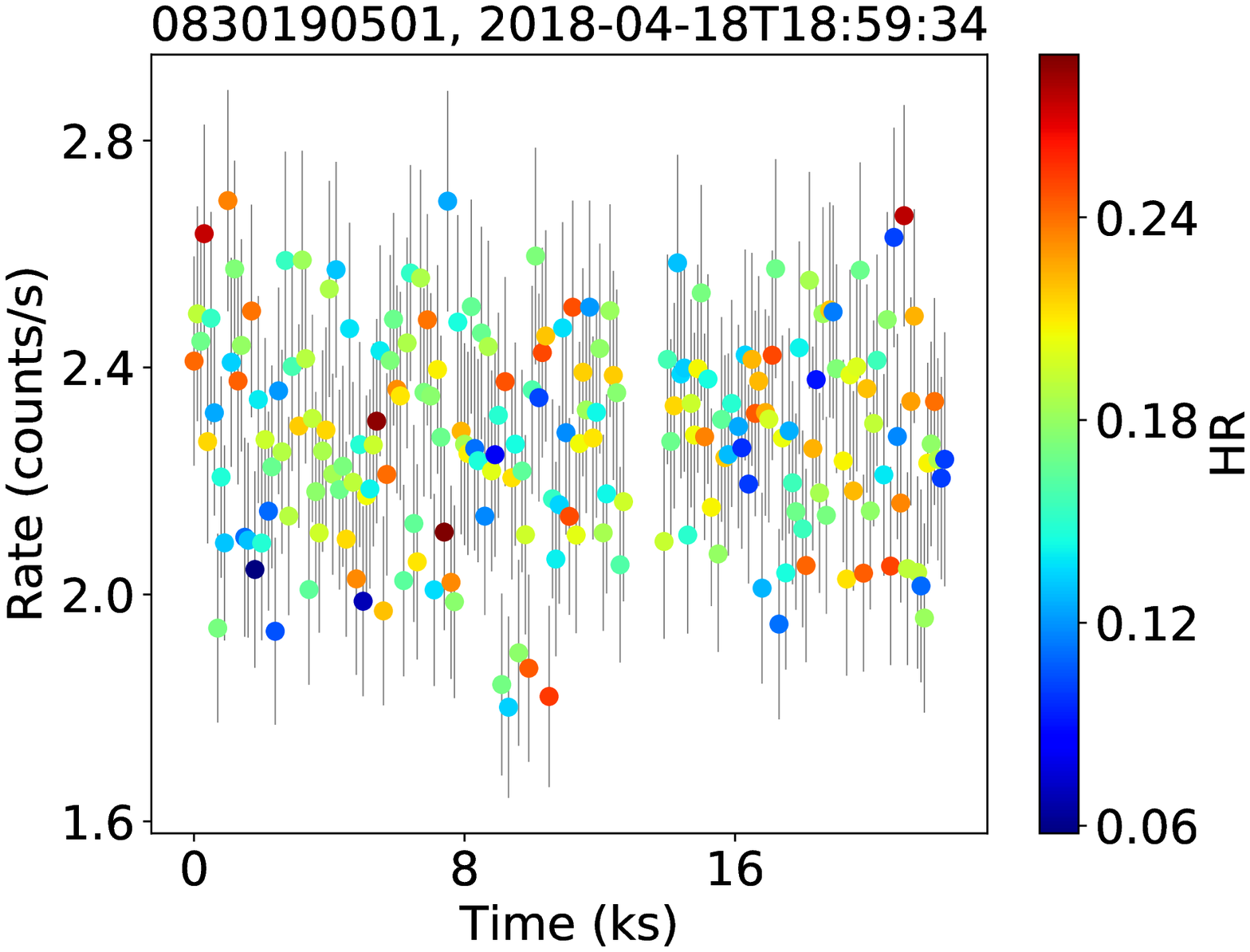}
	\end{minipage}
	\begin{minipage}{.3\textwidth} 
		\centering 
		\includegraphics[width=.99\linewidth]{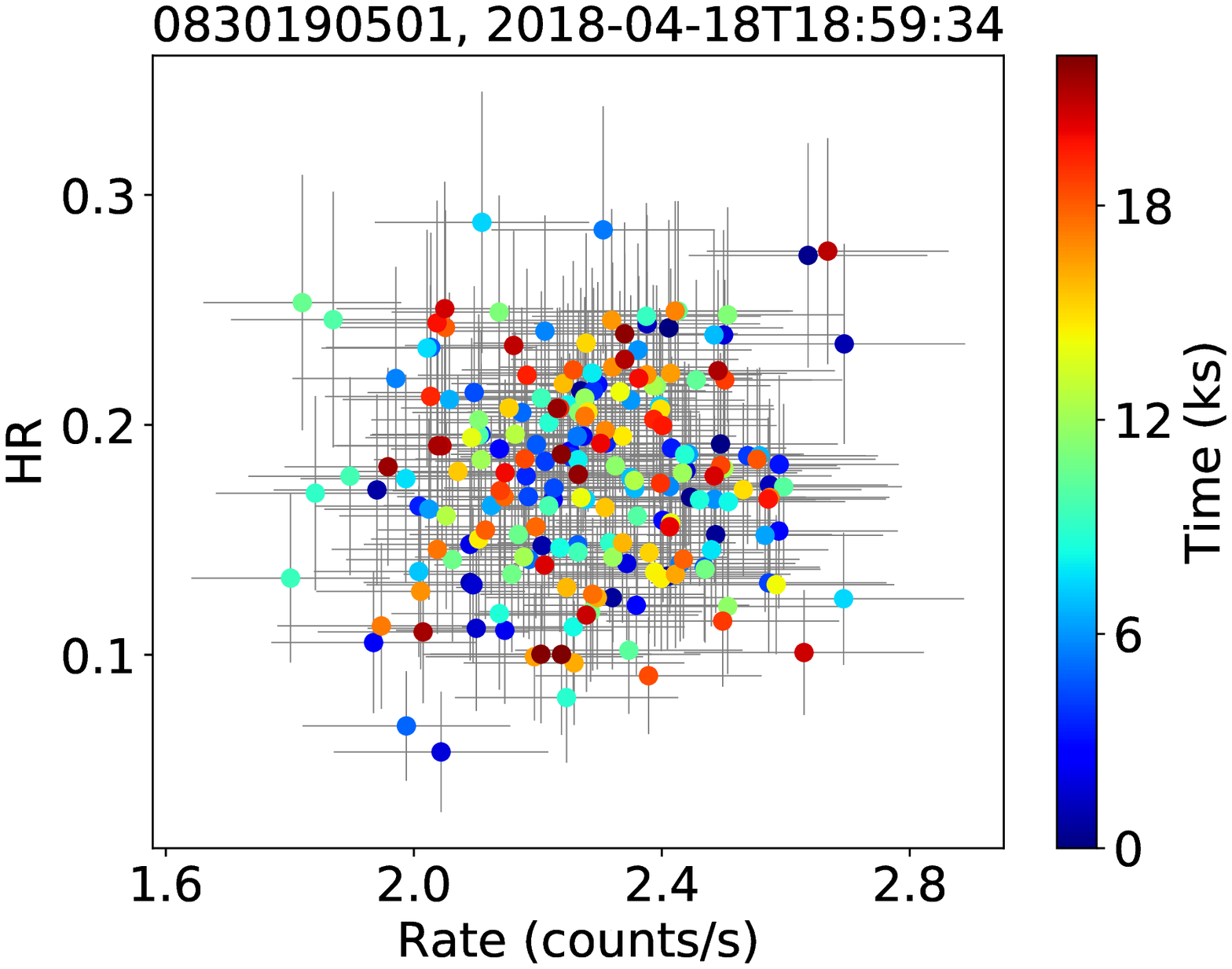}
	\end{minipage}
	\begin{minipage}{.3\textwidth} 
		\centering 
		\includegraphics[height=.99\linewidth, angle=-90]{PlotsPersonal/0830190501_PL.eps}
	\end{minipage}
	\begin{minipage}{.3\textwidth}
		\centering 
		\includegraphics[width=.99\linewidth]{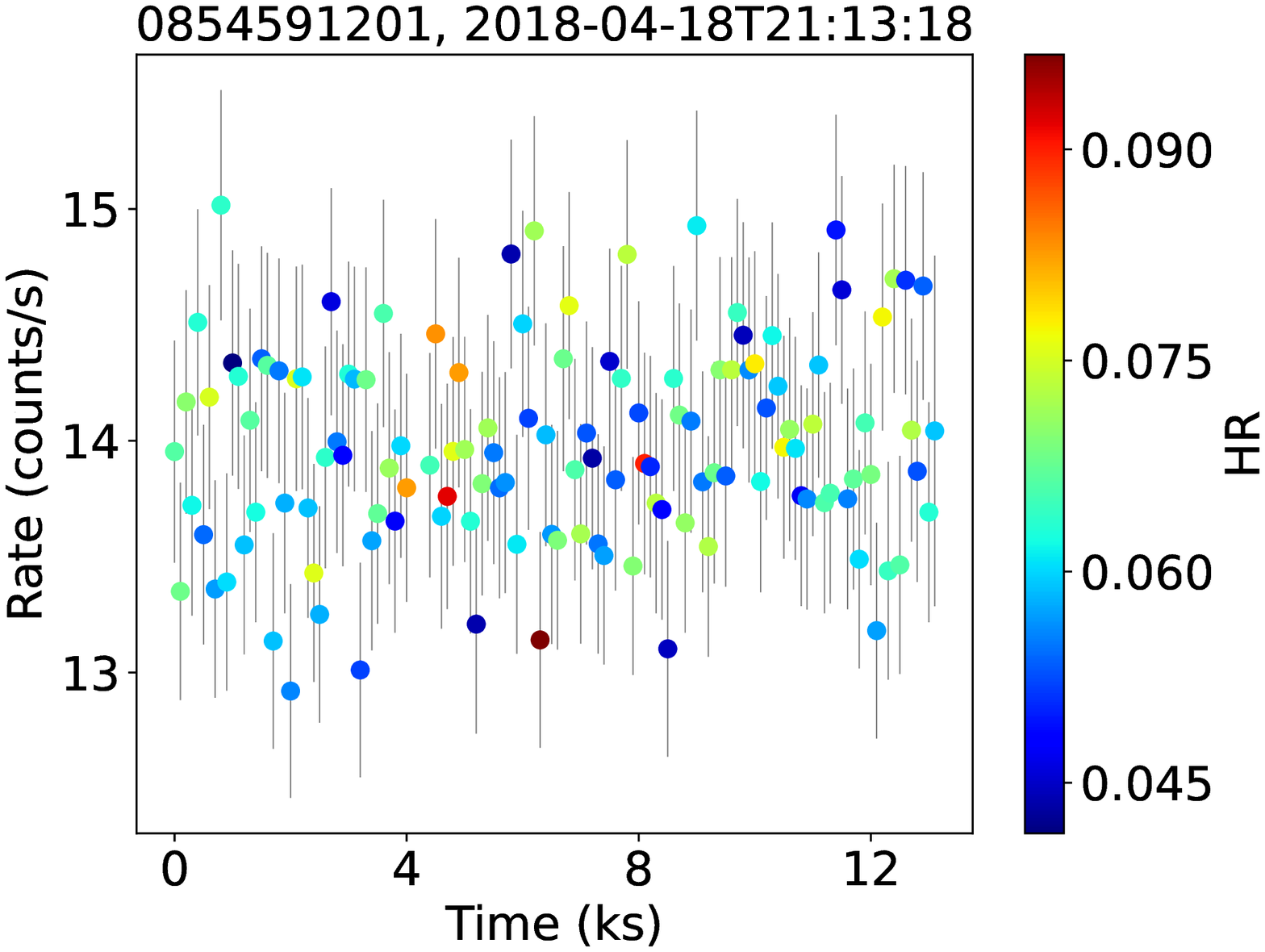}
	\end{minipage}
	\begin{minipage}{.3\textwidth}
		\centering 
		\includegraphics[width=.99\linewidth]{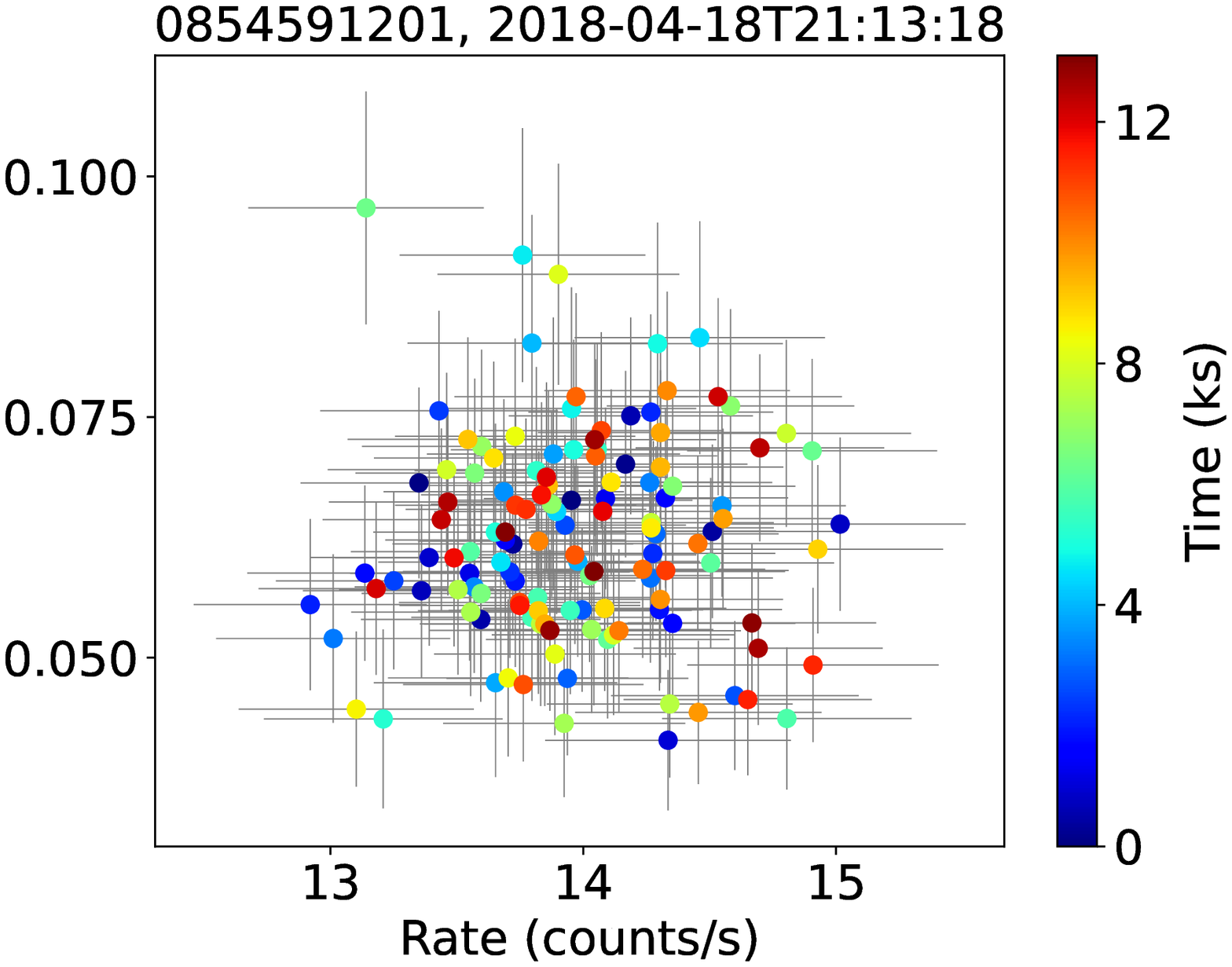}
	\end{minipage}
	\begin{minipage}{.3\textwidth}
		\centering 
		\includegraphics[height=.99\linewidth, angle=-90]{PlotsPersonal/0854591201_BBEPLP.eps}
	\end{minipage}
\end{figure*}

\begin{figure*}\label{app:s5}
	\centering
	\caption{LCs, HR plots and spectral fits derived from observations of S5 0716+714.}
	\label{fig:LC_S5}
	\begin{minipage}{.3\textwidth} 
		\centering 
		\includegraphics[width=.99\linewidth]{PlotsPersonal/0150495601+S003_lc.eps}
	\end{minipage}
	\begin{minipage}{.3\textwidth} 
		\centering 
		\includegraphics[width=.99\linewidth]{PlotsPersonal/0150495601+S003.eps}
	\end{minipage}
	\begin{minipage}{.3\textwidth} 
		\centering 
		\includegraphics[height=.99\linewidth, angle=-90]{PlotsPersonal/0150495601_EPLP.eps}
	\end{minipage}
	\begin{minipage}{.3\textwidth} 
		\centering 
		\includegraphics[width=.99\linewidth]{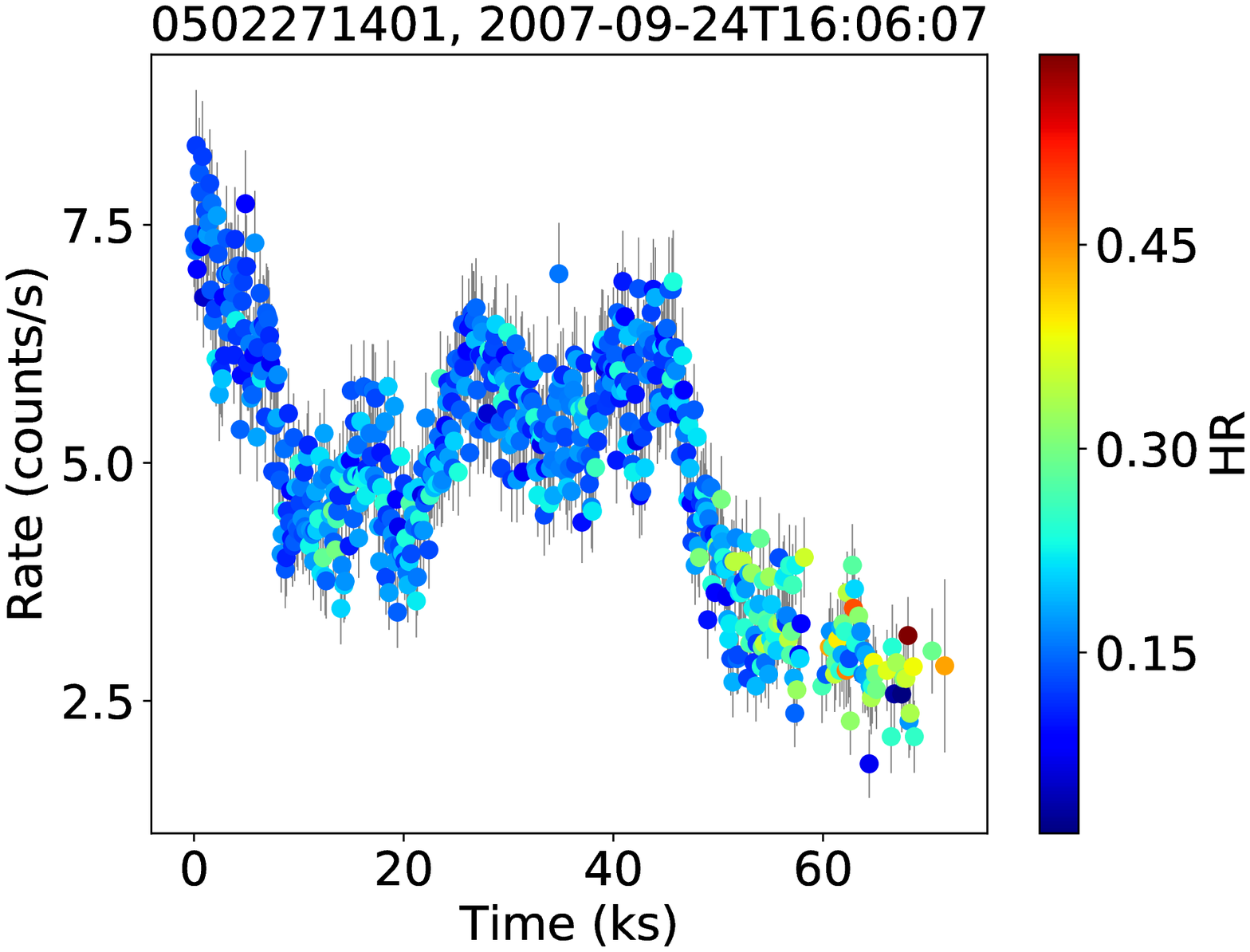}
	\end{minipage}
	\begin{minipage}{.3\textwidth} 
		\centering 
		\includegraphics[width=.99\linewidth]{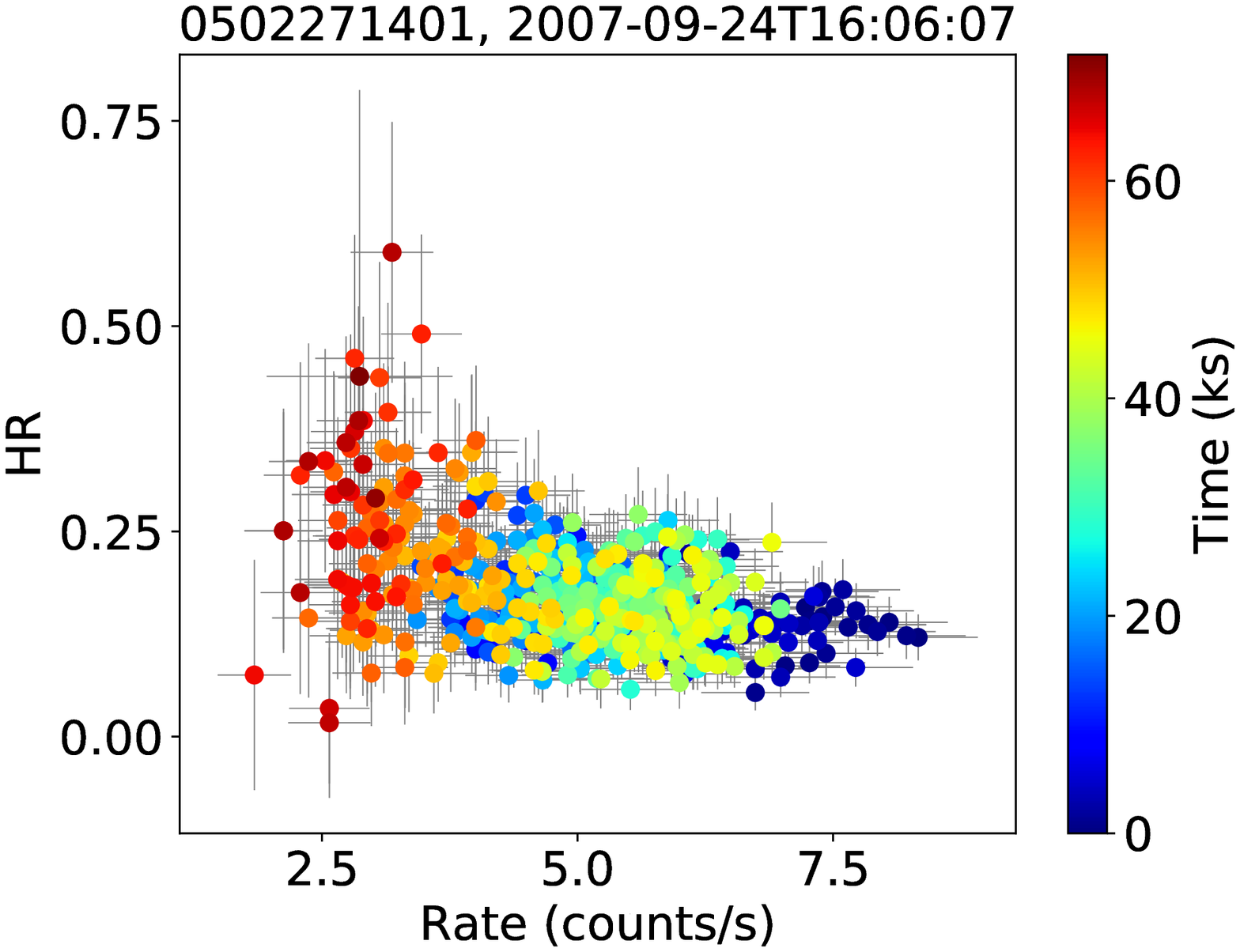}
	\end{minipage}
	\begin{minipage}{.3\textwidth} 
		\centering 
		\includegraphics[height=.99\linewidth, angle=-90]{PlotsPersonal/0502271401_EPLP.eps}
	\end{minipage}
\end{figure*}

\begin{figure*}\label{dfp:rbs}
	\centering
	\caption{DFPs of the observations of RBS 2070. Arbitrary unit is the rms-normalized power ($\mathrm{(rms/mean)^2\ Hz^{-1}}$).}
	\label{fig:DFT_RBS}
	\begin{minipage}{.43\textwidth} 
		\centering 
		\includegraphics[width=.9\linewidth]{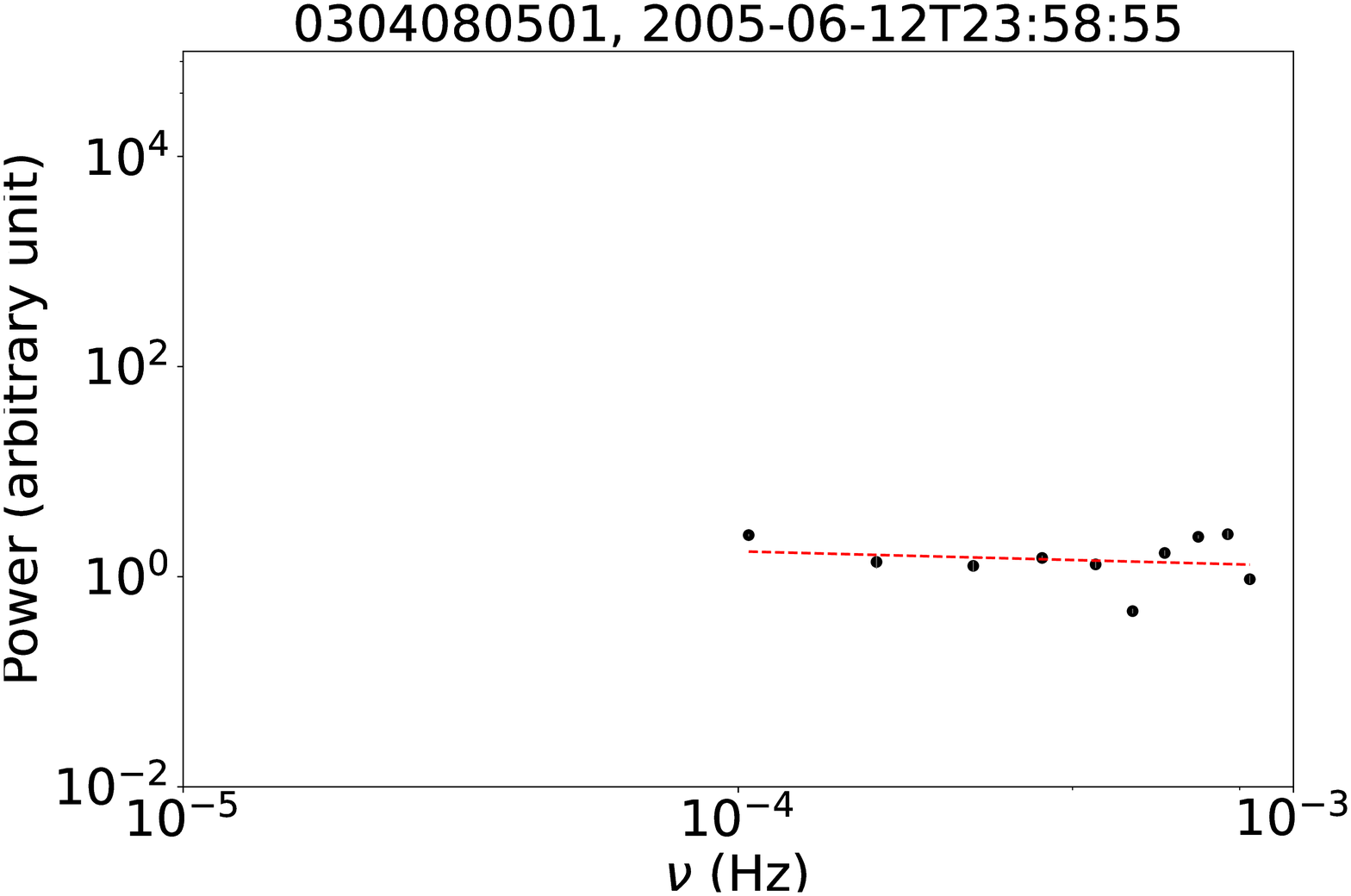}
	\end{minipage}
	\begin{minipage}{.43\textwidth} 
		\centering 
		\includegraphics[width=.9\linewidth]{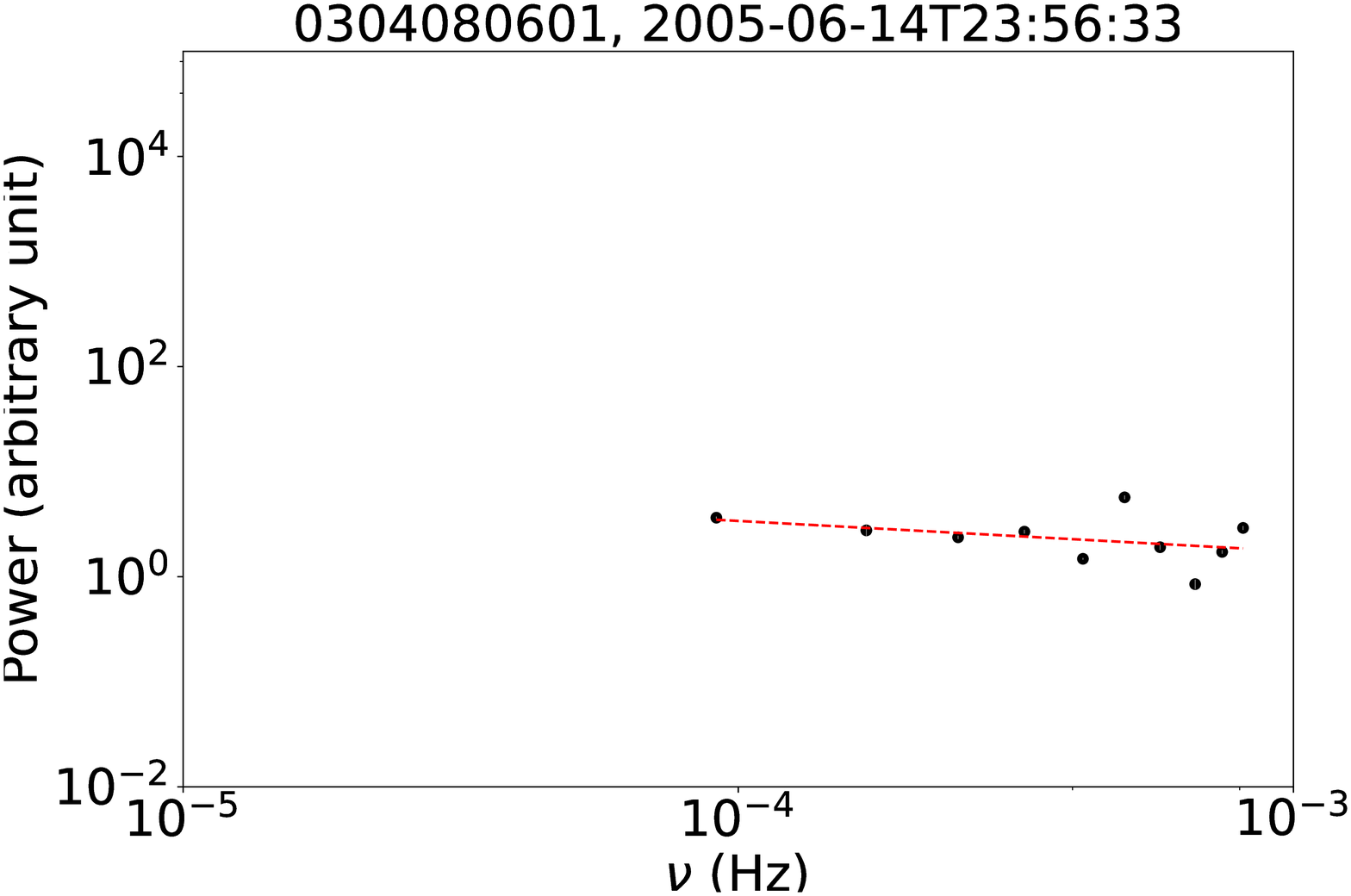}
	\end{minipage}
	\begin{minipage}{.43\textwidth} 
		\centering 
		\includegraphics[width=.9\linewidth]{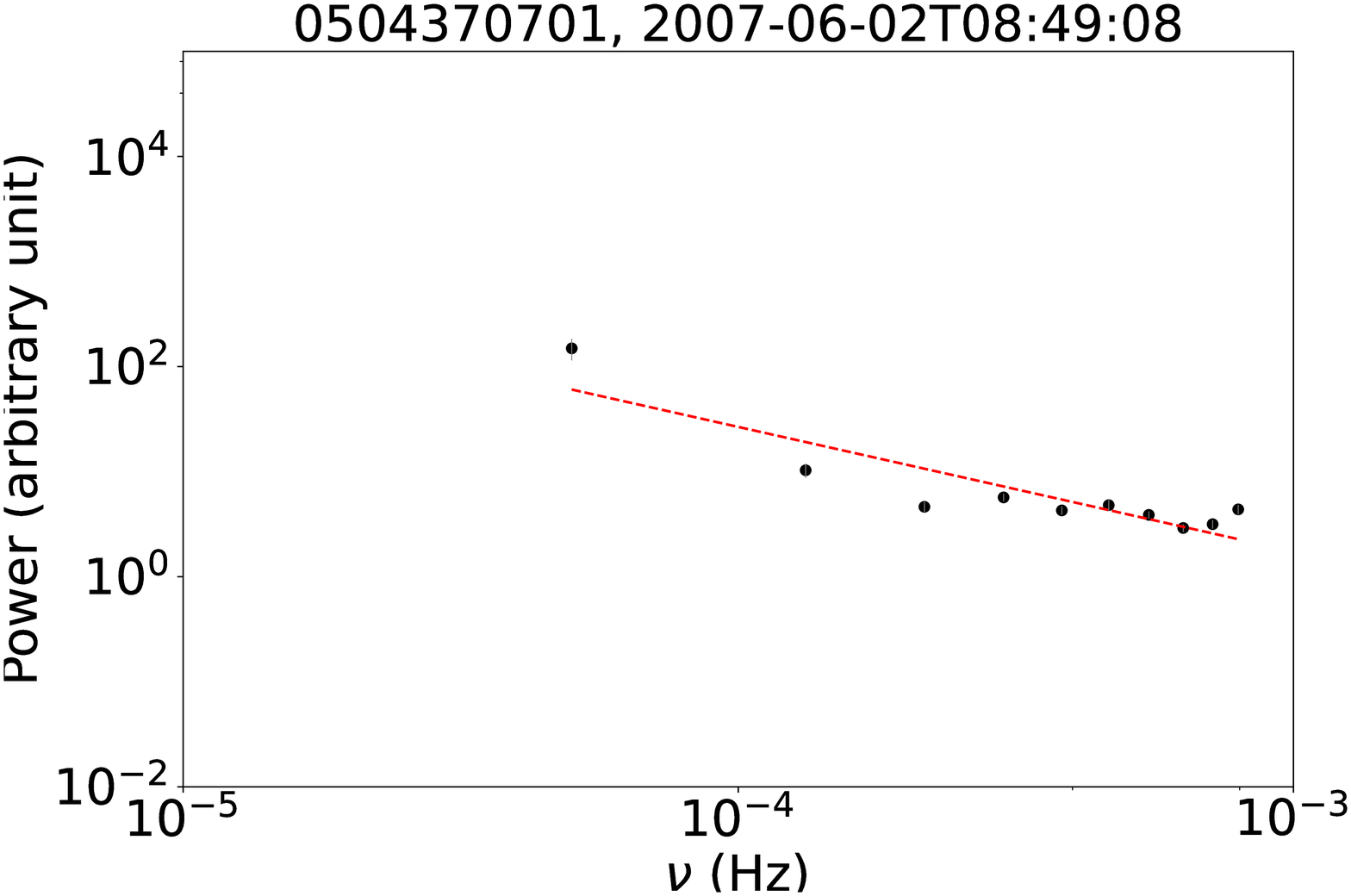}
	\end{minipage}
	\begin{minipage}{.43\textwidth} 
		\centering 
		\includegraphics[width=.9\linewidth]{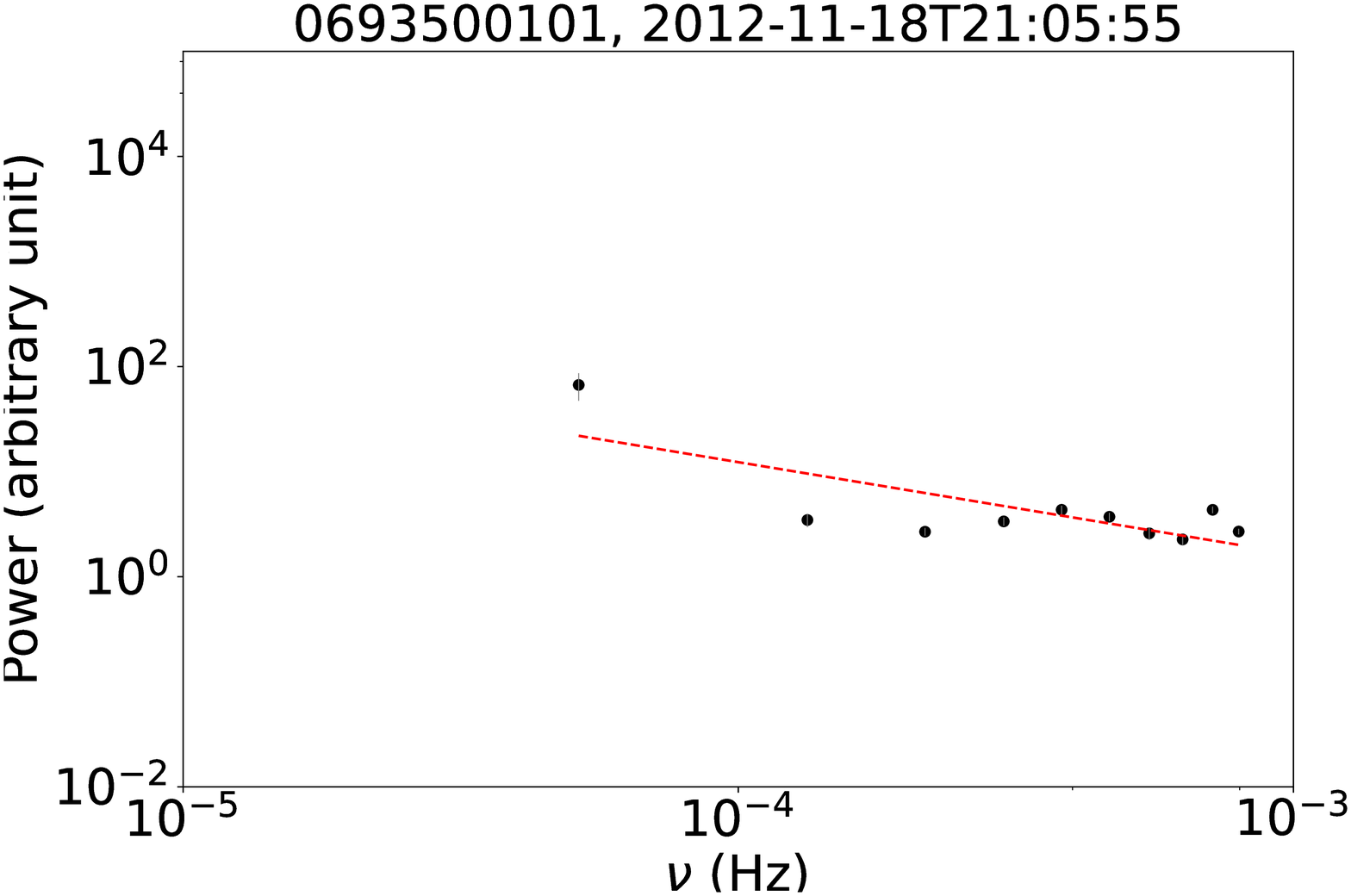}
	\end{minipage}
	\begin{minipage}{.43\textwidth} 
		\centering 
		\includegraphics[width=.9\linewidth]{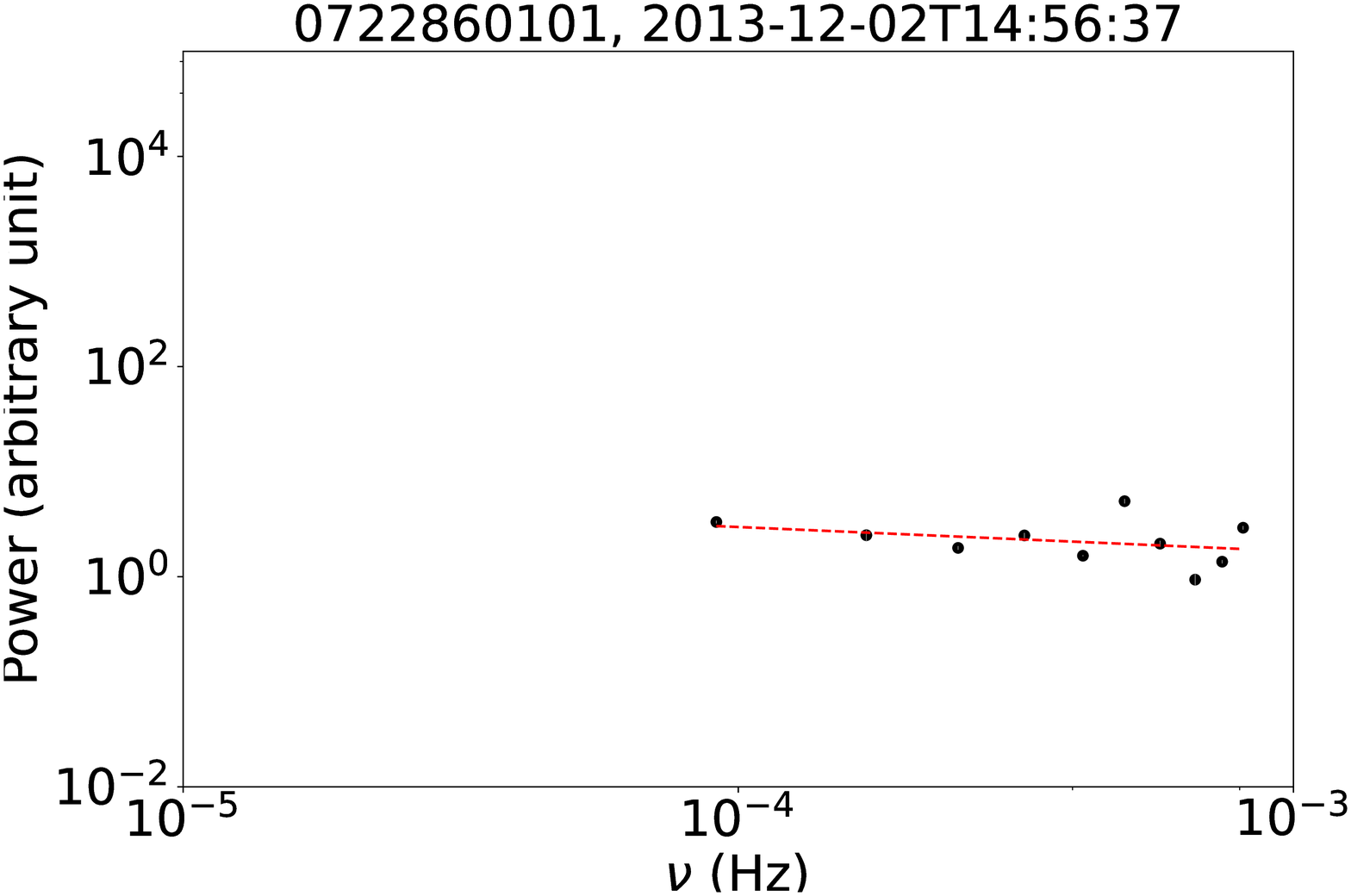}
	\end{minipage}
	\begin{minipage}{.43\textwidth} 
		\centering 
		\includegraphics[width=.9\linewidth]{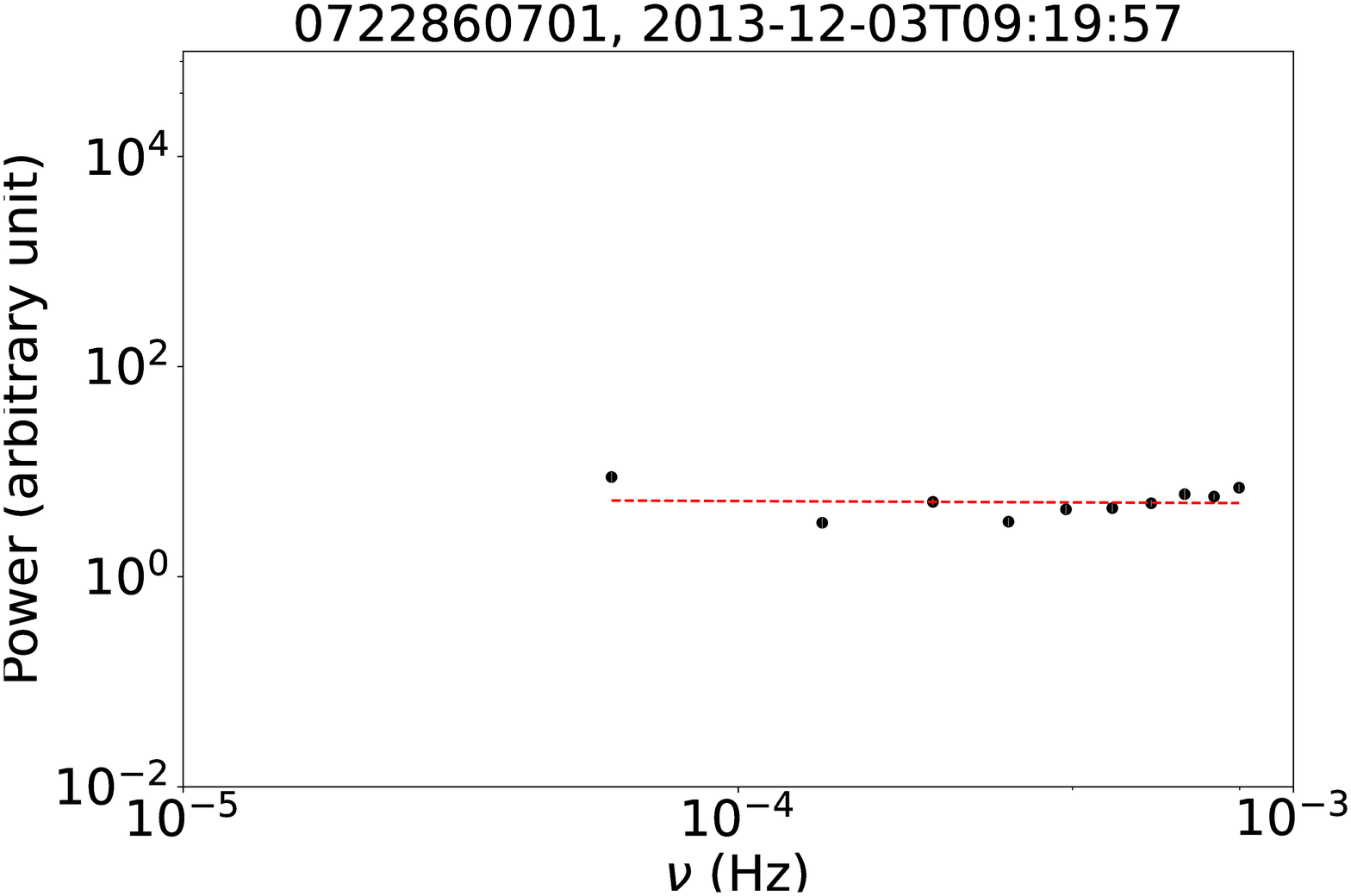}
	\end{minipage}
	\begin{minipage}{.43\textwidth} 
		\centering 
		\includegraphics[width=.9\linewidth]{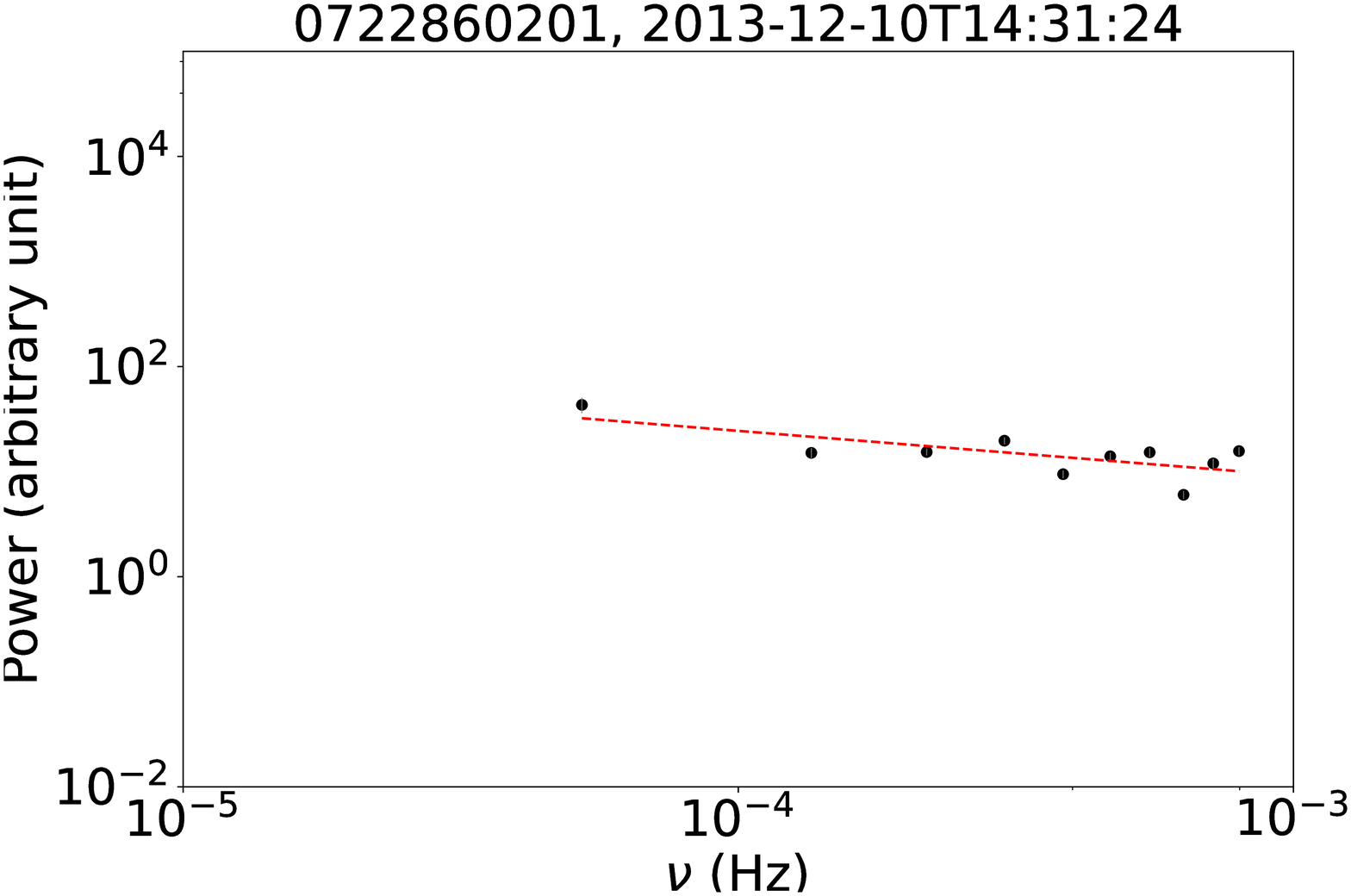}
	\end{minipage}
	\begin{minipage}{.43\textwidth} 
		\centering 
		\includegraphics[width=.9\linewidth]{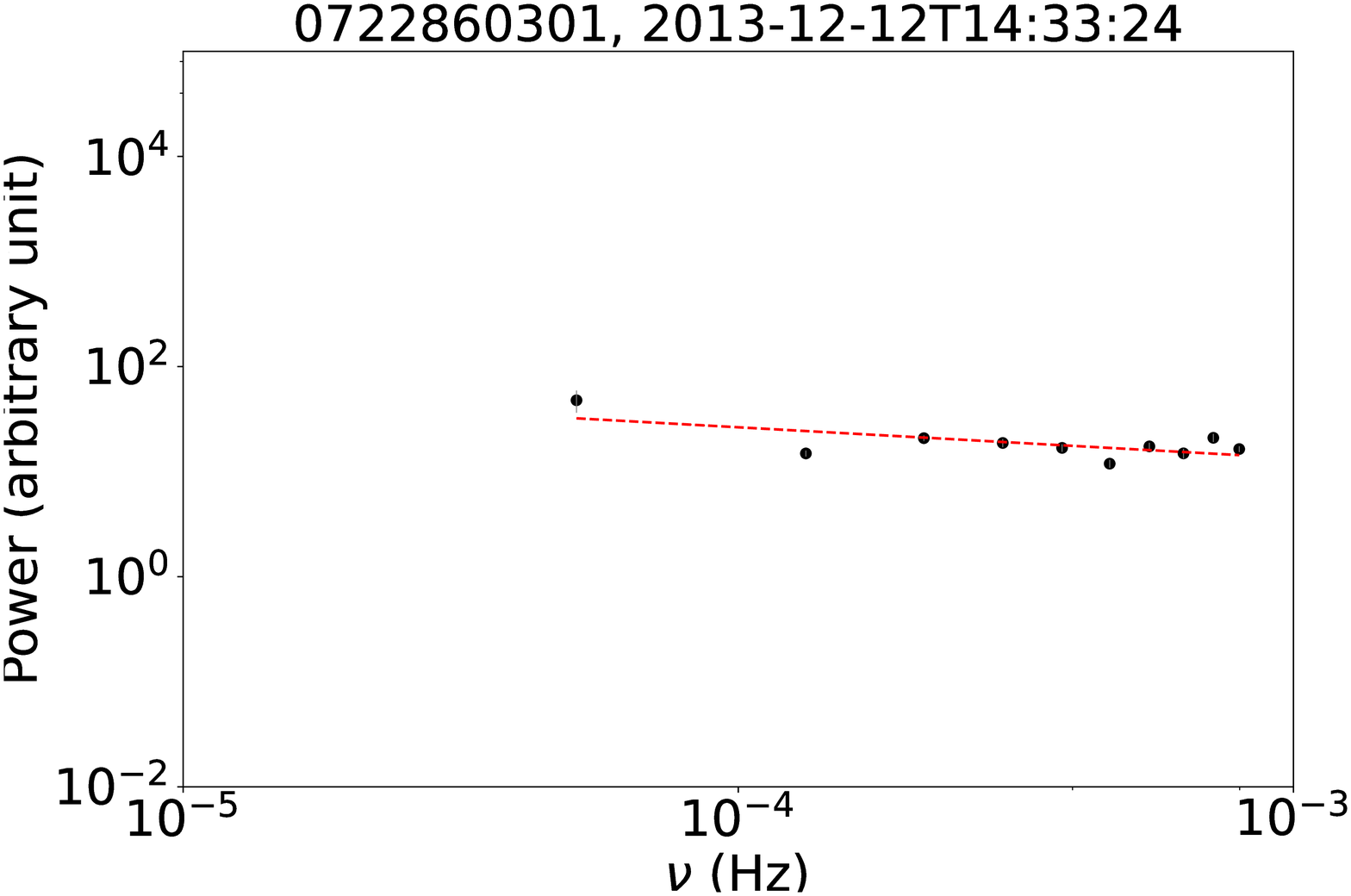}
	\end{minipage}
	\begin{minipage}{.43\textwidth} 
		\centering 
		\includegraphics[width=.9\linewidth]{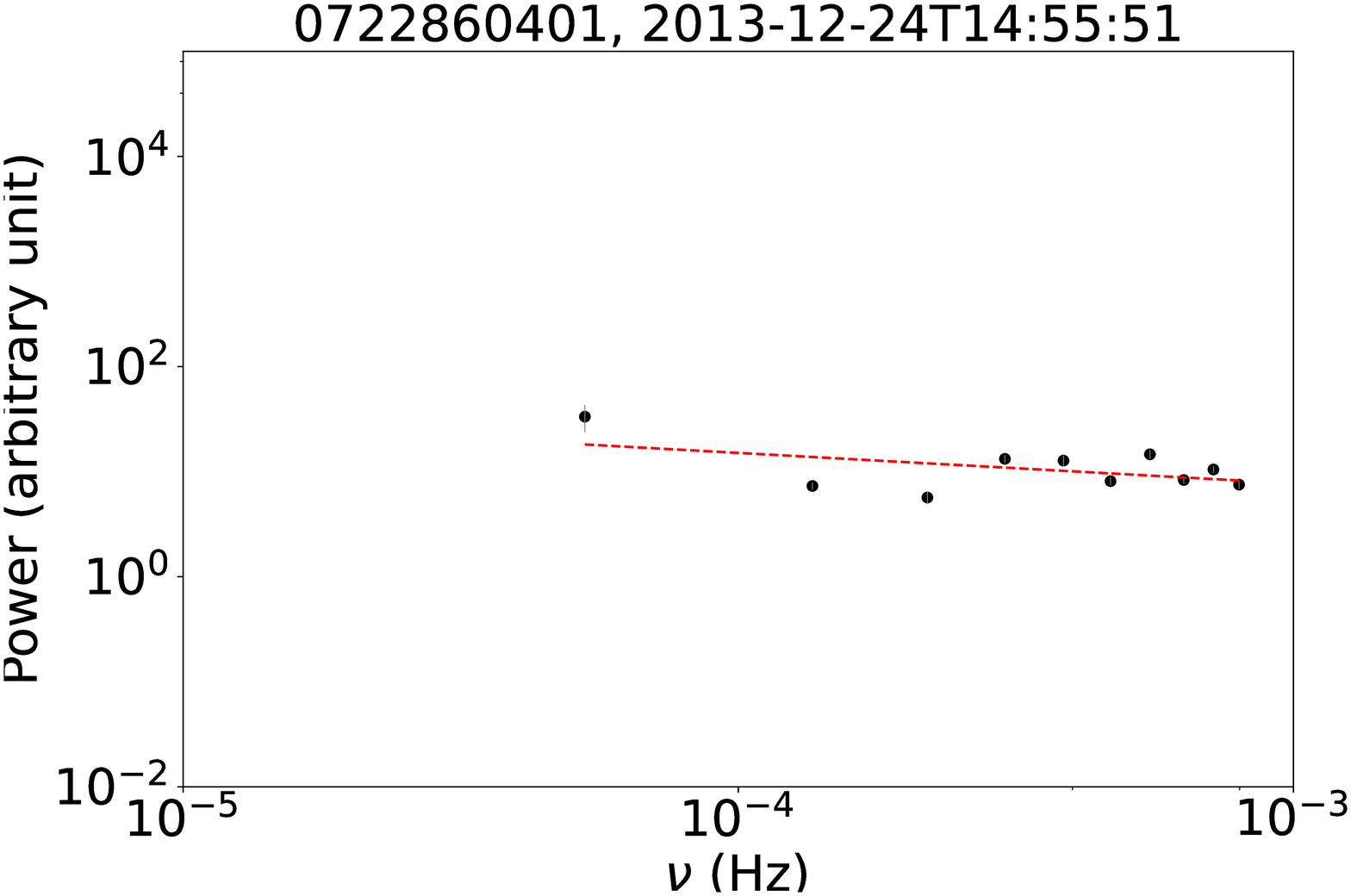}
	\end{minipage}
\end{figure*}

\begin{figure*}\label{dfp:oj}
	\centering
	\caption{DFPs of the observations of OJ 287. Arbitrary unit is the rms-normalized power ($\mathrm{(rms/mean)^2\ Hz^{-1}}$).}
	\label{fig:DFT_OJ}
	\begin{minipage}{.45\textwidth} 
		\centering 
		\includegraphics[width=.99\linewidth]{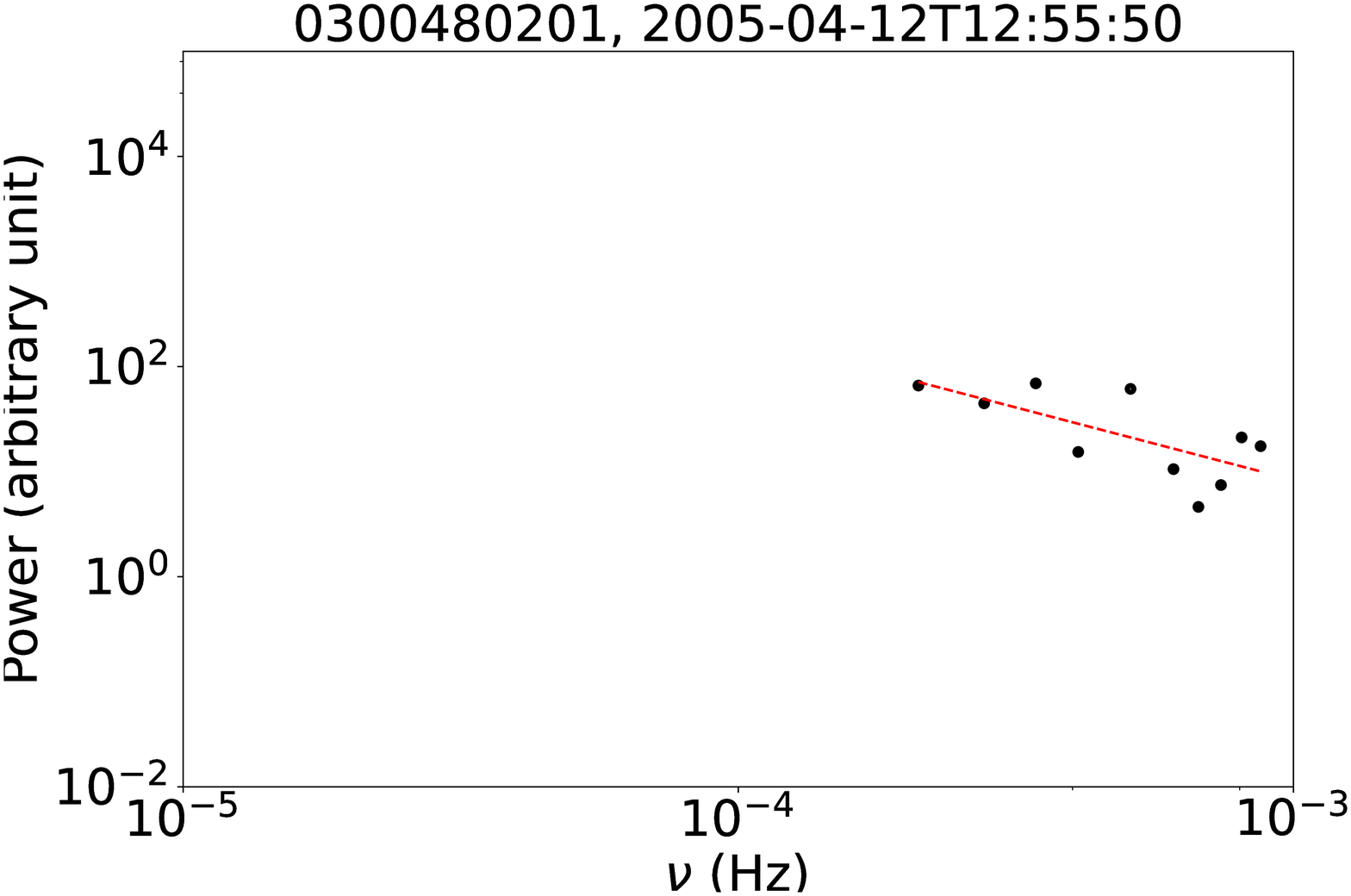}
	\end{minipage}
	\begin{minipage}{.45\textwidth} 
		\centering 
		\includegraphics[width=.99\linewidth]{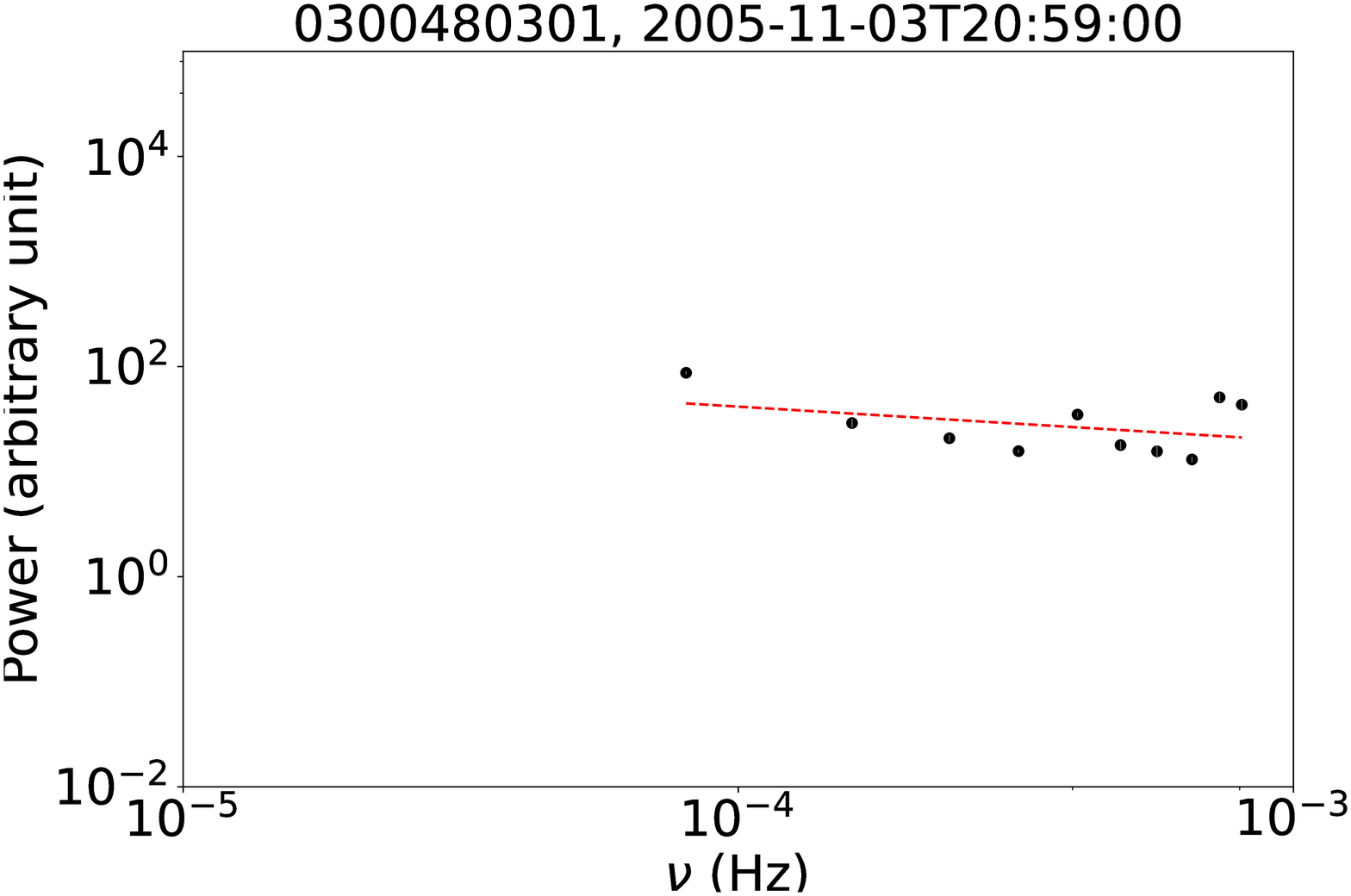}
	\end{minipage}
	\begin{minipage}{.45\textwidth} 
		\centering 
		\includegraphics[width=.99\linewidth]{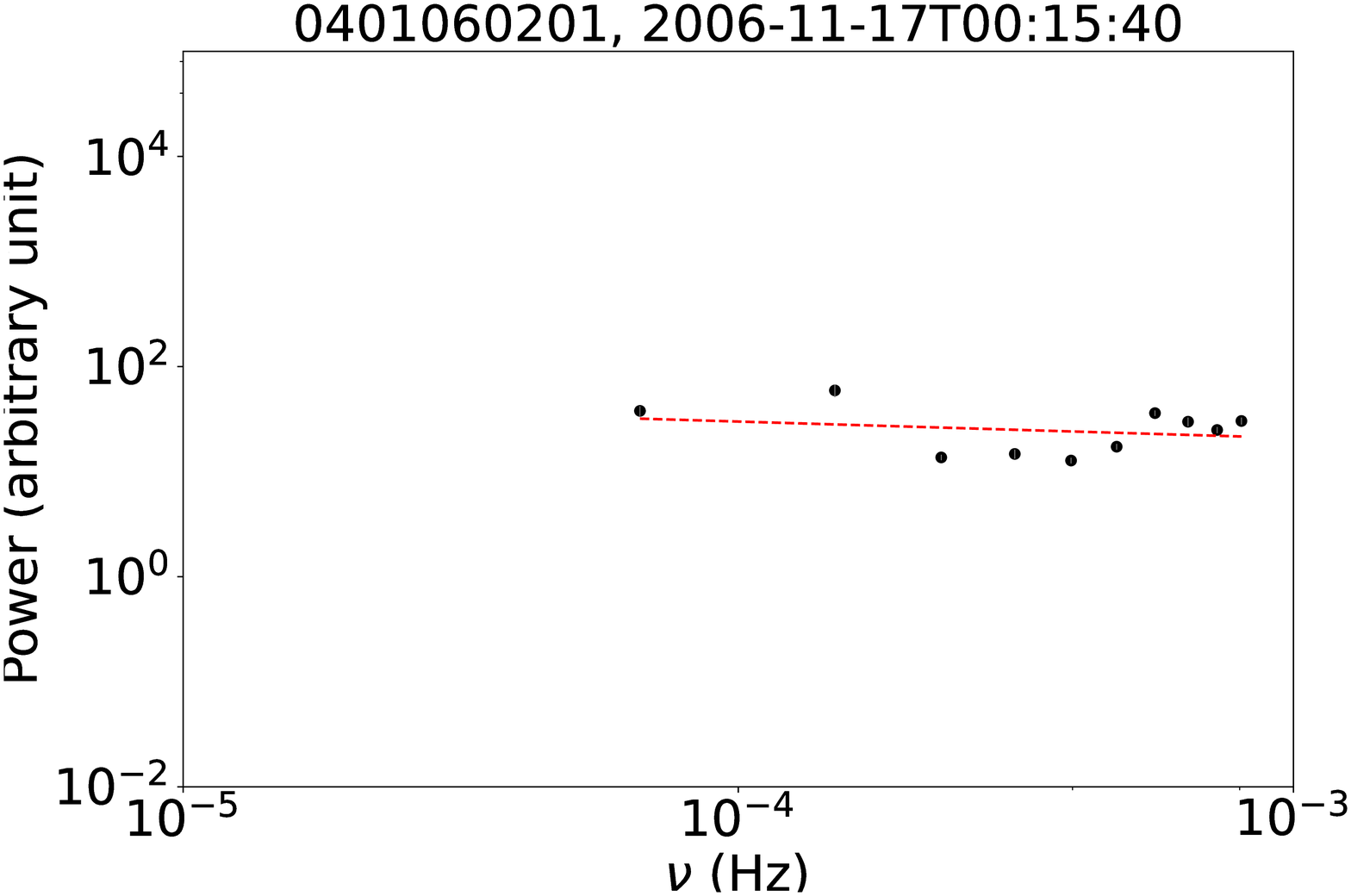}
	\end{minipage}
	\begin{minipage}{.45\textwidth} 
		\centering 
		\includegraphics[width=.99\linewidth]{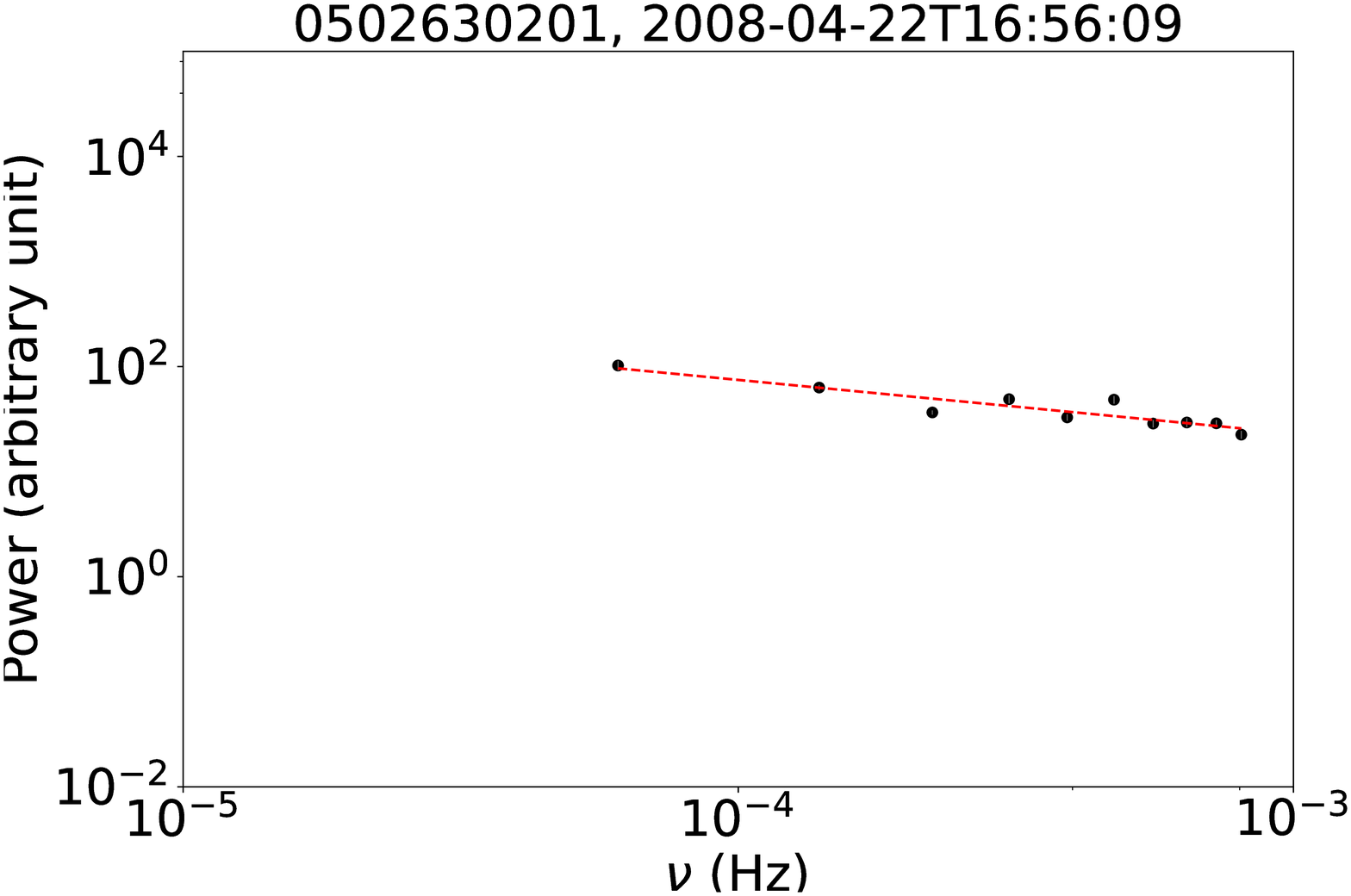}
	\end{minipage}
	\begin{minipage}{.45\textwidth} 
		\centering 
		\includegraphics[width=.99\linewidth]{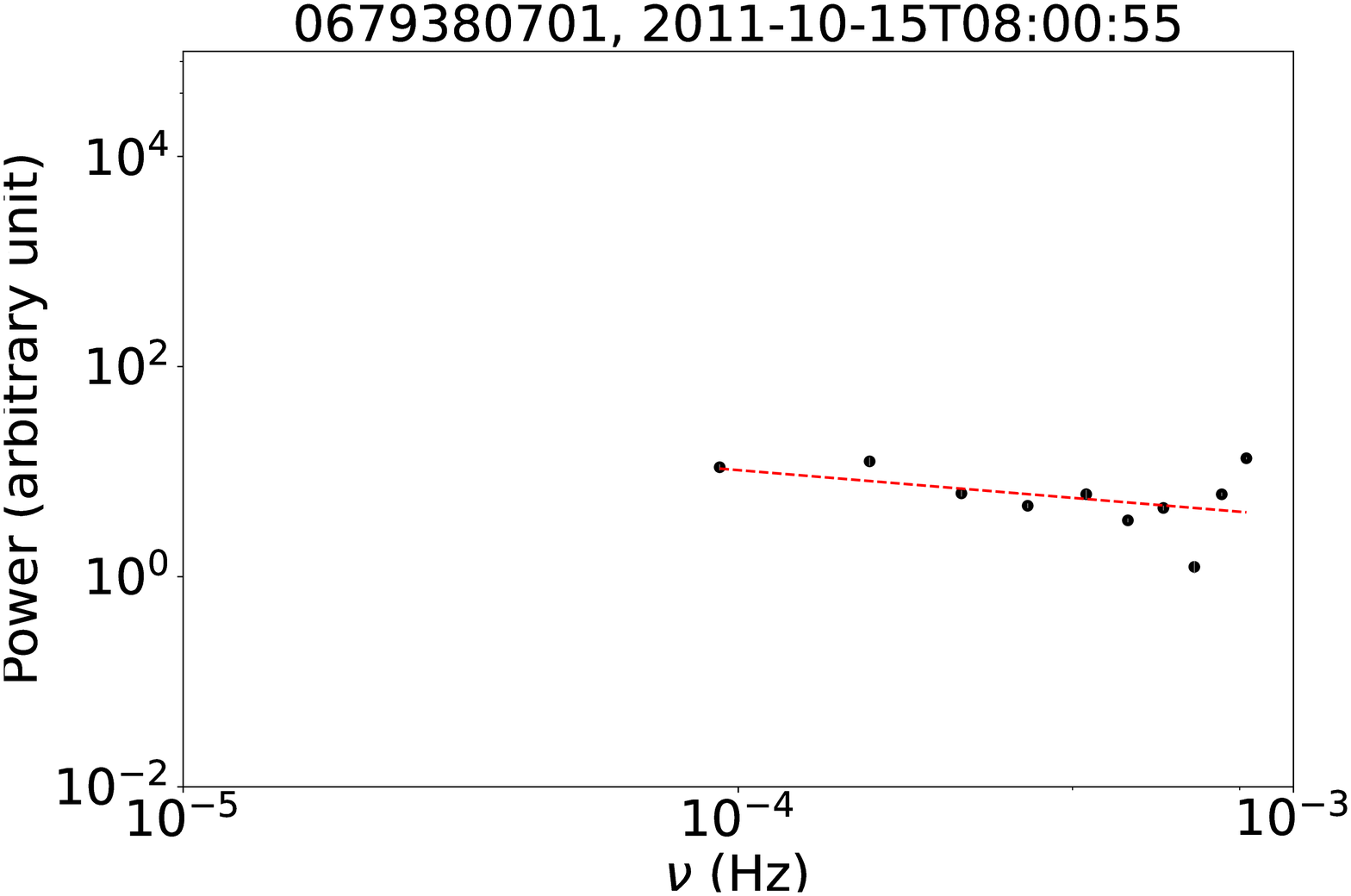}
	\end{minipage}
	\begin{minipage}{.45\textwidth} 
		\centering 
		\includegraphics[width=.99\linewidth]{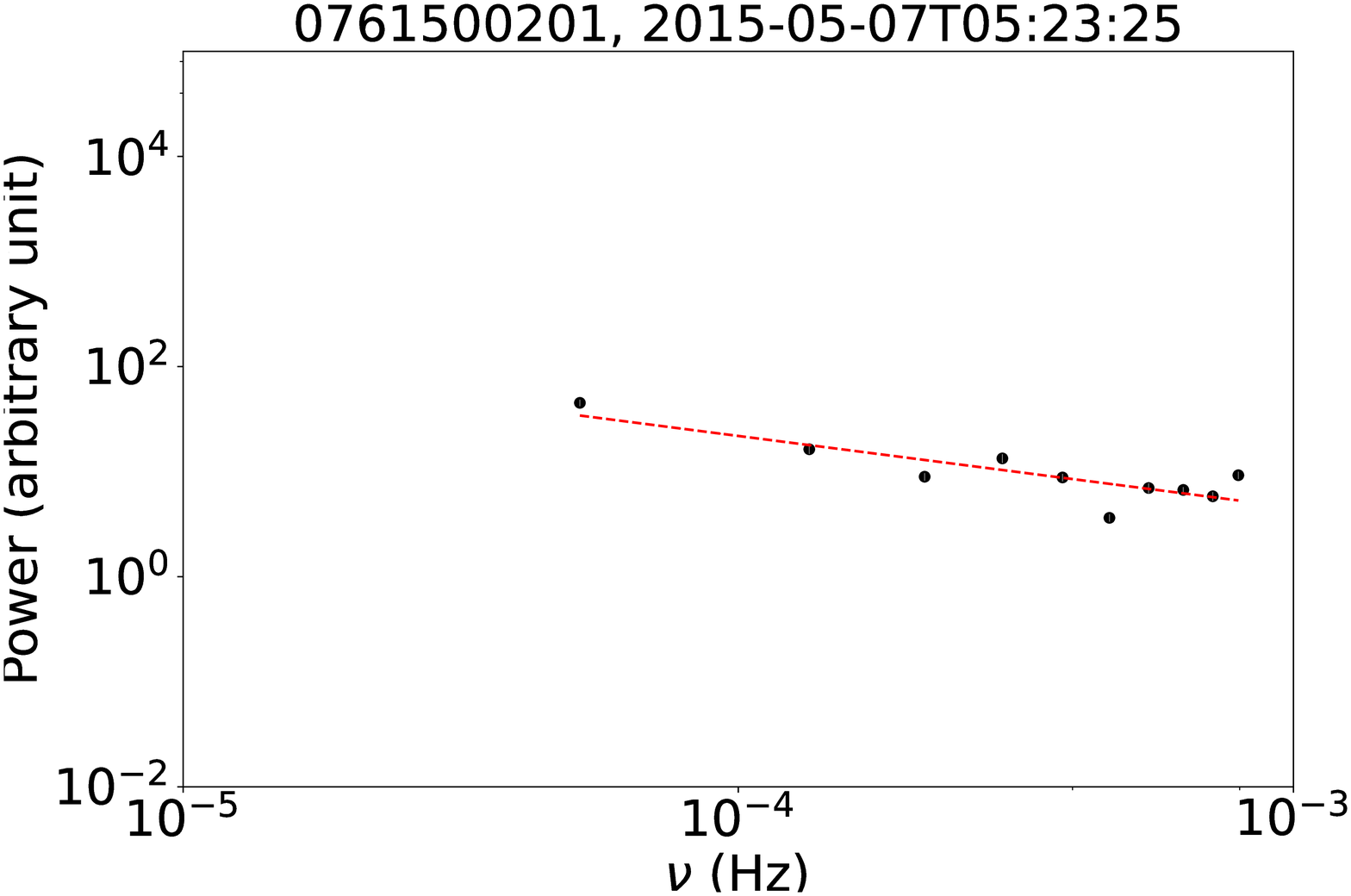}
	\end{minipage}
	\begin{minipage}{.45\textwidth} 
		\centering 
		\includegraphics[width=.99\linewidth]{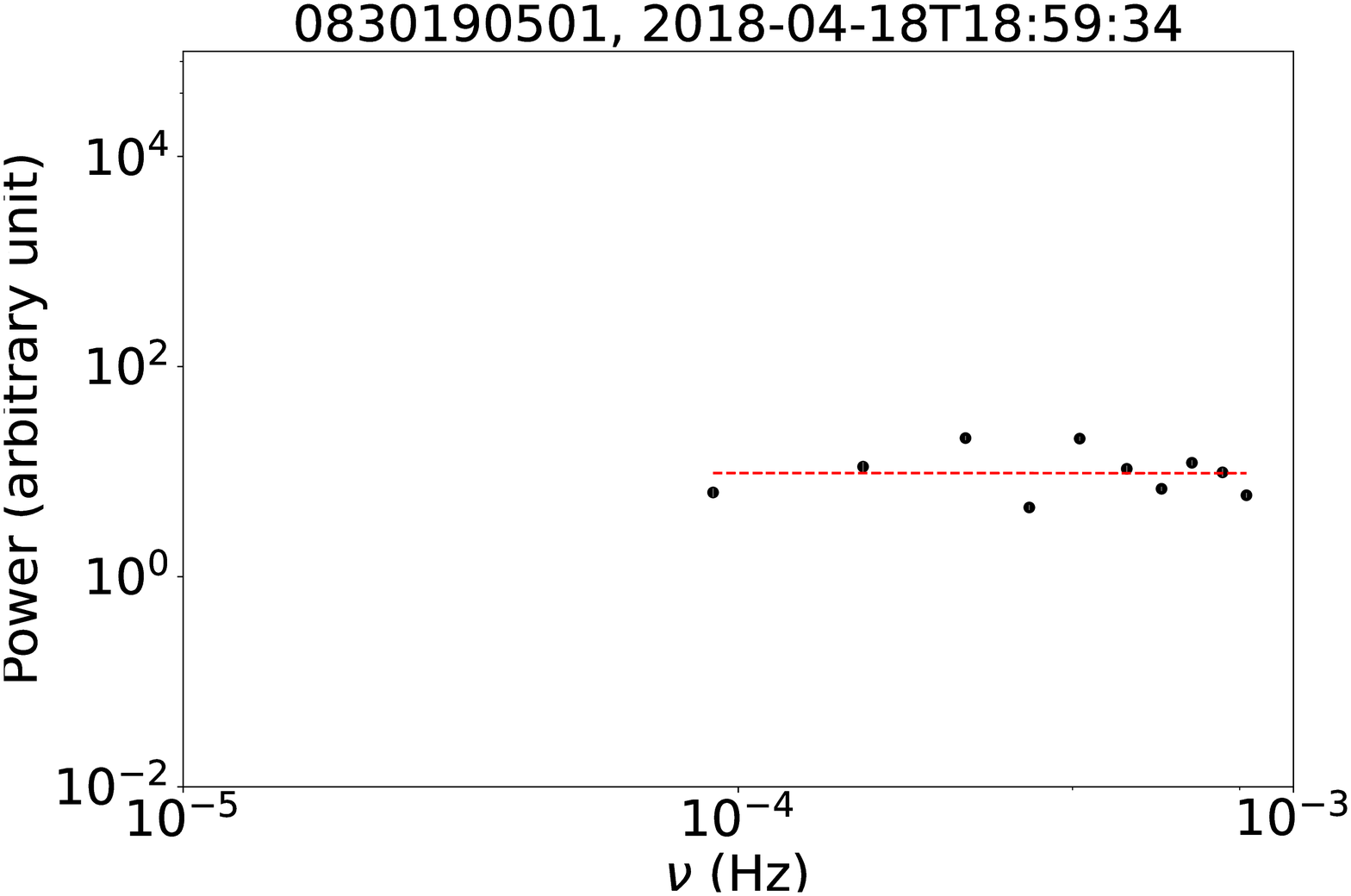}
	\end{minipage}
	\begin{minipage}{.45\textwidth} 
		\centering 
		\includegraphics[width=.99\linewidth]{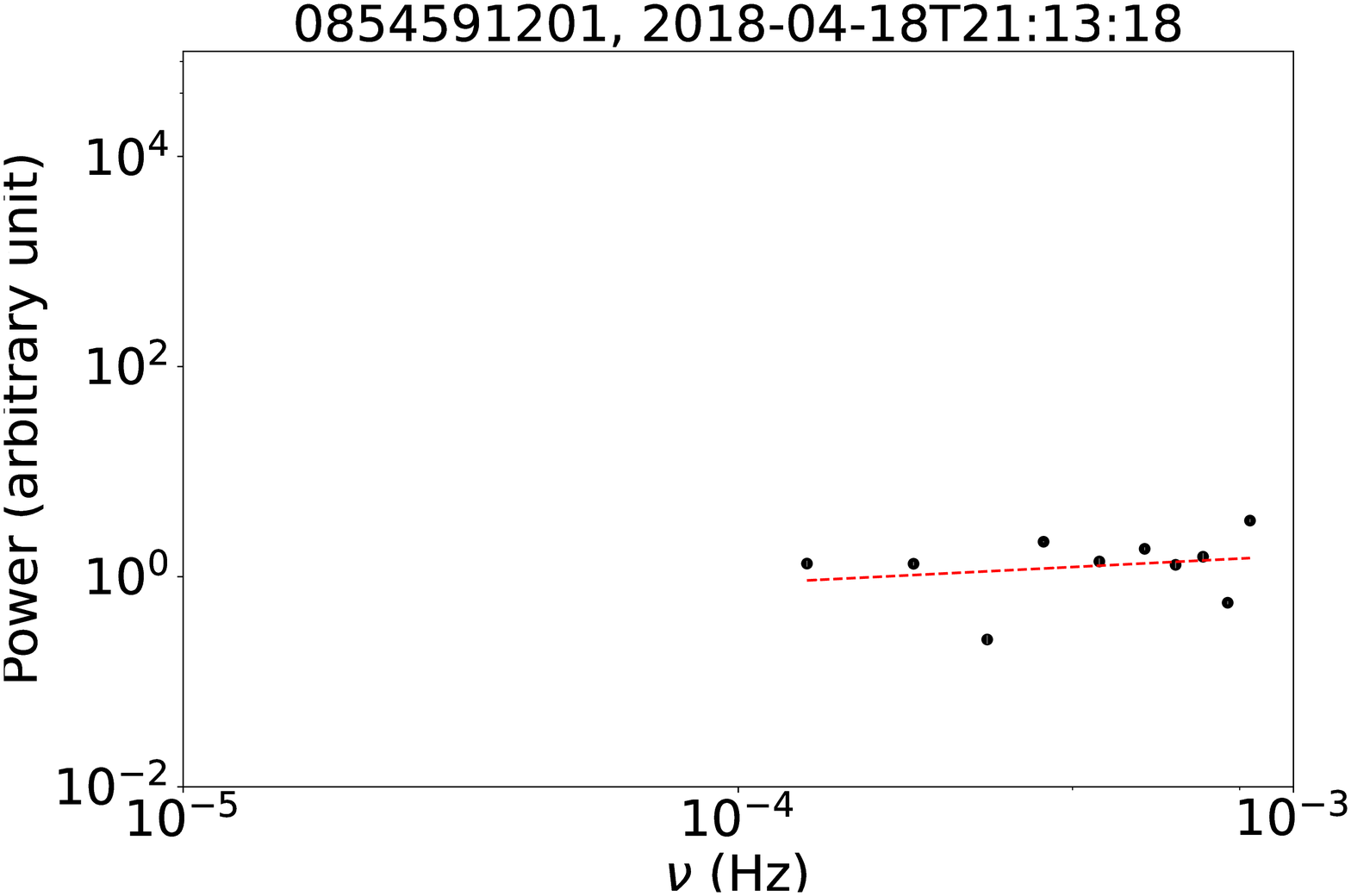}
	\end{minipage}
\end{figure*}

\begin{figure*}\label{dfp:mrk}
	\centering
	\caption{DFPs of the observations of Mrk 501. Arbitrary unit is the rms-normalized power ($\mathrm{(rms/mean)^2\ Hz^{-1}}$).}
	\label{fig:DFT_Mrk}
	\begin{minipage}{.45\textwidth} 
		\centering 
		\includegraphics[width=.99\linewidth]{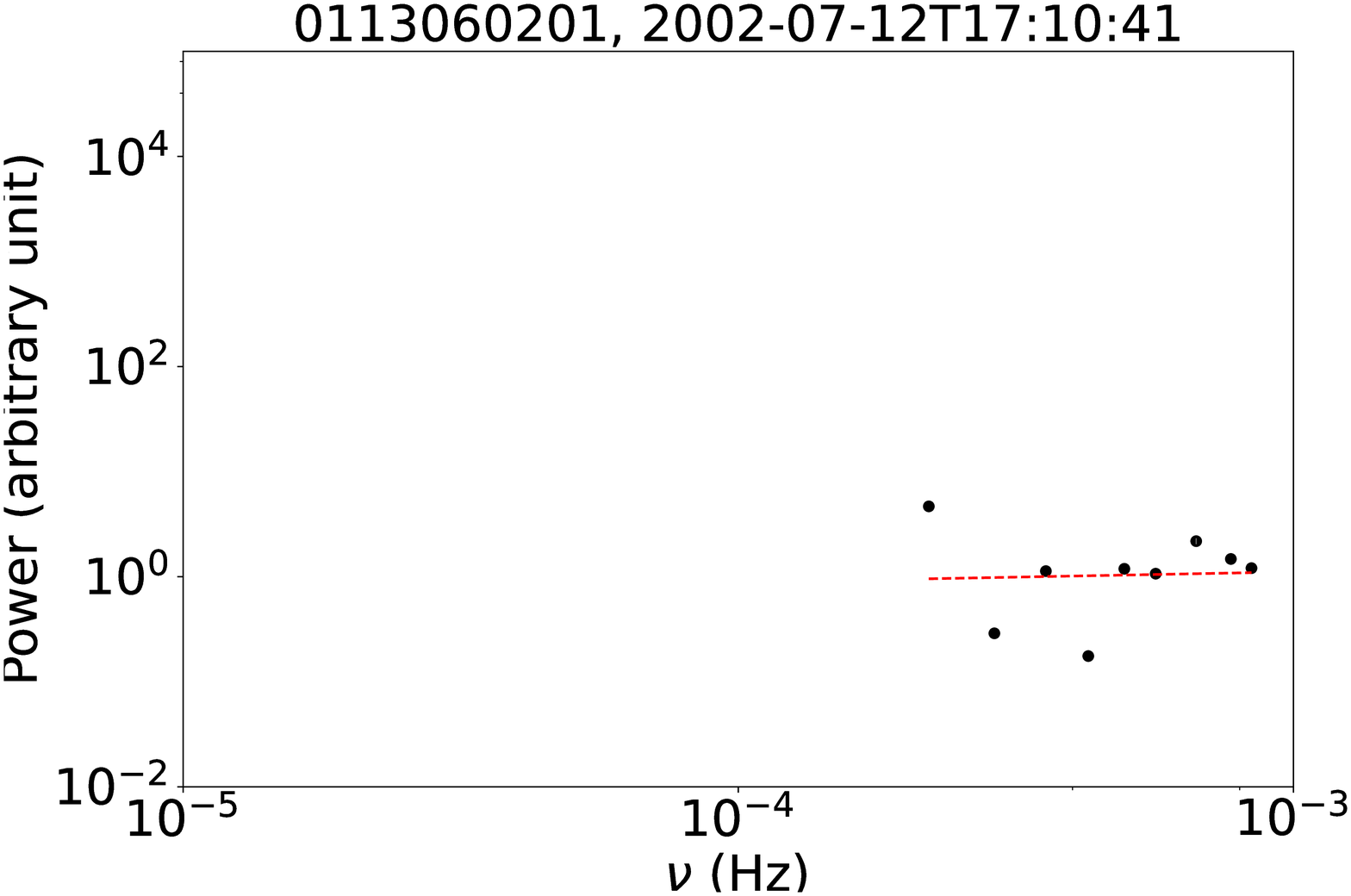}
	\end{minipage}
	\begin{minipage}{.45\textwidth} 
		\centering 
		\includegraphics[width=.99\linewidth]{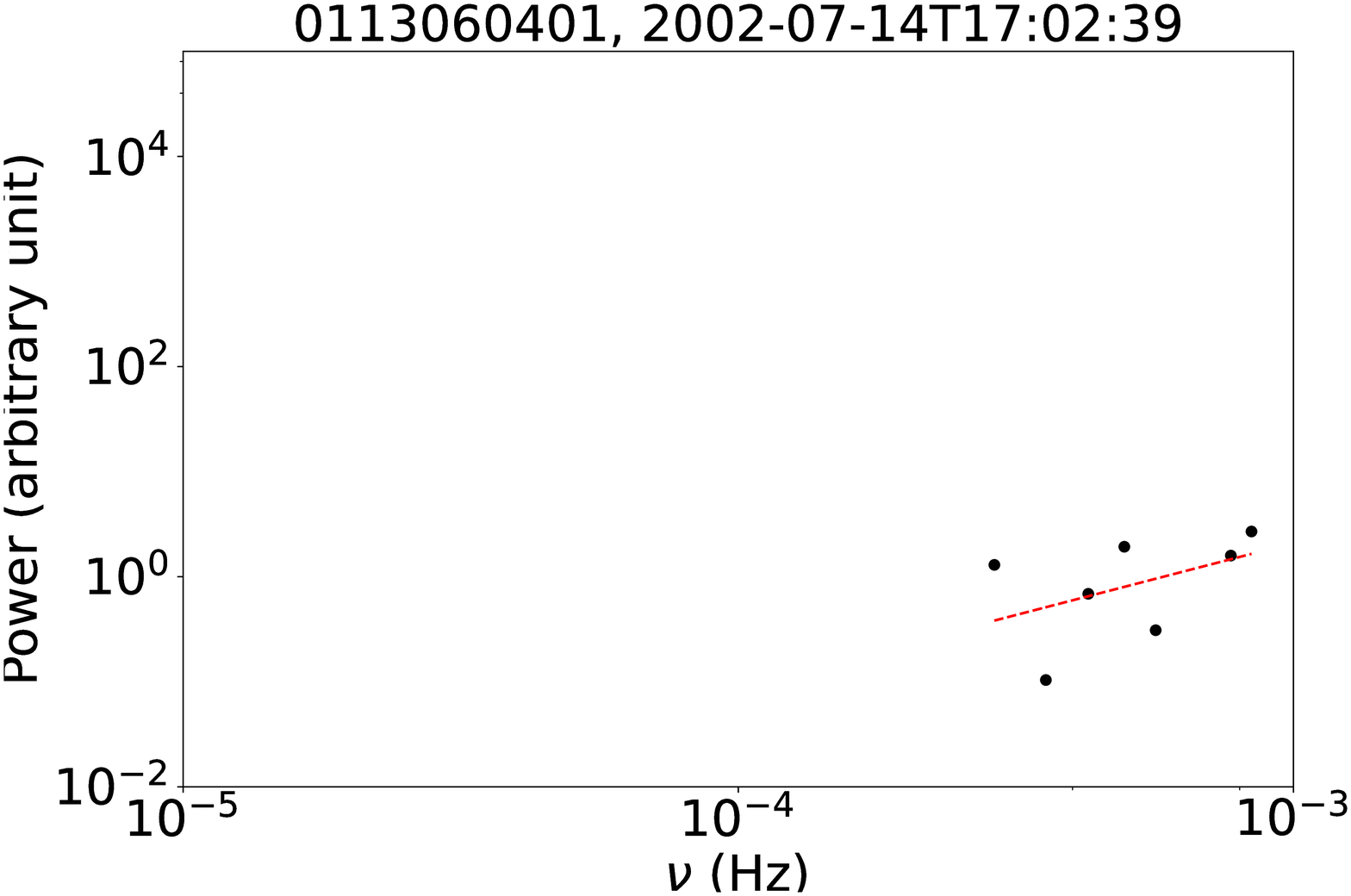}
	\end{minipage}
	\begin{minipage}{.45\textwidth} 
		\centering 
		\includegraphics[width=.99\linewidth]{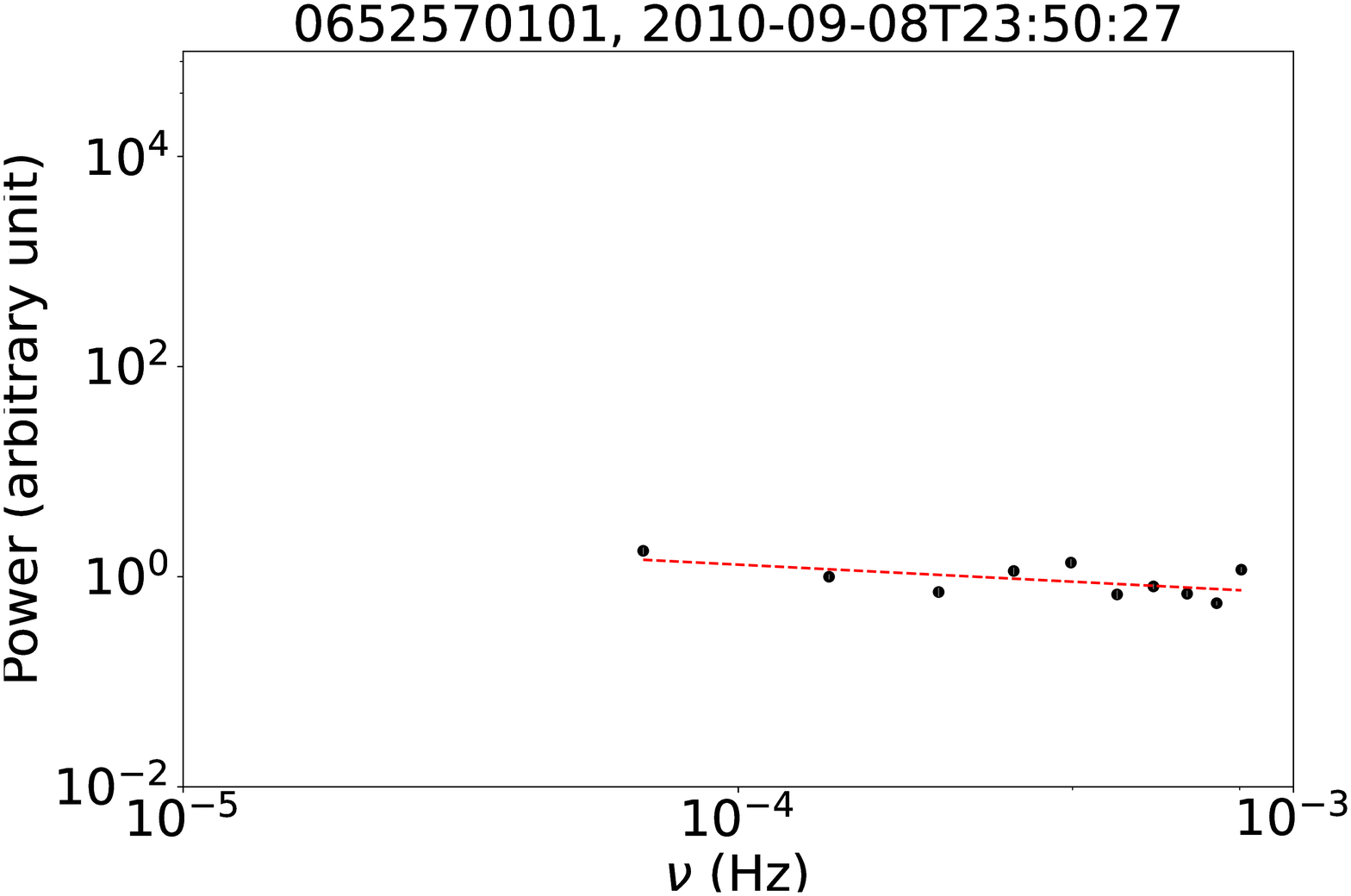}
	\end{minipage}
	\begin{minipage}{.45\textwidth} 
		\centering 
		\includegraphics[width=.99\linewidth]{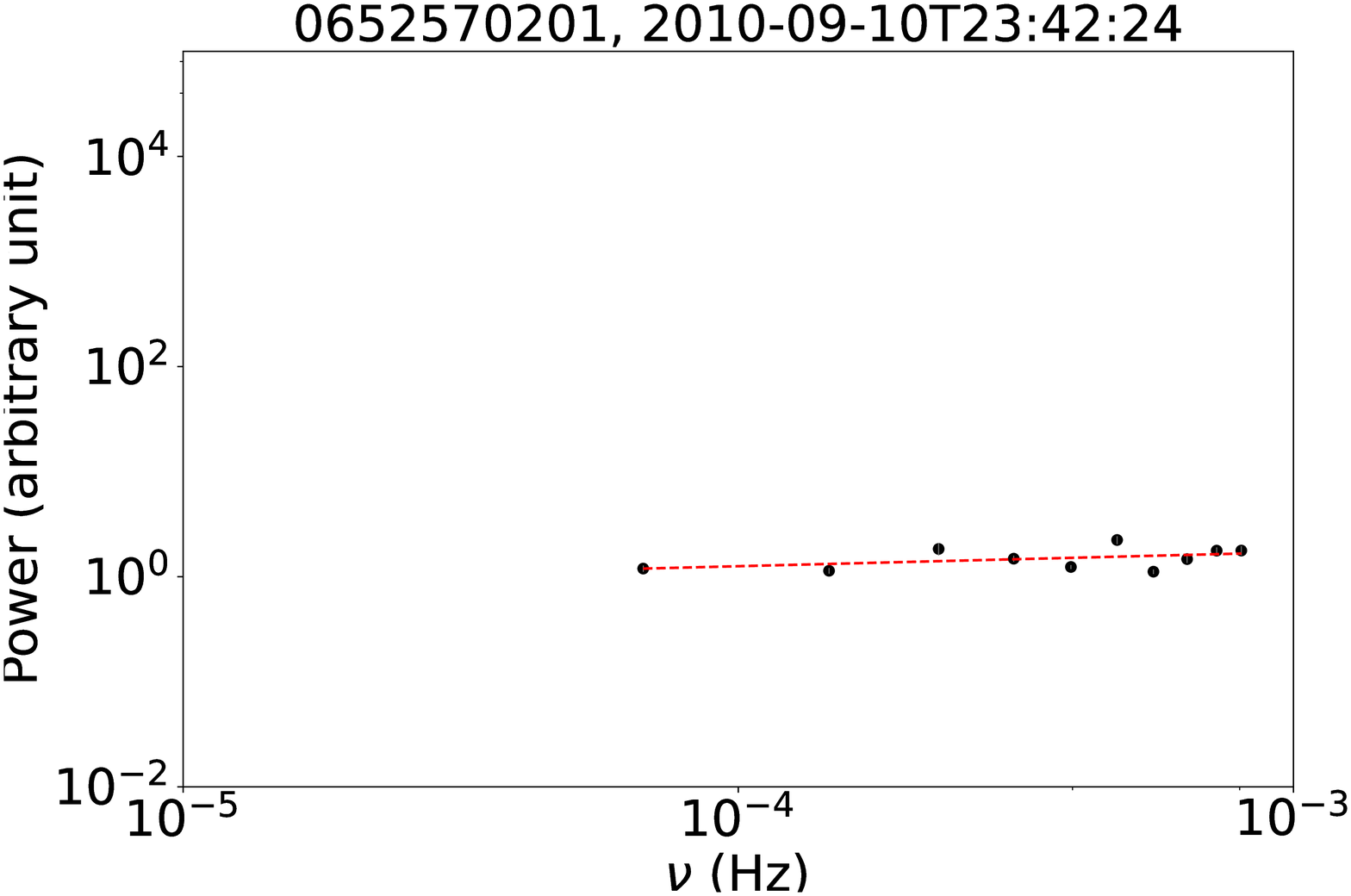}
	\end{minipage}
	\begin{minipage}{.45\textwidth} 
		\centering 
		\includegraphics[width=.99\linewidth]{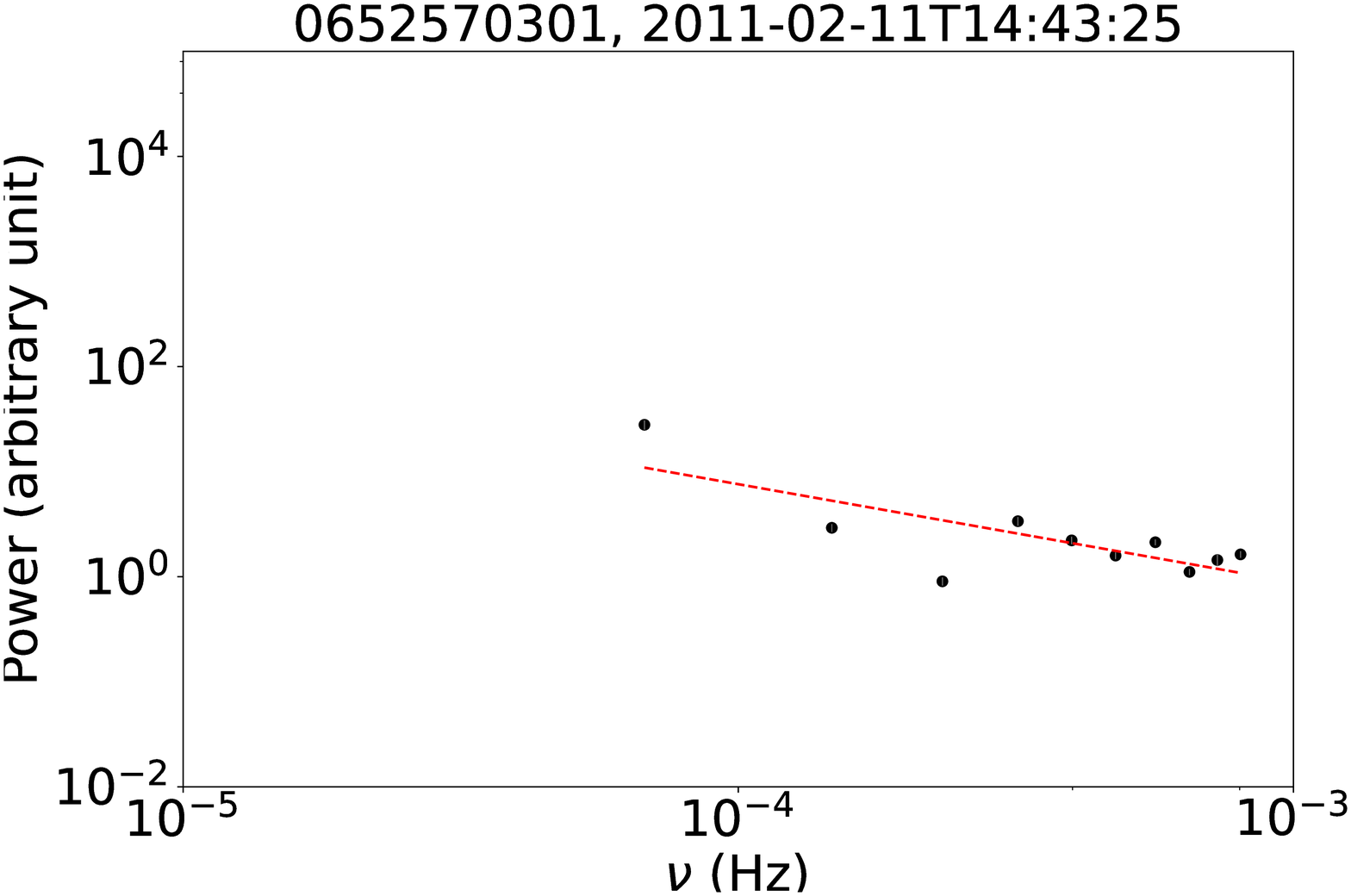}
	\end{minipage}
	\begin{minipage}{.45\textwidth} 
		\centering 
		\includegraphics[width=.99\linewidth]{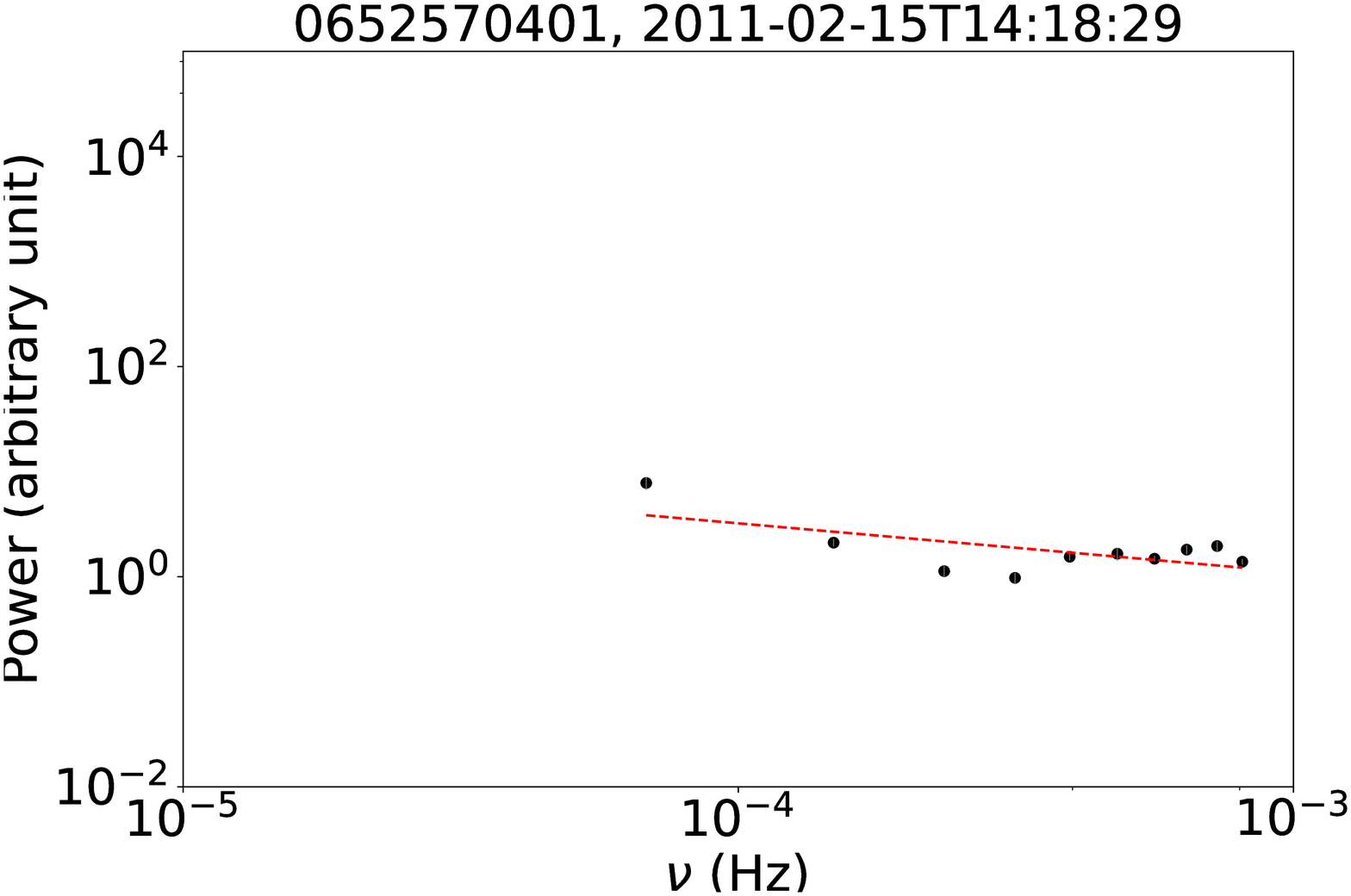}
	\end{minipage}
\end{figure*}

\begin{figure*}\label{dfp:s5}
	\centering
	\caption{DFPs of the observations of S5 0716+714. Arbitrary unit is the rms-normalized power ($\mathrm{(rms/mean)^2\ Hz^{-1}}$).}
	\label{fig:DFT_S5}
	\begin{minipage}{.45\textwidth} 
		\centering 
		\includegraphics[width=.99\linewidth]{PlotsPersonal/0150495601_DFT.eps}
	\end{minipage}
	\begin{minipage}{.45\textwidth} 
		\centering 
		\includegraphics[width=.99\linewidth]{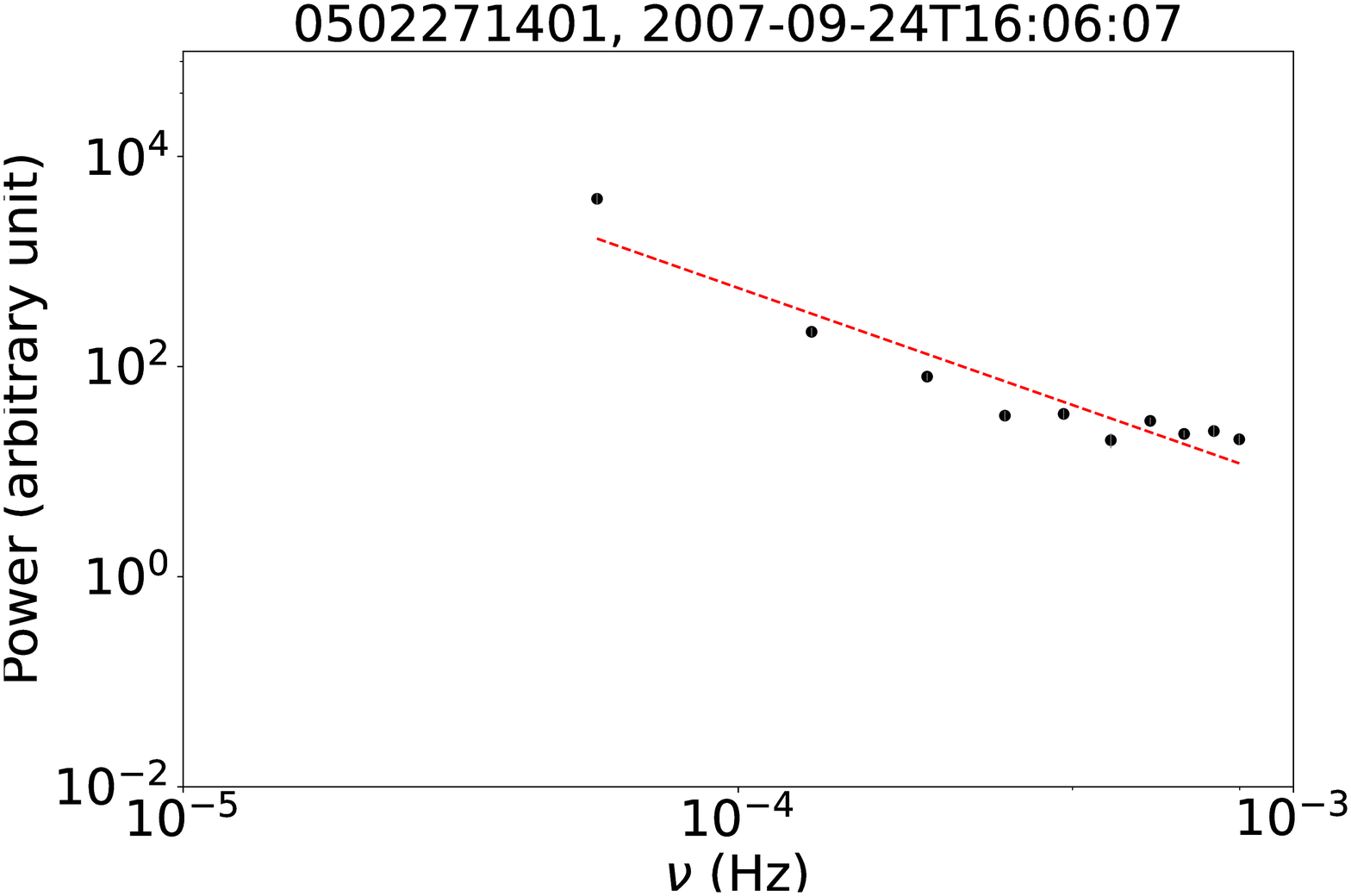}
	\end{minipage}
\end{figure*}

\bsp
\label{lastpage}
\end{document}